\documentclass{aa}
\usepackage{amsmath}
\usepackage{amssymb}
\usepackage[varg]{txfonts}
\usepackage{epsfig,graphicx}
\usepackage[colorlinks=true, linkcolor=blue, citecolor=blue, urlcolor=blue]{hyperref}  
\usepackage{natbib}
\bibpunct{(}{)}{;}{a}{}{,} 
\newcommand{\ser}{S\'ersic}
\usepackage{color}

\defcitealias{2015MNRAS.451.1728D}{DG15}
\defcitealias{1997A&A...325..135X}{X97}
\defcitealias{1999A&A...344..868X}{X99}
\defcitealias{2007A&A...471..765B}{B07}
\defcitealias{2016A&A...592A..71M}{M16}
\begin{document}

\title{\emph{HERschel} Observations of Edge-on Spirals (\textit{HER}OES)}
\subtitle{IV. Dust energy balance problem\thanks{{\it Herschel} is an ESA space observatory with science instruments provided by European-led Principal Investigator consortia and with important participation from NASA.} }

\titlerunning{Dust energy balance problem in the {\it HER}OES galaxies}

\author{
Aleksandr~V.~Mosenkov\inst{1,2,3}
\and Flor~Allaert\inst{1}
\and Maarten~Baes\inst{1}
\and Simone~Bianchi\inst{4}
\and Peter~Camps\inst{1}
\and Christopher~J.R.~Clark\inst{5}
\and Marjorie~Decleir\inst{1}
\and Gert~De Geyter\inst{1}
\and Ilse~De Looze\inst{6,1}
\and Jacopo~Fritz\inst{7}
\and Gianfranco~Gentile\inst{8}
\and Benne W. Holwerda\inst{9}
\and Thomas~M.~Hughes\inst{10,11,12,13}
\and Fraser~Lewis\inst{14,15}
\and Matthew~W.L.~Smith\inst{5}
\and Joris~Verstappen\inst{16}
\and Sam~Verstocken\inst{1}
\and S\'ebastien~Viaene\inst{1,17}}
\institute{Sterrenkundig Observatorium, Universiteit Gent, Krijgslaan 281, B-9000 Gent, Belgium\\  
\email{mosenkovAV@gmail.com}
\and
St.Petersburg State University, Universitetskij pr. 28, 198504, St. Petersburg, Stary Peterhof, Russia
\and
Central Astronomical Observatory of RAS, Pulkovskoye chaussee 65/1, 196140, St. Petersburg, Russia
\and INAF, Osservatorio Astrofisico di Arcetri, Largo E. Fermi 5,I-50125, Florence, Italy
\and School of Physics \& Astronomy, Cardiff University, Queen’s Buildings, The Parade, Cardiff, CF24 3AA, UK
\and Department of Physics and Astronomy, University College London, Gower Street, London WC1E 6BT, UK
\and Instituto de Radioastronom\'{i}a y Astrof\'{i}sica, CRyA, UNAM, Campus Morelia, A.P. 3-72, C.P. 58089, Michoac\'{a}n, Mexico 
\and Department of Physics and Astrophysics, Vrije Universiteit Brussel, Pleinlaan 2, 1050 Brussels, Belgium
\and University of Louisville, Department of Physics and Astronomy, 102 Natural Sciences Building, Louisville KY 40292, USA
\and Instituto de F\'{i}sica y Astronom\'{i}a, Universidad de Valpara\'{i}so, Avda. Gran Breta\~{n}a 1111, Valpara\'{i}so, Chile
\and CAS Key Laboratory for Research in Galaxies and Cosmology, Department of Astronomy, University of Science and Technology of China, Hefei 230026, China
\and School of Astronomy and Space Science, University of Science and Technology of China, Hefei 230026, China
\and Chinese Academy of Sciences South America Center for Astronomy, China-Chile Joint Center for Astronomy, Camino El Observatorio \#1515, Las Condes, Santiago, Chile
\and Faulkes Telescope Project, Cardiff University, The Parade, Cardiff CF24 3AA, Cardiff, Wales
\and Astrophysics Research Institute, Liverpool John Moores University, IC2, Liverpool Science Park, 146 Brownlow Hill, Liverpool L3 5RF, UK
\and Kapteyn Astronomical Institute, University of Groningen, Landleven 12, Groningen NL-9747AD, The Netherlands
\and Centre for Astrophysics Research, University of Hertfordshire, College Lane, Hatfield, AL10 9AB, UK
}
\date{}
\abstract{We present results of the detailed dust energy balance study for the seven large edge-on galaxies in the {\it HER}OES sample using 3D radiative transfer (RT) modelling. Based on available optical and near-infrared observations of the {\it HER}OES galaxies, we derive the 3D distribution of stars and dust in these galaxies. For the sake of uniformity, we apply the same technique to retrieve galaxy properties for the entire sample: we use a stellar model consisting of a \ser\ bulge and three double-exponential discs (a superthin disc for a young stellar population and thin and thick discs for old populations). For the dust component, we adopt a double-exponential disc with the new THEMIS dust-grain model. We fit oligochromatic radiative transfer (RT) models to the optical and near-infrared images with the fitting algorithm \textsc{fitskirt} and do panchromatic simulations with the \textsc{skirt} code at wavelengths ranging from ultraviolet to submillimeter. We confirm the previously stated dust energy balance problem in galaxies: for the {\it HER}OES galaxies, the dust emission derived from our RT calculations underestimates the real observations by a factor 1.5--4 for all galaxies except NGC\,973 and NGC\,5907 (apparently, the latter galaxy has a more complex geometry than we used). The comparison between our RT simulations and the observations at mid-infrared--submillimeter wavelengths shows that most of our galaxies exhibit complex dust morphologies (possible spiral arms, star-forming regions, more extended dust structure in the radial and vertical directions). We suggest that, in agreement with the results from Saftly et al. (2015), the large- and small-scale structure is the most probable explanation for the dust energy balance problem. 
}


\maketitle

\section{Introduction}
\label{sec:intro}
Cosmic dust severely obscures astronomical objects at ultraviolet and optical wavelengths and, thus, impedes our study of these objects in this waverange. However, with the beginning of the era of the infrared (IR) astronomy, dust began to play an important role in the study of astrophysical processes, taking place in the interstellar medium (ISM), and galaxy evolution.   

According to multiple studies \citep[see e.g.][and references therein]{2002MNRAS.335L..41P, 2016A&A...586A..13V}, approximately one third of the bolometric luminosity in normal spiral galaxies is attenuated by dust: dust reshapes the spectral energy distribution (SED) of galaxies by absorbing and scattering light at short wavelengths and re-emitting at longer wavelengths. Therefore, a dust component, being mainly concentrated in the galactic plane, in optical bands usually appears as a dimmed lane from the edge-on view (e.g. an outstanding example is our Milky Way Galaxy). This component attenuates galactic starlight which passes through it, and, in turn, looks like a very thin emitting disc at far-infrared-submillimeter (FIR-submm) wavelengths.   

Until recently, the poor resolution and limited wavelength coverage of infrared instruments was an obstacle to observe dust in even nearby galaxies. With the launch of the \textit{Herschel} Space Observatory \citep{2010A&A...518L...1P}, we are now able to observe the entire dust peak, in the 70 to 500~$\mu$m wavelength range (i.e. tracing both warm and cold dust). The high spatial resolution achieved by \textit{Herschel} allows us to examine the dust in higher detail than ever possible before.

Panchromatic radiative transfer modelling of galaxies provides a powerful tool to analyse different characteristics of dust in galaxies (i.e. optical properties, dust distribution, clumpiness, etc.) in a self-consistent way \citep[see e.g.][]{2000A&A...362..138P, 2008A&A...490..461B, 2010A&A...518L..39B, 2011A&A...527A.109P, 2011ApJ...741....6M, 2012MNRAS.419..895D, 2012A&A...541L...5H, 2016A&A...592A..71M}. The following method has become a standard practice to perform panchromatic dust energy balance studies. From optical observations, the properties and spatial distribution of stars and dust can be constrained using a radiative transfer code to model the propagation of stellar light and its interaction with dust particles in a galaxy. In its turn, the dust emission predicted from the radiative transfer simulations (based on the obtained RT model) can be compared to the observed thermal dust re-emission at IR-submm wavelengths. Such a complementary study imposes that dust features can easily be identified from optical as well as infrared observations. This requirement has limited the number of galaxies for which detailed dust energy balance studies have been attempted in the past. 
Edge-on spirals are considered ideal cases for studying since projection effects allow to resolve the distribution of different stellar populations \citep[see e.g.][]{1981A&A....95..116V,1997AJ....113.2061M,2002AJ....124.1328D,2005Ap.....48..221T,2005AJ....130.1574S,2018A&A...610A...5C} and dust vertically \citep[see e.g.][]{1987ApJ...317..637K,1999A&A...344..868X,2001A&A...372..775M, 2004A&A...425..109A, 2008A&A...490..461B, 2010A&A...518L..39B, 2011A&A...527A.109P, 2012MNRAS.419..895D, 2016A&A...592A..71M}. Thus, using such a multiwavelength technique, a complex 3D model of an edge-on galaxy can be built, which can include both stellar and dust constituents.

Interestingly, multiple dust energy balance studies of individual edge-on galaxies reveal an inconsistency between the predicted FIR/submm fluxes of their radiative transfer models and the observed emission in those wavebands (e.g. \citealt{2000A&A...362..138P, 2001A&A...372..775M, 2004A&A...425..109A, 2005A&A...437..447D, 2010A&A...518L..39B, 2011A&A...527A.109P, 2012MNRAS.419..895D, 2015MNRAS.451.1728D, 2016A&A...592A..71M}). Although radiative transfer models might successfully explain the observed optical attenuation, the modelled dust emission underestimates the observed thermal dust re-emission by a factor of 3--4.
In order to reconcile the results of the radiative transfer models with the observations, two scenarios have been proposed: either a significant underestimation of the FIR/submm dust emissivity has been argued \citep[see e.g.][]{2004A&A...425..109A, 2005A&A...437..447D, 2011ApJ...741....6M} or, alternatively, the distribution of a sizeable fraction of dust in clumps or a second inner dust disc having a negligible attenuation on the bulk of the starlight \citep[see e.g.][]{2000A&A...362..138P, 2001A&A...372..775M, 2008A&A...490..461B, 2011A&A...531L..11B, 2011ApJ...741....6M, 2012MNRAS.419..895D, 2012A&A...541L...5H}. 

Recently, \citet[][hereafter \citetalias{2015MNRAS.451.1728D}]{2015MNRAS.451.1728D} studied the dust energy balance in two edge-on galaxies with available {\it Herschel} observations: IC\,4225 and NGC\,5166. Using an earlier obtained oligochromatic\footnote{modelling based on a small number of images simultaneously.} model (using optical data), and adding a young stellar disc to match the ultraviolet (UV) fluxes, they concluded that for NGC\,5166 this additional component is necessary in order to reproduce the observed emission at longer wavelengths. However, for the other galaxy, IC\,4225, the modelled far-infrared emission still underestimates the observed fluxes by a factor of three. The proposed reasons for this discrepancy in IC\,4225 might include 1) that it is too small to properly fit the dust disc parameters, and 2) there is emission from obscured star-forming regions which are embedded in dense dust clouds and, thus, do not contribute noticeably to the observed UV flux but have a clear impact on the FIR emission \citep[see e.g.][]{2012MNRAS.419..895D, 2014A&A...571A..69D, 2015MNRAS.451.1728D}. It is likely that there is no single origin for the dust energy balance problem, and several mechanisms can be responsible for it. 

Several years ago, the {\it HER}OES project \citep{2013A&A...556A..54V} was initiated to study a set of seven edge-on spiral galaxies, with the aim to present a detailed, systematic and homogeneous study of the amount, spatial distribution and properties of the interstellar dust in these seven galaxies, and investigate the link between the dust component and stellar, gas and dark matter components. These galaxies are well-resolved (with the optical diameter $D>4$\,arcmin), and have {\it Herschel} observations in five PACS \citep{2010A&A...518L...2P} and SPIRE \citep{2010A&A...518L...3G} bands at 100, 160, 250, 350 and 500~$\mu$m. Also, these galaxies were selected to have a clear and regular dust lane, and they have already been fitted with a radiative transfer code based on their optical and near-infrared (NIR) data \citep{1997A&A...325..135X,1999A&A...344..868X,2007A&A...471..765B}, therefore an indirect comparison with the previous models can be done. However, we should notice that the used assumptions and fitting strategies in the mentioned studies are different, and, therefore, a uniform approach is necessary to compare the results of RT modelling for all {\it HER}OES galaxies. 

\citet[][hereafter \citetalias{2016A&A...592A..71M}]{2016A&A...592A..71M} studied the dust energy balance problem in one of the \textit{HER}OES galaxies IC\,2531. For the first time, we used a novel approach for a simultaneous fitting of a set of optical and near-infrared images (7 bands in total) to retrieve a dust model. In it, we showed that the dust properties of IC\,2531 are better constrained if near-infrared imaging is used, in addition to optical data. Based on our oligochromatic fit, we constructed a panchromatic\footnote{modelling based on a wide range of wavelengths simultaneously.} radiative transfer model for the stars and dust in IC\,2531, ranging from UV to submm wavelengths. Following \citetalias{2015MNRAS.451.1728D}, we added another disc component representing a young stellar population to match the \textit{FUV} flux. Upon comparison of our panchromatic simulations with the {\it Herschel} observations at FIR--submm wavelengths, we demonstrated an apparent excess of observed dust emission at those wavelengths for two different dust-grain models: the BARE-GR-S model \citep{2004ApJS..152..211Z} and the THEMIS model \citep{2017arXiv170300775J}. The BARE-GR-S model consists of a mixture of polycyclic aromatic hydrocarbons, graphite and silicate grains, while the new THEMIS model is composed of amorphous carbon and amorphous silicates. We showed that the THEMIS model better reproduces the observed emission at the FIR--submm peak. It is interesting to notice, that as in the case of IC\,4225 from \citetalias{2015MNRAS.451.1728D}, the inclusion of the additional young stellar population did not solve the dust energy problem in IC\,2531.

In this work, we aim to continue the work started in \citetalias{2016A&A...592A..71M} and analyse the dust characteristics in the six remaining \textit{HER}OES galaxies, following the same strategy as has been proposed in \citetalias{2016A&A...592A..71M}, in a uniform and self-consistent way. For all galaxies in our sample we use available optical and near-infrared observations to constrain their dust models and perform panchromatic simulations. We investigate to which degree our models can predict the galaxy fluxes in different bands, paying particular attention to the dust emission in the FIR-submm domain. This comparison will allow us to answer the question how common the dust energy balance problem might be in galaxies and will give us a clue on a possible reason of its existence.

This paper is organised as follows. Section \ref{sec:sample} reviews the sample as well as observing and data reduction strategies. Section \ref{sec:Fitting} presents a brief description of the approach we use in this work to create oligochromatic and panchromatic models. In Section \ref{sec:Results}, we present results of our study for each individual galaxy and compare our results with the literature. 
In Section \ref{sec:discussion}, the results of our radiative transfer modelling procedure are discussed in the frame of the dust energy balance problem. The main conclusions of our study are summed up in Section \ref{sec:summary}. In the Appendix, we provide our models and simulations for each galaxy.

\section{The sample} 
\label{sec:sample}

\begin{figure*}
\includegraphics[width=\textwidth , angle=0, clip=]{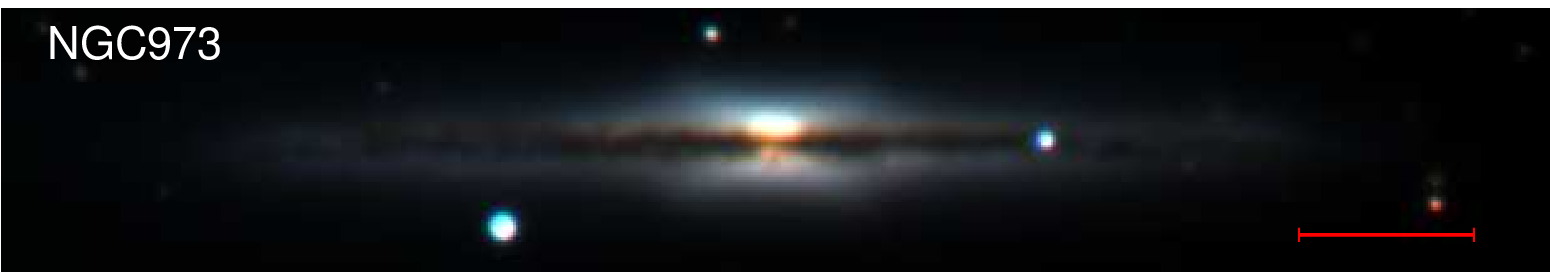}
\includegraphics[width=\textwidth , angle=0, clip=]{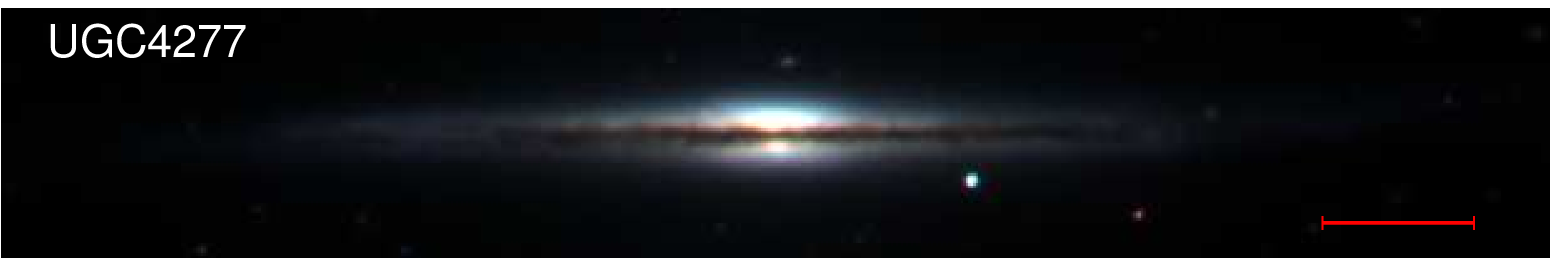}
\includegraphics[width=\textwidth , angle=0, clip=]{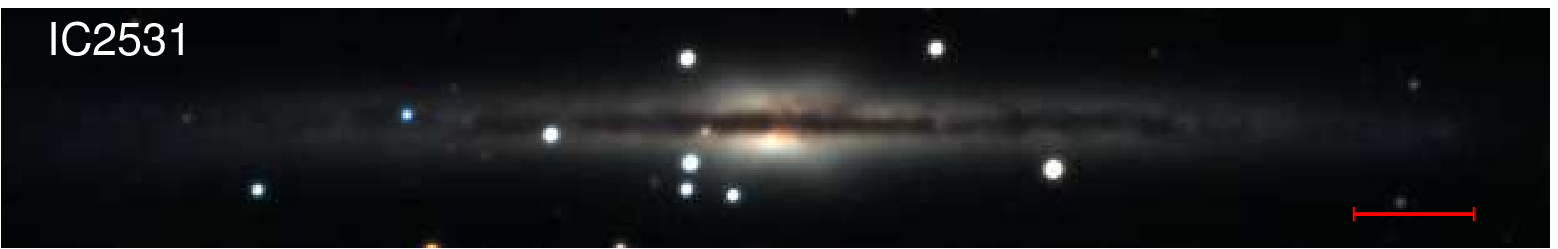}
\includegraphics[width=\textwidth , angle=0, clip=]{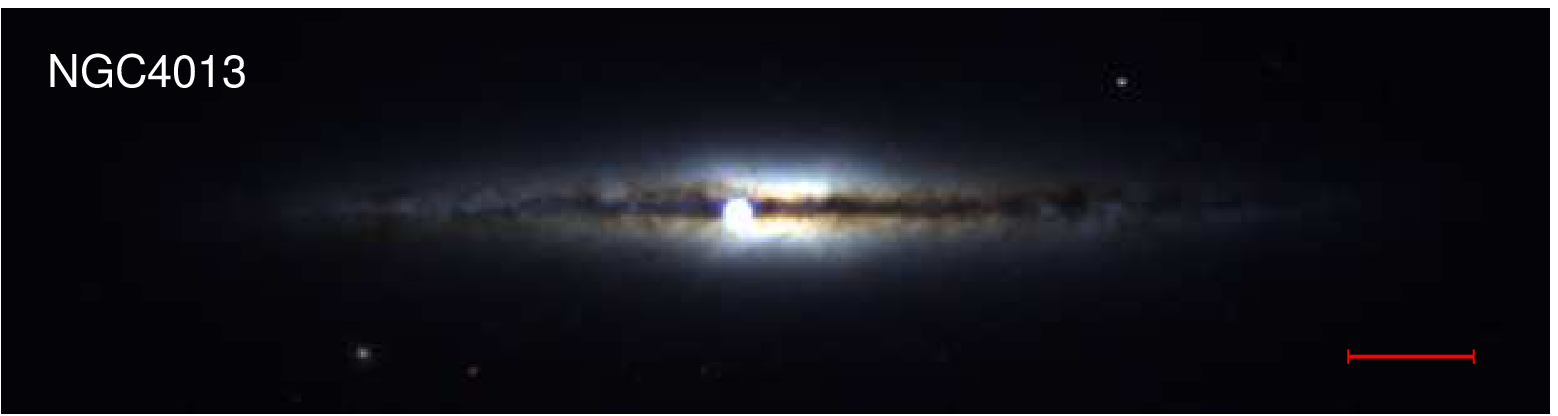}
\includegraphics[width=\textwidth , angle=0, clip=]{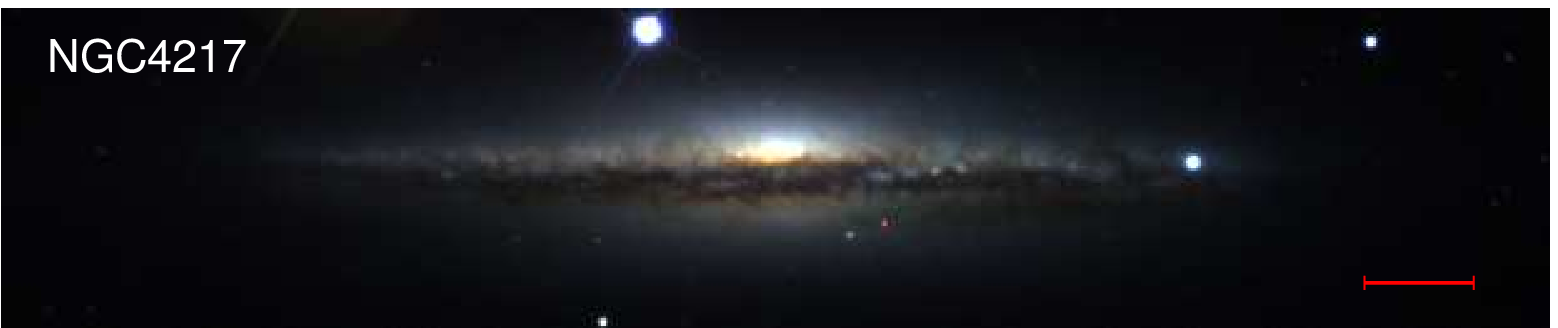}
\includegraphics[width=\textwidth , angle=0, clip=]{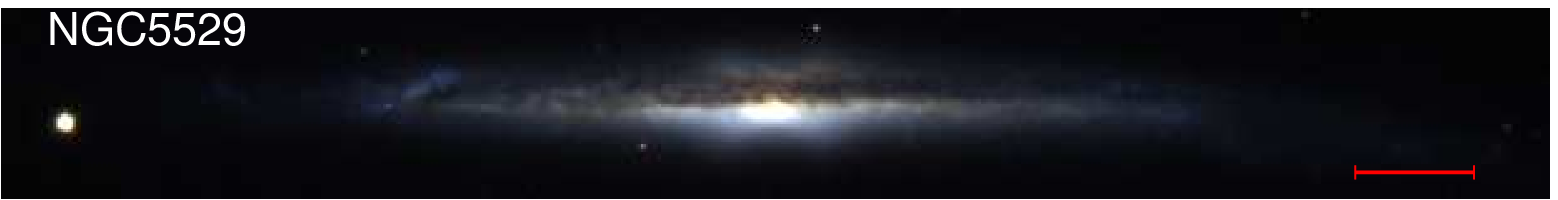}
\includegraphics[width=\textwidth , angle=0, clip=]{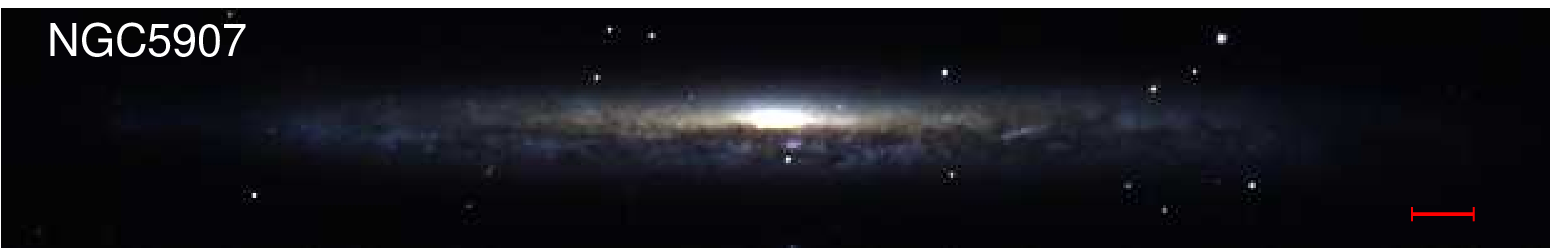}
\caption{Composite RGB-images of the $g$-, $r$- and $i$-passband (or $B$-, $V$-, and $R$-passband for IC\,2531) frames (see Sect.~\ref{sec:observations}). The length of the red bar in the bottom right corner of each panel is 30~arcsec.}
\label{figure1}
\end{figure*}

The construction of the sample of the \textit{HER}OES galaxies is described in detail in \citet{2013A&A...556A..54V}. Our sample consists of seven relatively nearby edge-on galaxies: NGC\,973, UGC\,4277, IC\,2531, NGC\,4013, NGC\,4217, NGC\,5529, and NGC\,5907 (see Fig.~\ref{figure1}, which shows their RGB images, and Table~\ref{tab:General_info}, which lists some basic characteristics of these objects). Their morphological types suggest that these are late-type galaxies with rather compact bulges, though some of them (NGC~973, IC~2531, NGC~4013, and NGC~5529) clearly exhibit extended boxy/peanut-shape (B/PS) bulges which are often related to bar structures \citep[see e.g.][]{1999AJ....118..126B,2004AJ....127.3192C,2008ApJ...679L..73M} and may reside in more than half of the disc galaxies in the Local Universe \citep{2000A&AS..145..405L, 2016ASSL..418...77L}.

The sample galaxies are rather bright and the difference in absolute magnitudes between the faintest and brightest galaxy is not larger than 1.6~mag. All galaxies have similar sizes (in kpc) except for NGC\,4013 and NGC\,4217 which are about half the size of the others. 

As follows from Table~\ref{tab:General_info}, all \textit{HER}OES galaxies are oriented almost exactly edge-on; only NGC\,5529 and NGC\,5907 have slightly lower inclination (their dust lane is more offset from the major axis). Interestingly, these galaxies differ significantly by apparent axis ratio $smb/sma$ --- IC\,2531 and NGC\,5907 seem rather thin (0.14 for both) whereas NGC\,4013 and NGC\,5529 have very thick outermost isophotes (0.38 and 0.30, respectively). Accordingly, the vertical-to-radial extent ratio of stellar discs in these galaxies should differ by a factor 2 to 3. Taking into account that the galaxies are observed at different distances, the angular extent of their dust lanes may also be different. Although the \textit{HER}OES galaxies were selected by their distinct, regular dust lanes, several objects exhibit somewhat clumpy and patchy dust inclusions. The dust lanes in some galaxies are rather extended and thick (NGC\,4217, NGC\,5907), in others they are relatively thin and warped (UGC\,4277, IC\,2531).

All \textit{HER}OES galaxies are group members \citep{1979ApJS...40..527P,1992ApJS...81..413M, 1998Afz....41..308M}, except for UGC\,4277 which is classified as an isolated galaxy \citep{1973AISAO...8....3K}.

For the \textit{HER}OES galaxies radiative transfer modelling has been done before (though using different modelling strategies): NGC\,973 is described in \citet[][hereafter \citetalias{1997A&A...325..135X}]{1997A&A...325..135X}; UGC\,4277 -- in \citet[][hereafter \citetalias{2007A&A...471..765B}]{2007A&A...471..765B}; NGC\,4013 -- in \citetalias{1999A&A...344..868X}, \citetalias{2007A&A...471..765B}, and \citet{2013A&A...550A..74D}; NGC\,4217 -- in \citetalias{2007A&A...471..765B}; NGC\,5529 -- in \citetalias{1999A&A...344..868X}, \citetalias{2007A&A...471..765B}; NGC\,5907 -- in \citetalias{1999A&A...344..868X} and \citet{2001A&A...372..775M}. 
The galaxy IC\,2531 has been analysed in great detail in \citetalias{2016A&A...592A..71M} (it was also modelled by \citetalias{1999A&A...344..868X}), therefore in this paper we study the remaining six \textit{HER}OES galaxies. For completeness, the results of our modelling for IC\,2531 are also provided.

\begin{table*}[t] 
  \centering
  \caption{Basic properties of the \textit{HER}OES galaxies.} 
  \label{tab:General_info}
  \begin{tabular}{lcccccccccc}
    \hline \hline\\[-2ex]
    Galaxy & RA & Dec & Type & $m_\mathrm{3.6/W1}$ & $sma$ & $smb$ & $D$ & Scale & $M_\mathrm{3.6/W1}$ & $i$ \\
    & (J2000) & (J2000) & & (mag) & (arcmin) & (arcmin) & (Mpc) & (pc/arcsec) & (mag) & (deg) \\
    \hline\\[-1ex]
    NGC\,973 & 02:34:20 & +32:30:20 & Sb & 11.03 & 2.14 & 0.41 & 63.5 & 308 & -22.98 &  89.6 \\
    UGC\,4277 & 08:13:57 & +52:38:55 & Scd & 12.42 & 1.66 & 0.33 & 76.5 & 371 & -22.00 & 88.9 \\
    IC\,2531     & 09:59:56 & --29:37:01 & Sc & 10.94 & 3.16 & 0.43 & 36.8 & 178 & -21.89 & 89.6 \\
    NGC\,4013 & 11:58:31 & +43:56:48 & Sb & 9.98 & 2.86 & 1.09 & 18.6 & 90 & -21.37 & 89.7 \\
    NGC\,4217 & 12:15:51 & +47:05:30 & Sb & 9.91 & 3.5 & 0.95 & 19.6 & 95 & -21.55 & 88.0 \\
    NGC\,5529 & 14:15:34 & +36:13:36 & Sc & 10.98 & 2.19 & 0.66 & 49.5 & 240 & -22.50 & 87.4 \\
    NGC\,5907 & 15:15:54 & +56:19:44 & Sc & 9.11 & 5.75 & 0.80 & 16.3 & 79 & -21.96 & 87.2 \\[0.5ex]
     
    \hline\\
  \end{tabular}
\parbox[t]{180mm}{ {\bf Notes.} The celestial coordinates and morphologies are taken from NASA/IPAC Extragalactic Database (NED). The distances $D$ with the corresponding scales are taken from \cite{2013A&A...556A..54V}, which were found to be the average values of the redshift-independent distance measurements (mostly based on the Tully-Fisher relation). The apparent magnitudes $m_{\mathrm{3.6}/W1}$, the semi-major and semi-minor axes ($sma$ and $smb$) are measured for the isophote of 25 mag/arcsec$^2$. The absolute magnitudes $M_{\mathrm{3.6}/W1}$ correspond to the adopted distances $D$ and apparent magnitudes $m_{\mathrm{3.6}/W1}$. For UGC\,4277, we used its $WISE$ W1 image to retrieve its magnitudes and semi-major and semi-minor axes. The inclinations of the galaxies are taken from the radiative transfer models from \citetalias{1997A&A...325..135X} (NGC\,973), \citetalias{1999A&A...344..868X} (IC\,2531, NGC\,4013, NGC\,5529, NGC\,5907), and \citetalias{2007A&A...471..765B} (UGC\,4277, NGC\,4217). Notice that these inclinations are used in our fitting (see Sect.~\ref{sec:Fitting}) as a first guess. The fit values of $i$ are given in Table~\ref{tab:FitSKIRT_res}.}
\end{table*} 

\subsection{Observations} 
\label{sec:observations}
To investigate the dust energy balance problem in a galaxy, we need to collect data over a large wavelength range -- from far-ultraviolet to far-infrared/submm regions. For this purpose, we used available databases of optical and NIR sky surveys to extract images of the \textit{HER}OES galaxies. Here we should notice that we split up our imaging into two sets. The first set of observations was constructed to find \textsc{fitskirt} oligochromatic models, by fitting optical and NIR images (we will call them the reference images) simultaneously. The second set of observations was prepared to supplement galaxy SEDs with fluxes at UV and mid-infrared--far-infrared (MIR--FIR) wavelengths with $\lambda > 4\,\mu$m and compare the predicted models from our \textsc{skirt} simulations with these observations (see Sect.~\ref{sec:Fitting}).

Optical observations for NGC\,4013, NGC\,4217, NGC\,5529, and NGC\,5907 were retrieved from SDSS DR12 \citep{2011AJ....142...72E} in all five $ugriz$ bands, with a pixel size of 0.396 arcsec/pixel and an average seeing of 1.1 arcsec. For NGC\,973 and UGC\,4277, the optical observations were executed on the William Herschel Telescope (WHT) on La Palma, equipped with the Auxiliary-port CAMera \citep[ACAM,][]{1998NewAR..42..503B}. The observing run yielded a total of 9 hours, during 3 moonless nights from the 25th till the 27th of December, 2011. 
The instrument provided a large 8.3 arcmin field of view, with 0.25 arcsec/pixel. The average seeing for the performed observations is 1.32 arcsec. ACAM is equipped with Sloan $ugriz$ filters, therefore we used optical observations in the same filters for almost all \textit{HER}OES galaxies, except for IC\,2531 which is located in the southern hemisphere and was observed at the \textit{Faulkes Telescope South} in the $B$-, $V$- and $R$-band filters \citepalias[for details, see][]{2016A&A...592A..71M}.

Four galaxies (NGC\,973, UGC\,4277, NGC\,4013, and NGC\,4217) have been observed at the Telescopio Nazionale Galileo (TNG), a 3.58m telescope of the Roque de Los Muchachos Observatory on La Palma. The observations in the $K$ band were executed during August and December of 2011 with an average seeing of 1.09 arcsec. The pixel size is 0.25 arcsec/pixel. In addition to the $K$ band and for the remaining galaxies we used 2MASS observations in all three $JHK_s$ bands with a PSF FWHM of 2 to 3$\arcsec$ (depending on the atmospheric blurring). 

All galaxies, except for UGC\,4277, have \textit{Spitzer Space Telescope} \citep{2004ApJS..154....1W} observations made by the Infrared Array Camera \citep[IRAC,][]{2004ApJS..154...10F} at 3.6\,$\mu$m (IRAC~1). For NGC\,973, the mosaic *maic.fits and uncertainties *munc.fits files were taken from post-Basic Calibrated Data, available through the \textit{Spitzer} Heritage Archive\footnote{http://sha.ipac.caltech.edu/}. For the other galaxies the observations were downloaded from the \textit{Spitzer} Survey of Stellar Structure in Galaxies archive \citep[S4G,][]{2010PASP..122.1397S, 2013ApJ...771...59M, 2013AAS...22123002R, 2015ApJS..219....3M, 2015ApJS..219....4S}\footnote{http://irsa.ipac.caltech.edu/data/SPITZER/S4G/}. For UGC\,4277, we used the \textit{Wide-field Infrared Survey Explorer} \citep[WISE,][]{2010AJ....140.1868W} in the $W1$ (3.4\,$\mu\mathrm{m}$) band. These optical and NIR observations (from 7 to 9 bands, depending on the galaxy) form our set of reference images. After some processing described in Sect.~\ref{sec:preparation} they were used to obtain the galaxy models which can be found in Sect.~\ref{sec:Fitting}.  

The supplementary data include observations from the {\it Galaxy Evolution Explorer} ({\it GALEX}, \citealp{2005ApJ...619L...1M, 2014AdSpR..53..900B}) at 0.152\,$\mu\mathrm{m}$ (\textit{FUV}) and 0.227\,$\mu\mathrm{m}$ (\textit{NUV}), which were downloaded from the All Sky Imaging Survey reachable trough the GalexView service\footnote{http://galex.stsci.edu/GalexView/}. WISE observations are available for the entire sky, hence we make use of them in our study as well, in all four $W1$ (3.4\,$\mu\mathrm{m}$), $W2$ (4.6\,$\mu\mathrm{m}$), $W3$ (12\,$\mu\mathrm{m}$) and $W4$ (22\,$\mu\mathrm{m}$) passbands, as an important source of information about the stellar and dust distribution, as well as the star formation rate in galaxies. NGC\,973, NGC\,5907, and NGC\,4013 have additional IRAC\,2 (4.5\,$\mu$m), IRAC\,3 (5.8\,$\mu$m) and IRAC\,4 (8.0\,$\mu$m) observations, while NGC\,4217 and NGC\,5529 have only IRAC\,2 images. For NGC\,4013 and NGC\,5907, \textit{Multiband Imaging Photometer for Spitzer} \citep[MIPS,][]{2004ApJS..154...25R} observations in all three bands (24, 70, and 160\,$\mu$m) were downloaded from the \textit{Spitzer} Heritage Archive.  

Additionally, we used {\it Herschel} PACS and SPIRE imaging to study galaxy SEDs at MIR and FIR wavelengths. To retrieve the \textit{Herschel} photometer observations, the \textit{Herschel} Science Archive\footnote{http://archives.esac.esa.int/hsa/whsa/} was queried and the observations in all five PACS and SPIRE bands were downloaded.

\subsection{Data preparation} 
\label{sec:preparation}

All the collected observations described in Sect.~\ref{sec:observations} have been initially reduced and prepared for the subsequent analysis, as it is required by the fitting codes we use in this work (see Sect.~\ref{sec:Fitting}). The galaxy image processing included several steps: astrometry correction (so that all the frames would have the same astrometry), resampling (rebinning to the frame with the smallest pixel size), determination and subtraction of the sky background, rotating the frames (to align the galactic plane along the horizontal axis), cropping the frames to the minimal size to cover the whole galaxy and at the same time have little empty space. The last step was masking out background and foreground objects, the light of which can contaminate the light of the galaxy. Each step is described in detail in \citetalias{2016A&A...592A..71M}.

In short, all these steps are realised with the \textsc{python} Toolkit for \textsc{skirt} (PTS\footnote{http://www.skirt.ugent.be/pts/index.html}, \citealt{2015A&C.....9...20C}, Verstocken et al. \textit{in prep.}), which contains a lot of functionalities and useful routines. The sky background was fitted by a two-dimensional polynomial of the second order using the unmasked pixels (the initial masks were created by one of the PTS scripts) and then subtracted from the original frames. The borders of the cropped galaxy image were chosen so that the outermost galaxy isophotes, at AB surface brightness (SB) levels of 25.5 mag/arcsec$^2$, in all optical and NIR bands would be encompassed. To prepare the images with final masks, we used the segmentation maps created with the \textsc{sextractor} package \citep{1996A&AS..117..393B} and then revisited the created masks by hand.
 
In addition, creating the accurate point spread function (PSF) for every galaxy image in each band is important since atmosphere and telescope blurring can seriously affect the light coming from the galaxy and, thus, change the real SB distribution of the observed object. To determine the PSF in each frame (if applicable), we selected several good  stars (not crowded, with a high signal-to-noise ratio). Then they were fitted with a Moffat \citep{1969A&A.....3..455M} (for the SDSS, TNG and WHT observations) or Gaussian (for the 2MASS and WISE $W1$ images) function. For the IRAC images, we used in-flight point response function (PRF) images\footnote{http://irsa.ipac.caltech.edu/data/SPITZER/docs/irac/calibrationfiles/} for the centre of the IRAC 3.6~$\mu$m\ fields downsampled and re-rotated to correspond to the analysed galaxy frames. For the other bands, we used the kernels provided by \citet{2011PASP..123.1218A}. The final PSF images were rebinned and rotated to match the final galaxy images.

The reduction of the {\it Herschel} PACS and SPIRE data was performed in \cite{2013A&A...556A..54V}, however, we repeated it following the up-to-date recipe described in \citet{2017PASP..129d4102D} and \citet{2018A&A...609A..37C} for the DustPedia sample. For SPIRE, we applied HIPE~v13\footnote{http://www.cosmos.esa.int/web/herschel/hipe-download} and the Bright Galaxy Adaptive Element ({\sc BriGAdE}; \citealp{2012ApJ...748..123S}) pipeline. Final SPIRE maps were produced using the {\sc HIPE}\,v13 na\"ive map-maker, with pixel sizes of 6, 8, and 12 arcsec at 250, 350, and 500\,$\mu$m\, respectively. For PACS data, {\sc HIPE}\,v13 and {\sc Scanamorphous}\,v24 pipeline \citep{2013PASP..125.1126R} were used. The final PACS maps have pixel sizes of 3 and 4 arcsec at 100 and 160\,$\mu$m\, respectively.

All the images from the reference and supplement sets were prepared in the same way as described above. 

Integrated flux densities for all wavelengths were determined by means of aperture photometry from the \textsc{photutils} \textsc{python} package, which was applied to the sky subtracted frames. For each galaxy, we determined an elliptical aperture which is 1.5 times larger than the outermost isophote (at the $2\sigma$ level) among all the prepared galaxy images. 
For each galaxy, we used a uniform mask for all reference images and individual masks for supplemental observations. We calculated the total flux inside the aperture by summing the flux value of each pixel; the masked areas were filled with the interpolated values using a Gaussian weights kernel. 

To estimate the uncertainties, we calculated the total uncertainty by summing in quadrature the calibration uncertainty (for the surveys we used the values listed in tab.~1 from \citealt{2018A&A...609A..37C}, for the other images we used a calibration uncertainty of 5\%) and the uncertainty on the sky level.

We also added \textit{IRAS} \citep{1990BAAS...22Q1325M, 2003AJ....126.1607S, 2007A&A...462..507L} flux densities at 25, 60 and 100~$\mu$m\ taken from NED, and \textit{Planck} \citep{2014A&A...571A...1P} APERFLUX values at 353, 545 and 857~GHz (which correspond to 850\,$\mu\mathrm{m}$, 550\,$\mu\mathrm{m}$ and 350\,$\mu\mathrm{m}$, respectively) from the \textit{Planck} Legacy Archive\footnote{http://pla.esac.esa.int/pla}. The two brightest galaxies in our sample, NGC\,5907 and NGC\,4217, were detected by ISO as part of the ISOPHOT 170\,$\mu$m Serendipity Survey \citep{2004A&A...422...39S}. For NGC\,5907, we found two fluxes at 450\,$\mu\mathrm{m}$ and 850\,$\mu\mathrm{m}$ from the Submillimetre Common-User Bolometer Array \citep[SCUBA,][]{1999MNRAS.303..659H, 2005MNRAS.357..361S}. We did not include the Akari/FIS \citep{2007PASJ...59S.369M} 160\,$\mu$m flux densities found for four \textit{HER}OES galaxies in our study because the comparison with the PACS\,160 \,$\mu\mathrm{m}$ fluxes is very poor (to estimate the fluxes they used PSF photometry with a FWHM of about 60 arcsec at 160\,$\mu\mathrm{m}$, while all the galaxies in our sample are much larger).

We provide all flux densities (measured and taken from literature) and their uncertainties in Table~\ref{tab:Fluxes}.

\begin{sidewaystable*}
\caption{Observed flux densities $F_\nu$ in Jy and their corresponding errors. Correction for Galactic extinction (where applicable) has been applied according to \cite{2011ApJ...737..103S}. The masked objects were replaced by the interpolated values using a Gaussian weights kernel. No colour corrections to the retrieved fluxes were applied. The asterisk denotes the reference data.}
\label{tab:Fluxes}
\centering
\begin{tabular}{lcccccccc}
\hline
\hline\\[-2ex]
Survey & $\lambda$ & NGC\,973 & UGC\,4277 & IC\,2531 & NGC\,4013 & NGC\,4217 & NGC\,5529 & NGC\,5907 \\[-0.5ex]
       & ($\mu\mathrm{m}$) &  & & & & & & \\[0.5ex]
\hline\\[-1ex]
$FUV$ & 0.15 & --- &  --- & $0.001 \pm 0.0005$ &  $0.0008 \pm 0.0003$ &  $0.0007 \pm 0.0003$ &  $0.0009 \pm 0.0003$ &  $0.008 \pm 0.001$ \\
$NUV$ & 0.23 & --- &  $0.0003 \pm 0.0001$ & $0.002 \pm 0.0004$  &   $0.002 \pm 0.0003$ &  $0.002 \pm 0.0004$ &  $0.002 \pm 0.0002$ &  $0.01 \pm 0.001$ \\
SDSS $u$ & 0.36 & --- &  --- & --- &  $0.02 \pm 0.002^{*}$ &  $0.02 \pm 0.002^{*}$ &  $0.008 \pm 0.0005^{*}$ &  $0.05 \pm 0.004^{*}$ \\
WHT $u$ & 0.36 & $0.005 \pm 0.0003^{*}$ &  $0.004 \pm 0.0002^{*}$ & --- &  --- &  --- &  --- &  --- \\
FTS $B$ & 0.45 & --- & --- & $0.040 \pm 0.004^{*}$ & --- & --- & --- & --- \\
SDSS $g$ & 0.47 & --- &  --- & --- &  $0.07 \pm 0.005^{*}$ &  $0.08 \pm 0.009^{*}$ &  $0.03 \pm 0.002^{*}$ &  $0.18 \pm 0.009^{*}$ \\
WHT $g$ & 0.47 & $0.02 \pm 0.001^{*}$ &  $0.01 \pm 0.001^{*}$ & --- & --- &  --- &  --- &  --- \\
FTS $V$ & 0.55 & --- & --- & $0.070 \pm 0.007^{*}$ & --- & --- & --- & --- \\
SDSS $r$ & 0.62 & --- &  --- & --- &  $0.15 \pm 0.01^{*}$ &  $0.17 \pm 0.02^{*}$ &  $0.07 \pm 0.004^{*}$ &  $0.34 \pm 0.01^{*}$ \\
WHT $r$ & 0.62 & $0.04 \pm 0.002^{*}$ &  $0.02 \pm 0.001^{*}$ & --- & --- &  --- &  --- &  --- \\
FTS $R$ & 0.66 & --- & --- & $0.098 \pm 0.009^{*}$ & --- & --- & --- & --- \\
SDSS $i$ & 0.75 & --- &  --- & --- &  $0.23 \pm 0.02^{*}$ &  $0.26 \pm 0.03^{*}$ &  $0.1 \pm 0.006^{*}$ &  $0.51 \pm 0.02^{*}$ \\
WHT $i$ & 0.75 & $0.06 \pm 0.003^{*}$ &  $0.03 \pm 0.001^{*}$ & --- &  --- &  --- &  --- &  --- \\
SDSS $z$ & 0.89 & --- &  --- & --- &  $0.3 \pm 0.02^{*}$ &  $0.33 \pm 0.03^{*}$ &  $0.14 \pm 0.008^{*}$ &  $0.64 \pm 0.03^{*}$ \\
WHT $z$ & 0.89 & $0.09 \pm 0.005^{*}$ &  $0.04 \pm 0.001^{*}$ & --- &  --- &  --- &  --- &  --- \\
2MASS $J$ & 1.25 & $0.17 \pm 0.009^{*}$ &  $0.08 \pm 0.003^{*}$ & $0.250 \pm 0.016^{*}$ &  $0.51 \pm 0.04^{*}$ &  $0.51 \pm 0.04^{*}$ &  $0.23 \pm 0.01^{*}$ &  $1.09 \pm 0.04^{*}$ \\
2MASS $H$ & 1.65 & $0.24 \pm 0.01^{*}$ &  $0.1 \pm 0.003^{*}$ & $0.347 \pm 0.020^{*}$ &  $0.69 \pm 0.05^{*}$ &  $0.71 \pm 0.05^{*}$ &  $0.32 \pm 0.02^{*}$ &  $1.48 \pm 0.05^{*}$ \\
2MASS $K$ & 2.20 & $0.2 \pm 0.005$ &  $0.09 \pm 0.003$ & $0.291 \pm 0.015^{*}$ &  $0.59 \pm 0.01$ &  $0.62 \pm 0.02$ &  $0.29 \pm 0.02^{*}$ &  $1.36 \pm 0.05^{*}$ \\
TNG $K$ & 2.13 & $0.25 \pm 0.02^{*}$ &  $0.11 \pm 0.006^{*}$ & --- &  $0.66 \pm 0.06^{*}$ &  $0.73 \pm 0.06^{*}$ &  --- &  --- \\
\textit{WISE} $1$ & 3.4 & $0.11 \pm 0.006$ &  $0.05 \pm 0.002^{*}$ & $0.127 \pm 0.005$ &  $0.28 \pm 0.02$ &  $0.33 \pm 0.04$ &  $0.12 \pm 0.009$ &  $0.65 \pm 0.03$ \\
IRAC $1$ & 3.6 & $0.14 \pm 0.007^{*}$ &  --- & $0.156 \pm 0.007^{*}$ &  $0.38 \pm 0.03^{*}$ &  $0.41 \pm 0.03^{*}$ &  $0.15 \pm 0.01^{*}$ &  $0.85 \pm 0.03^{*}$ \\
IRAC $2$ & 4.5 & $0.11 \pm 0.004$ &  --- & --- &  $0.25 \pm 0.02$ &  $0.29 \pm 0.03$ &  $0.1 \pm 0.007$ &  $0.57 \pm 0.03$ \\
\textit{WISE} $2$ & 4.60 & $0.1 \pm 0.005$ &  $0.03 \pm 0.002$ & $0.088 \pm 0.004$ &  $0.21 \pm 0.02$ &  $0.25 \pm 0.03$ &  $0.08 \pm 0.006$ &  $0.47 \pm 0.03$ \\
IRAC $3$ & 5.8 & $0.14 \pm 0.005$ &  --- & --- &  $0.42 \pm 0.01$ &  --- &  --- &  $1.17 \pm 0.07$ \\
IRAC $4$ & 8.0 & $0.2 \pm 0.008$ &  --- & --- &  $0.92 \pm 0.04$ &  --- &  --- &  $2.36 \pm 0.13$ \\
\textit{WISE} $3$ & 12.1 & $0.18 \pm 0.01$ &  $0.06 \pm 0.005$ & $0.218 \pm 0.007$ &  $0.62 \pm 0.05$ &  $1.22 \pm 0.13$ &  $0.25 \pm 0.03$ &  $1.88 \pm 0.13$ \\
\textit{IRAS} $12$ & 12 & $0.08 \pm 0.02$ &  --- & --- &  $0.54 \pm 0.04$ &  $1.26 \pm 0.07$ &  $0.26 \pm 0.03$ &  $1.29 \pm 0.07$ \\
\textit{WISE} $4$ & 22.2 & $0.32 \pm 0.03$ &  $0.07 \pm 0.008$ & $0.188 \pm 0.007$ &  $0.75 \pm 0.07$ &  $1.48 \pm 0.22$ &  $0.27 \pm 0.03$ &  $2.19 \pm 0.18$ \\
MIPS $1$ & 24 & --- &  --- & --- &  $0.66 \pm 0.04$ &  --- &  --- &  $1.73 \pm 0.13$ \\
\textit{IRAS} $25$ & 25 & $0.3 \pm 0.02$ &  --- & --- &  $0.77 \pm 0.05$ &  $1.5 \pm 0.08$ &  $0.24 \pm 0.03$ &  $1.44 \pm 0.08$ \\
\textit{IRAS} $60$ & 60 & $1.69 \pm 0.12$ &  $0.35 \pm 0.04$ & --- &  $7.01 \pm 0.35$ &  $11.6 \pm 0.58$ &  $1.95 \pm 0.17$ &  $9.14 \pm 0.46$ \\
MIPS $2$ & 70 & --- &  --- & --- &  $10.01 \pm 1.1$ &  --- &  --- &  $23.21 \pm 2.66$ \\
\textit{IRAS} $100$ & 100 & $3.5 \pm 0.39$ &  $1.12 \pm 0.12$ & --- &  $24.4 \pm 1.23$ &  $41.2 \pm 2.06$ &  $7.73 \pm 0.55$ &  $37.4 \pm 1.87$ \\
PACS $100$ & 100 & $4.07 \pm 0.36$ &  $1.64 \pm 0.17$ & $5.82\pm0.51$ & $24.17 \pm 1.37$ &  $38.69 \pm 1.94$ &  $8.52 \pm 0.6$ &  $60.28 \pm 3.2$ \\
MIPS $3$ & 160 & --- & --- &  --- &  $36.92 \pm 5.5$ &  --- &  --- &  $86.16 \pm 11.75$ \\
PACS $160$ & 160 & $5.8 \pm 0.35$ &  $2.74 \pm 0.31$ & $9.94\pm0.65$ &  $32.44 \pm 1.8$ &  $56.79 \pm 2.84$ &  $12.3 \pm 0.74$ &  $88.29 \pm 4.69$ \\
ISO $170$ & 170 & --- & --- &  --- &  --- &  $62.27 \pm 9.34$ &  --- &  $35.8 \pm 5.37$ \\
SPIRE $250$ & 250 & $3.6 \pm 0.26$ &  $2.1 \pm 0.16$ & $6.72\pm0.48$ &  $17.67 \pm 1.24$ &  $28.79 \pm 2.07$ &  $8.04 \pm 0.57$ &  $53.21 \pm 3.73$ \\
SPIRE $350$ & 350 & $1.7 \pm 0.13$ &  $1.11 \pm 0.09$ & $3.51\pm0.26$ &  $7.52 \pm 0.53$ &  $12.27 \pm 0.94$ &  $3.95 \pm 0.28$ &  $24.45 \pm 1.72$ \\
\textit{Planck} $350$ & 350 & --- &  $1.58 \pm 0.94$ & $3.50\pm0.55$ &  $8.21 \pm 0.41$ &  --- &  $4.46 \pm 0.32$ &  $25.44 \pm 0.42$ \\
SCUBA $450$ & 450 & --- &  --- & --- &  --- &  --- &  --- &  $13.2 \pm 0.91$ \\
SPIRE $500$ & 500 & $0.64 \pm 0.06$ &  $0.43 \pm 0.04$ & $1.42\pm0.12$ &  $2.52 \pm 0.19$ &  $4.25 \pm 0.39$ &  $1.54 \pm 0.12$ &  $9.08 \pm 0.64$ \\
\textit{Planck} $550$ & 550 & --- &  --- & $1.18\pm0.29$ &  $2.27 \pm 0.22$ &  --- &  $1.61 \pm 0.18$ &  $7.9 \pm 0.24$ \\
\textit{Planck} $850$ & 850 & --- &  --- & $0.31\pm0.17$ &  $0.62 \pm 0.09$ &  --- &  $0.52 \pm 0.09$ &  $2.12 \pm 0.13$ \\
SCUBA $850$ & 850 & --- &  --- & --- &  --- &  --- &  --- &  $1.96 \pm 0.03$ \\
\hline\\
 \end{tabular}
\end{sidewaystable*}

\section{Fitting strategy and simulations}
\label{sec:Fitting}

Our strategy is described in detail in \citetalias{2016A&A...592A..71M}. The general idea is as follows. First, we fit a model, which consists of several stellar components, e.g. a \ser\ \citep{ser1968} bulge and thin and thick double-exponential discs, to a NIR image. IRAC~3.6\,$\mu\mathrm{m}$ observations are ideally suited for solving this task since images at these wavelengths serve as good relative proxies for stellar mass \citep{1984ApJS...54..127E, 1993PASP..105..651G, 1995ApJ...447...82R,2012ApJ...744...17M,2014ApJ...788..144M,2015ApJS..219....5Q} and the influence of dust is here significantly diminished. For UGC\,4277, there are no \textit{Spitzer} observations. Therefore, in this case we use the TNG $K$-band image, which also offers a good opportunity to trace the distribution of old stellar populations. The visual inspection of these NIR images (in the 3.6$\,\mu$m and $K$ bands) did not reveal any apparent traces of a dust lane in these galaxies. Therefore, we can consider, as a first approximation, the attenuation at these wavelengths to be zero.

Once the stellar model is obtained, we can use the \textsc{fitskirt}\footnote{http://www.skirt.ugent.be/skirt/index.html} code \citep{2013A&A...550A..74D, 2014MNRAS.441..869D}. It allows one to find structural properties of the dust component by means of \textsc{skirt} \citep{2011ApJS..196...22B, 2015A&C.....9...20C} oligochromatic radiative transfer simulations and with help of a genetic algorithm-based optimisation \citep[]{1989gaso.book.....G}. At this stage, a set of optical and NIR observations can be fitted simultaneously, together with a dust component which is often represented by a double-exponential disc. Here, we keep fixed the geometrical sizes of the stellar components to the
values retrieved in the first step, but allow to change their relative luminosities. This is a relatively fast and effective way to fit the dust model and maximally avoid the problem of the local minimum of $\chi^2$, by reducing the number of free parameters when searching for the best model fit.

The third step is performing \textsc{skirt} panchromatic radiative transfer simulations using the obtained \textsc{fitskirt} models. Having them in hand, we can create mock observations (we call them further `panchromatic simulations') and, therefore, predict how the galaxy would look like at different wavelengths, from \textit{FUV} to submm, and then compare the real and model SEDs, as well as their real and mock images, and conclude how accurately our model is able to describe the observations at specific wavelengths.

Here we should clarify why we use this three-step approach instead of directly fitting all available FUV-submm observations simultaneously to derive the parameters for both stellar and dust components. First, in the present version of \textsc{fitskirt}, all wavelengths are considered to be independent. Therefore, the code is only applicable to wavelengths up to NIR. Further, with increasing wavelength, the dust emission becomes an important part of the simulation, and, hence, this assumption is no longer valid. Second, the computational cost of such multiwavelength fitting is very high because not only the number of free model parameters substantially increases, but also the amount of data, which should be analysed simultaneously during the fitting.

\subsection{Decomposition of the IRAC 3.6\,$\mu\mathrm{m}$ images} 
\label{sec:IRACdecomp}

The surface brightness profile of a stellar bulge is usually fitted with a \ser\ model:
\begin{equation}
I(r) = I_\mathrm{e,b} \: \mathrm{e}^{ -b_\mathrm{n} \left[ \left( \frac{r}{r_\mathrm{e,b}} \right)^{1/n_\mathrm{b}} \! - \: 1 \right] }\,
\end{equation}
where $I_\mathrm{e,b}$ is the effective surface brightness, i.e. the surface brightness at the half-light radius of the bulge $r_\mathrm{e,b}$, and $b_\mathrm{n}$ is a function of the \ser\ index $n_\mathrm{b}$ \citep{1993MNRAS.265.1013C,1999A&A...352..447C}. An additional free parameter is the apparent axis ratio of the bulge $q_\mathrm{b}$. We also apply a \ser\ profile for describing SB of a halo for one \textit{HER}OES galaxy, NGC\,4013. 

The three-dimensional axisymmetric stellar disc can be described by a double-exponential function $j(R,z)$, which in a cylindrical coordinate system $(R, z)$ aligned with the disc (where the disc mid-plane has $z = 0$) is given by
\begin{equation}
j(R,z) = j_{0} \; \mathrm{e}^{-R/h_\mathrm{R}-|z|/h_\mathrm{z}}\,,
\label{exp_disc}
\end{equation}
where $j_{0}$ is the central luminosity density of the disc, $h_\mathrm{R}$ is the disc scale length, and $h_\mathrm{z}$ is the vertical scale height. In this work we use a modified profile that includes a break radius (it can also be called truncation/anti-truncation radius). \citet{2005ApJ...626L..81E,2008AJ....135...20E}, \citet{2013ApJ...771...59M}, and \citetalias{2016A&A...592A..71M} have shown that for some galaxies this profile better describes the apparent SB distribution  \citep{2005ApJ...626L..81E, 2015ApJ...799..226E}:

\begin{equation}
j(R,z) = S \, j_{0} \, 
\text{e}^{-\frac{R}{h_{\text{R,inn}}}-\frac{|z|}{h_\text{z}}} 
\left(1 + \text{e}^{\frac{s\,(R-R_\text{b})}{h_{\text{R,out}}}}\right)^{ \frac{1}{s}
\left(\frac{h_{\text{R,out}}}{h_{\text{R,inn}}} - 1\right)
}\,.
\label{eq:broken_disc}
\end{equation}
In this formula, $s$ parametrises the sharpness of the transition between the inner and outer profiles with the break radius $R_\text{b}$. Large values of the sharpness parameter ($s \gg 1$) correspond to a sharp transition, and small values ($s \sim 1$) set a very gradual break. The dimensionless quantity $S$ is a scaling factor, given by
\begin{equation}
S = \left(1 + \text{e}^{-\frac{s \, R_\text{b}}{h_{\text{R,out}}}}\right)^{-\frac{1}{s} \left(\frac{h_{\text{R,out}}}{h_{\text{R,inn}}} - 1\right)}\,.
\label{eq:broken_disc_S}
\end{equation}
In total, this model contains five free parameters: the scale length of the inner disc $h_{\text{R,inn}}$, the scale length of the outer disc $h_{\text{R,out}}$, the scale height $h_\mathrm{z}$, the break radius $R_\text{b}$, and the sharpness of the break $s$.

Visual inspection of all minor-axis profiles of the \textit{HER}OES galaxies (see Appendix~\ref{Appendix_IMFITs}) suggests the presence of both a thin and a thick disc. Almost all the galaxies (except for UGC\,4277) show breaks on their radial profiles, therefore for them we used the model (\ref{eq:broken_disc}), and for the remaining galaxies the simple model (\ref{exp_disc}) was applied. As such, our uniform model generally consists of a thin and a thick broken disc and a central \ser\ bulge, except that for NGC\,4013 the model also includes a \ser\ stellar halo (see Sect.~\ref{sec:NGC4013}).
 
In our study we make use of the \textsc{imfit}\footnote{http://www.mpe.mpg.de/$\sim$erwin/code/imfit/} code \citep{2015ApJ...799..226E} which works with different 3D geometries (including the model of a broken double-exponential disc (\ref{eq:broken_disc}), which is realised in the \textsc{imfit} {\it BrokenExponentialDisk3D} function with $n=100$ in Eqs.~(40) and (41) in \citealt{2015ApJ...799..226E}).
To minimise the number of free parameters, we fixed the inclinations for both the thin and thick discs. These inclinations were initially found from the radiative transfer modelling of the \textit{HER}OES galaxies with a simple `disc+bulge' stellar model and a dust disc (described in the PhD thesis by \citealt{phd}). As shown in Table~\ref{tab:FitSKIRT_res}, which lists the results of our RT modelling (see Sect.\ref{sec:fitskirt}), the inclinations adopted here are essentially the same as found in our subsequent oligochromatic fitting.
Also, we fixed the sharpness of the breaks at $s=5$, since by varying its value we did not find a significant difference in the decomposition results. In addition, the break radii should have the same value for both the thin and the thick disc.

We wrote a special \textsc{python} wrapper to apply genetic algorithms in our process of searching for the best-fit model. To estimate uncertainties of the free parameters in our model, we applied the genetic algorithm ten times and took the scatter in the fitted parameters as their uncertainties. 

Table~\ref{tab:IRAC_dec_res} summarises the results of the \textsc{imfit} decomposition for all \textit{HER}OES galaxies. The detailed description of the obtained \textsc{imfit} models  is given in Sect.~\ref{sec:Results}.

\begin{table*}[t]
\caption{Results for the \textsc{imfit} decomposition of the $K$-band image for UGC\,4277 and the IRAC 3.6\,$\mu\mathrm{m}$ images for the rest of the {\it HER}OES galaxies.}
\label{tab:IRAC_dec_res}
\centering
\resizebox{18 cm}{!}{
    \begin{tabular}{ccccccccc}
    \hline \hline\\[-2ex] 
    Parameter & unit & NGC\,973 & UGC\,4277 & IC\,2531 & NGC\,4013 & NGC\,4217 & NGC\,5529 & NGC\,5907 \\
    \hline\\[-1ex]
    $i$                           & deg & $89.5$         & $88.6$        & $89.6$        & $89.8$        & $87.4$        & $87.2$         &   $85.1$        \\[+0.5ex]
    \hline\\[-1ex]
    1. Thin disc:\\[+0.5ex]
    $h_\mathrm{R,inn}^\mathrm{t}$ & kpc & $12.34\pm0.95$ & $7.03\pm0.28$ & $8.0\pm0.54$  & $5.34\pm0.89$ & $3.31\pm0.42$ & $9.53\pm0.13$  & $4.75\pm0.51$ \\[+0.5ex]
    $h_\mathrm{R,out}^\mathrm{t}$ & kpc & $3.88\pm0.27$  & ---           & $3.33\pm0.58$ & $0.86\pm0.10$ & $0.01\pm0.1$  & $1.08\pm0.18$  & $3.45\pm0.31$ \\[+0.5ex]
    $h_\mathrm{z}^\mathrm{t}$     & kpc & $0.35\pm0.03$  & $0.49\pm0.03$ & $0.61\pm0.04$ & $0.18\pm0.01$ & $0.11\pm0.01$ & $0.31\pm0.03$  & $0.16\pm0.02$ \\[+0.5ex]
    $R_\mathrm{b}^\mathrm{t}$                & kpc & $23.09\pm2.20$ & ---           & $21.41\pm3.57$& $8.79\pm1.72$ & $11.19\pm1.20$& $23.27\pm2.21$ & $19.80\pm2.58$\\[+0.5ex] 
    $L_\mathrm{t}/L_\mathrm{tot}$ & --- & $0.05\pm0.01$  & $0.41\pm0.04$ & $0.66\pm0.07$ & $0.25\pm0.08$ & $0.36\pm0.07$ & $0.45\pm0.04$  & $0.85\pm0.05$ \\[+0.5ex]
    \hline\\[-1ex]
    2. Thick disc:\\[+0.5ex]
    $h_\mathrm{R,inn}^\mathrm{T}$ & kpc & $17.88\pm1.63$ & $9.32\pm0.77$ & $24.87\pm0.77$& $2.44\pm0.22$ & $3.53\pm0.1$  & $7.65\pm0.22$  & $5.59\pm0.49$ \\[+0.5ex]
    $h_\mathrm{R,out}^\mathrm{T}$ & kpc & $5.62\pm0.71$  & ---           & ---           & $2.12\pm0.58$ & $3.39\pm0.18$ & ---            &        ---           \\[+0.5ex]
    $h_\mathrm{z}^\mathrm{T}$     & kpc & $1.34\pm0.14$  & $2.08\pm0.16$ & $1.57\pm0.18$ & $0.49\pm0.02$ & $0.93\pm0.03$ & $1.20\pm0.12$  & $1.88\pm0.25$ \\[+0.5ex]
    $R_\mathrm{b}^\mathrm{T}$                & kpc & $23.09\pm2.20$ & ---           & ---           & $8.79\pm1.72$ & $11.19\pm1.20$& --- & --- \\[+0.5ex] 
    $L_\mathrm{T}/L_\mathrm{tot}$ & --- & $0.39\pm0.06$  & $0.39\pm0.05$ & $0.15\pm0.03$ & $0.55\pm0.07$ & $0.60\pm0.07$ & $0.43\pm0.06$ & $0.10\pm0.03$ \\[+0.5ex]
    \hline\\[-1ex]
    3. Bulge:\\[+0.5ex]
    $r_\mathrm{e,b}$              & kpc & $1.20\pm0.07$  & $1.68\pm0.08$ & $1.86\pm0.11$ & $0.20\pm0.03$ & $0.40\pm0.03$ & $1.49\pm0.16$ & $0.56\pm0.09$ \\[+0.5ex]
    $n_\mathrm{b}$                & --- & $5.48\pm0.18$  & $2.64\pm0.2$  & $2.26\pm0.4$  & $1.54\pm0.3$  & $1.54\pm0.3$  & $3.33\pm0.21$ & $0.97\pm0.30$ \\[+0.5ex]
    $q_\mathrm{b}$                & --- & $0.49\pm0.01$  & $0.67\pm0.02$ & $0.85\pm0.03$ & $1.00\pm0.04$ & $0.95\pm0.02$ & $0.70\pm0.02$ & $0.37\pm0.02$ \\[+0.5ex]
    $L_\mathrm{b}/L_\mathrm{tot}$ & --- & $0.56\pm0.06$  & $0.20\pm0.03$ & $0.19\pm0.05$ & $0.05\pm0.01$ & $0.04\pm0.01$ & $0.12\pm0.02$ & $0.05\pm0.01$ \\[+0.5ex]
    \hline\\[-1ex]
    4. Halo:\\[+0.5ex]
    $r_\mathrm{e,h}$              & kpc & ---  & --- & --- & $7.44\pm1.22$ & --- & --- & --- \\[+0.5ex]
    $n_\mathrm{h}$                & --- & ---  & --- & --- & $1.57\pm0.23$ & --- & --- & --- \\[+0.5ex]
    $q_\mathrm{h}$                & --- & ---  & --- & --- & $0.39\pm0.05$ & --- & --- & --- \\[+0.5ex]
    $L_\mathrm{h}/L_\mathrm{tot}$ & --- & ---  & --- & --- & $0.15\pm0.01$ & --- & --- & --- \\[+0.5ex]
    \hline\\[-1ex]
    Total:\\[+0.5ex]     
    $L_\mathrm{tot}$              & AB-mag                        & $-22.99\pm0.13$ & $-22.95\pm0.10$ & $-21.92\pm0.19$ & $-21.36\pm0.17$ & $-21.53\pm0.14$ & $-22.49\pm0.09$& $-21.96\pm0.23$ \\[+0.5ex]
    $M_\star$                     & $10^{10}\,\mathrm{M}_{\odot}$ & $21.21\pm2.50$   & $13.5\pm1.23$   & $7.91\pm1.69$   & $4.73\pm0.71$   & $5.52\pm0.72$   & $13.42\pm1.10$  & $8.2\pm1.7$     \\[+0.5ex]    
    \hline\\[-0.5ex]
    \end{tabular}
}    
\parbox[t]{180mm}{ {\bf Notes.} The inclination angle $i$ was fixed for both the thin and thick discs. We used the {\it BrokenExponentialDisk3D} function for the thin and/or thick disc if the respective $R_\mathrm{b}$ is defined in the table, otherwise the {\it ExponentialDisk3D} was used. The bulges and the halo were fitted with the \ser\ function. To compute the total stellar mass $M_\star$ from the total luminosity $L_\mathrm{tot}$ in the respective NIR band, we relied on the mass-to-light ratio at 3.6\,$\mu\mathrm{m}$ of \cite{2012AJ....143..139E}, except for UGC\,4277 for which the mass-to-light ratio in the $K$ band $0.8$~$\mathrm{M}_\odot / L_\odot$  from \cite{2000ApJ...533L..99M} was used.}
\end{table*}

\subsection{{\sc{fitskirt}} fitting} 
\label{sec:fitskirt}

\textsc{fitskirt} is a fitting code that combines
the output of \textsc{skirt} with a genetic algorithm optimisation library to
retrieve best-fitting model parameters for the stellar and dust components in
a galaxy. 
The results retrieved from the 3D \textsc{fitskirt} model should reproduce the fitted images of the galaxy, taking the effects of dust
obscuration and multiple anisotropic scattering fully into account.

The automated fitting routine \textsc{fitskirt} has been tested on a mock (\textsc{skirt} simulated) galaxy image \citep{2013A&A...550A..74D}. Also, its capabilities have been validated by applying it to a dozen of real edge-on galaxies \citep[see][]{2012MNRAS.427.2797D, 2014MNRAS.441..869D, 2015A&A...576A..31S, 2015A&A...579A.103V, 2016A&A...592A..71M}. In these works, it was found that \textsc{fitskirt} is able to give reasonable constraints on all free parameters describing the stellar disc, stellar \ser\ bulge, dust disc, and even dust ring. 

\citet{2014MNRAS.441..869D} showed that oligochromatic fitting, i.e. fitting applied to a number of bands simultaneously, has clear advantages over monochromatic fitting in terms of accuracy. In particular, the parameters which describe the dust distribution, have a smaller spread as the oligochromatic fitting method is less prone to degeneracies in the free parameters. In addition to that, \citetalias{2016A&A...592A..71M} demonstrated for the example of one \textit{HER}OES galaxy IC\,2531 that the optical+NIR data based model is much
better constrained than the model where only the optical images are fitted. This is why in this study we collected available optical and NIR data in a broad range of wavelengths.

To describe the dust constituent in our fitting, we apply a model of a double-exponential dust disc \citep[see e.g.][]{1999A&A...344..868X, 2010A&A...518L..39B, 2013A&A...556A..54V, 2013A&A...550A..74D, 2014MNRAS.441..869D}:
\begin{equation}
  \rho(R,z)
  =
  \frac{M_\mathrm{d}}{4\pi\,h_{R,\text{d}}^2\,h_{\text{z,d}}}
  \mathrm{e}^{-\frac{R}{h_{\text{R,d}}}
  -\frac{|z|}{h_{z,\text{d}}}}\,,
\end{equation}
where $M_{\text{d}}$ is the total dust mass, $h_{\text{R,d}}$ is the radial scale length, and $h_{\text{z,d}}$ is the vertical scale height of the dust disc. To calculate the central face-on optical depth, one should use the following expression:
\begin{equation}
  \tau_{\lambda}^\mathrm{f}
  \equiv
  \int_{-\infty}^{\infty}
  \kappa_{\lambda}\,\rho(0,z)\,{\mathrm{d}}z
  =
  \frac{\kappa_{\lambda}\,M_{\mathrm{d}}}{2\pi\,h_{\text{R,d}}^2}\,,
\label{tauf}
\end{equation}
where $\kappa_{\lambda}$ is the extinction coefficient of the dust. The central edge-on optical depth can be calculated as $\tau_{\lambda}^\mathrm{e}=h_{\text{R,d}}/h_{\text{z,d}} \cdot \tau_{\lambda}^\mathrm{f}$. If the stellar disc has a truncation at some radius, the dust disc is also truncated at this radius.

To find the dust model parameters, we fix the geometry of the \textit{HER}OES galaxies built upon the IRAC~3.6 data models, which were described in Sect.~\ref{sec:IRACdecomp}. We allow only the dust disc parameters, the alignment of the galaxy centre, the inclination angle $i$, and the galaxy position angle $PA$ to change during the fitting. At the same time, the luminosities of the stellar components are determined individually at each wavelength. This simulates the wavelength-dependent behaviour of luminosity ratios of different stellar components (e.g. the bulge-to-total luminosity ratio), which are included in the model \citep{2013A&A...550A..74D, 2014MNRAS.441..869D}. 

We should stress here that we do not make any assumptions about the characteristics of the stellar populations, and about how the emission in different wavebands is linked. We merely fit the stellar emission in every band individually (but assume the same dust distribution at each wavelength). By doing so, we significantly reduce the number of free parameters and, at the same time, use a complex model to describe the stellar and dust components of the galaxy.  

In our \textsc{fitskirt} fitting we use 
the new THEMIS model, which is described in detail in \cite{2017arXiv170300775J} and implemented in \textsc{skirt} by \citet{2015A&A...580A..87C}.
This model is completely built on the basis of interstellar dust analogue material synthesised, characterised and analysed in the laboratory. In this model, there are two families of dust particles: amorphous carbon and amorphous silicates. For the silicates, it is assumed that 50\% of the mass is amorphous enstatite, and that the remaining half is amorphous forsterite. This model is becoming a new paradigm, and is being developed in the frame of the DustPedia project.

The \textsc{fitskirt} computations were done using a multi-core computer and the high performance cluster of the Flemish Supercomputer Center (Vlaams Supercomputer Centrum). For each galaxy, we used about 50 processor units when running \textsc{fitskirt} for one galaxy with the following specification: $5\times 10^5$ photon packages, 200 individuals in a population, and 100 generations (this has been tested to be optimal values by \citealt{2012MNRAS.427.2797D, 2014MNRAS.441..869D}). To estimate the uncertainties on the parameters, we repeated the fitting five times. Taking into account that the average processing time of one fit was about 70 hours (up to 1 week, depending on the dimensions of the reference images), the total time for the whole sample was approximately 122500 CPU hours, or 2450 hours of work spread among 50 cores.  

The results of the fits for all seven \textit{HER}OES galaxies can be found in Table~\ref{tab:FitSKIRT_res}.

\begin{table*}[t] 
  \centering
  \caption{Results of \textsc{fitskirt} fitting for the {\it HER}OES galaxies.} 
  \label{tab:FitSKIRT_res}
  \begin{tabular}{ccccccccc}
    \hline \hline\\[-2ex]
    Parameter & unit & NGC\,973 & UGC\,4277 & IC\,2531 & NGC\,4013 & NGC\,4217 & NGC\,5529 & NGC\,5907 \\
    \hline\\[-1ex]
    $i$ & deg & $89.5\pm0.1$ & $88.7\pm0.2$ & $89.5\pm0.1$ & $89.8\pm0.1$ & $87.2\pm0.2$ & $87.4\pm0.1$ & $84.9\pm0.3$ \\
    $h_\mathrm{R,d}$ & kpc & $8.28\pm0.81$ & $8.85\pm1.00$ & $8.44\pm0.29$ & $3.09\pm0.15$ & $6.27\pm0.14$ & $11.69\pm0.99$ & $7.13\pm0.64$\\
    $h_\mathrm{z,d}$ & kpc & $0.36\pm0.01$ & $0.17\pm0.03$ & $0.25\pm0.01$ & $0.15\pm0.01$ & $0.34\pm0.01$ & $0.26\pm0.02$ & $0.25\pm0.10$ \\
    $M_\mathrm{d}$   & $10^7~M_{\odot}$ & $ 8.17\pm0.92$ & $5.71\pm0.74$ & $4.08\pm0.22$ & $1.08\pm0.09$ & $3.33\pm0.14$ & $5.68\pm0.70$ & $8.97\pm1.31$ \\
    $\tau_\mathrm{V}^{f}$ & --- & $1.35 \pm 0.11$ & $0.83\pm0.08$ & $0.57\pm0.01$ & $1.29\pm0.01$ & $0.97\pm0.01$ & $0.47\pm0.02$ & $2.01\pm0.33$ \\
    $\tau_\mathrm{V}^{e}$ & --- & $31.44 \pm 5.60$ & $42.18\pm2.23$ & $19.26\pm0.18$ & $26.71\pm0.23$ & $17.79\pm0.47$ & $21.41\pm1.01$ & $57.97\pm9.22$\\
    \hline\\
  \end{tabular}
\end{table*} 

\subsection{{\sc{skirt}} simulations} 
\label{sec:skirt}

To perform a uniform and detailed energy balance study of the \textit{HER}OES galaxies, one important step is needed --- we should extend our oligochromatic models from Sect.~\ref{sec:fitskirt} to panchromatic models. We did this to not only reproduce the images at optical and near-infrared wavelengths, but also to restore the entire UV--submm SED and create panchromatic simulations which can be compared to the real ones. As such, our approach consists of two stages.

First, we directly use the oligochromatic \textsc{fitskirt} model obtained earlier in Sect.~\ref{sec:fitskirt} in order to predict the emission of the galaxy in the entire UV to submm domain. 

Panchromatic simulations imply that the properties for both the stars and the dust need to be set over the entire wavelength domain. For the stellar components, we assume a \cite{2003MNRAS.344.1000B} single stellar population SED with a \cite{2003PASP..115..763C} initial mass function and a solar metallicity ($Z=0.02$). For all galaxies in our sample we use the same ages of the stellar components as we determined for IC\,2531: for both the thin and the thick disc, they are about 5\,Gyr, and 8\,Gyr for the bulge (we discuss this in Sect.~\ref{sec:discussion_oligo}). The stellar halo for NGC\,4013 was assumed to be 10 Gyr old.

As follows from \citet{2000A&A...362..138P,2012MNRAS.419..895D,2012MNRAS.427.2797D}, \citetalias{2015MNRAS.451.1728D} and \citetalias{2016A&A...592A..71M}, an additional source of UV luminosity, i.e., a young stellar component, is needed in order to match the observed and model SEDs. As UV radiation is easily absorbed by interstellar dust, the addition of a young population will also affect the dust emission at MIR to submm wavelengths. Therefore, at the second stage we add a young stellar disc, which is completely obscured by dust in the reference images, with the following properties. In our models, we assume a scale height of $1/3$ of the scale height of the dust disc and the same scale length as for the thin disc. However, as shown in \citet{2014A&A...571A..69D}, \citetalias{2015MNRAS.451.1728D}, \citetalias{2016A&A...592A..71M}, and \cite{2017A&A...599A..64V} varying the thickness of the young stellar disc between one third of the dust scale height to one dust scale height barely affects the resulting SED. 

The observed {\it GALEX} \textit{FUV} image traces unobscured star formation regions of the galaxy. The stellar emission spectrum of these stars is described through a \textsc{starburst}99 SED template which represents a stellar population with a constant, continuous star formation rate (SFR) and an evolution up to 100 Myr \citep{1999ApJS..123....3L}. In this case, the initial mass function is a \cite{1955ApJ...121..161S} IMF with masses between 1 and 100\,M$_{\odot}$ and with solar metallicity. The luminosity of this component is constrained by the {\it GALEX} \textit{FUV} (or \textit{NUV} in the case of UGC\,4277) flux density.

The simulated SEDs for both models with (except for NGC\,973, which has very noisy \textit{GALEX} observations) and without a young stellar population are shown in Appendix~\ref{Appendix_SEDs}. The panchromatic simulations, together with the real observations, are presented in Appendix~\ref{Appendix__simulations}.

Below we discuss the results individually for each galaxy.

\section{Results}
\label{sec:Results}

\subsection{NGC\,973} 
\label{sec:NGC973}

\begin{figure}
\centering
\includegraphics[width=8cm, angle=0, clip=]{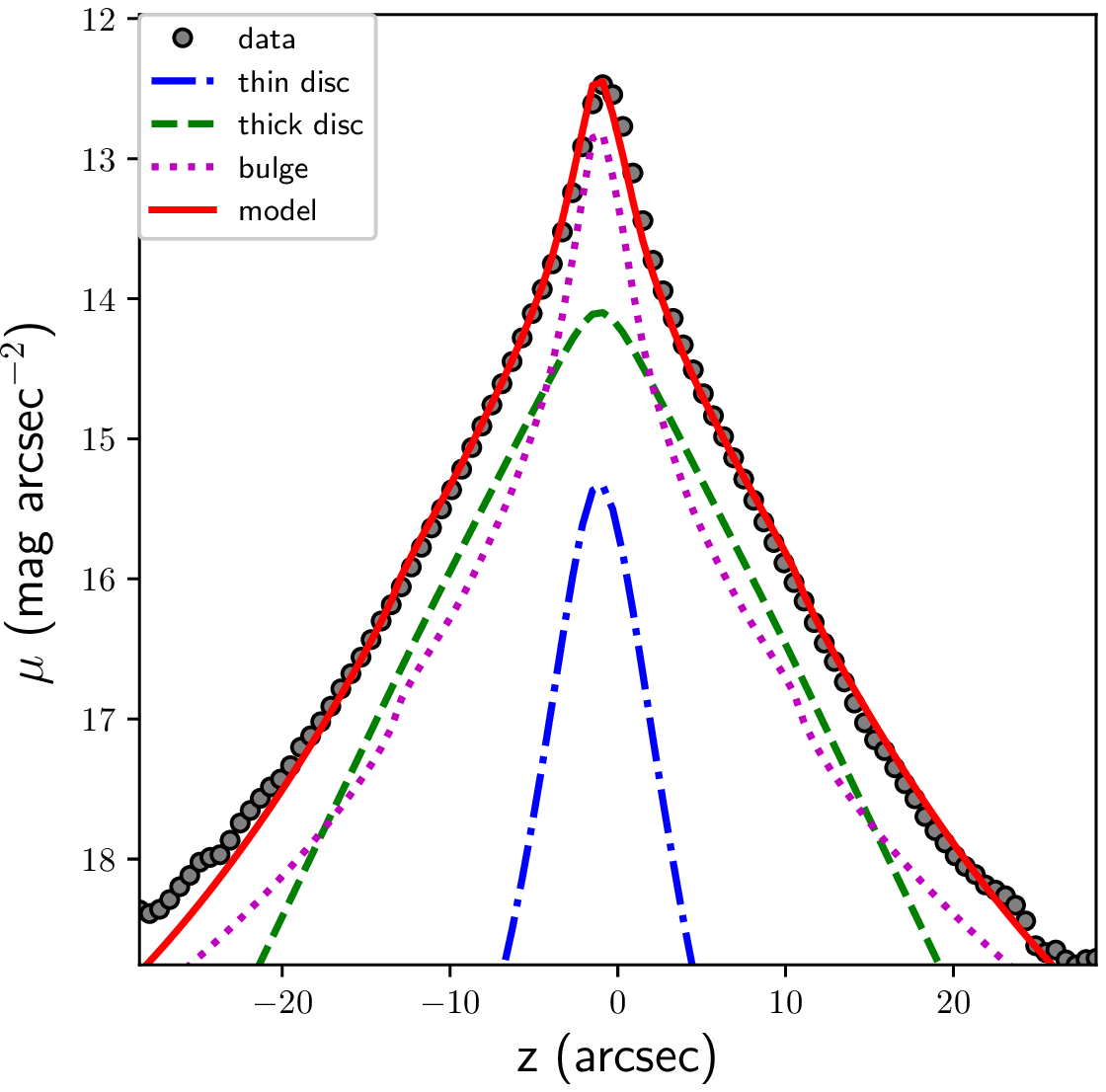}
\includegraphics[width=8cm, angle=0, clip=]{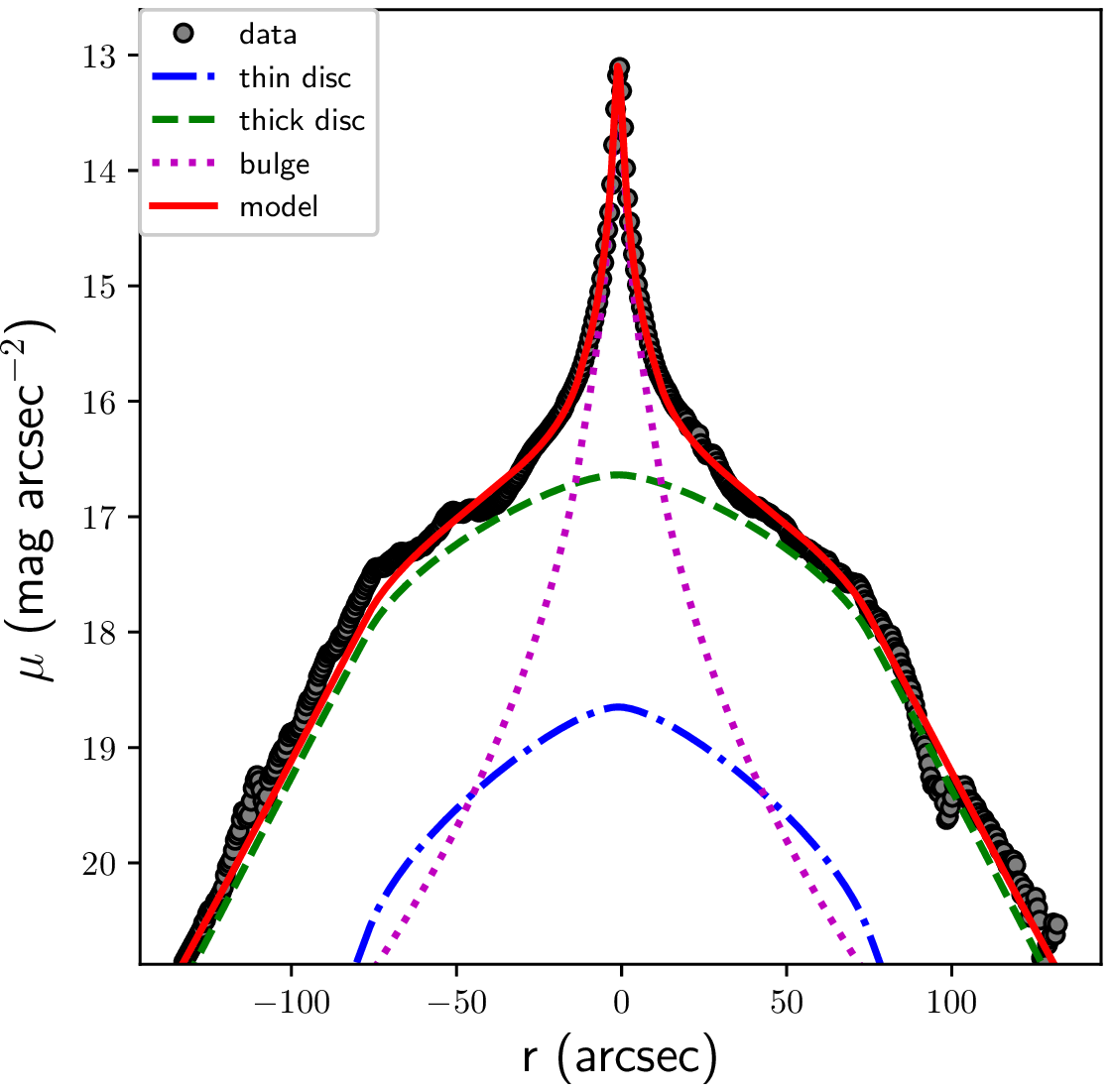}
\caption{Cumulative vertical (lefthand) and horizontal (righthand) profiles of NGC\,973 plotted for its IRAC 3.6\,$\mu$m image, with its overlaid \textsc{imfit} model.}
\label{imfit_NGC973}
\end{figure}

This galaxy classified as an Sb spiral (NED) is the most massive galaxy in our sample, with a radius of the outermost isophote of 25 mag/arcsec$^2$ up to 40~kpc. It is seen almost exactly edge-on; the dust lane is fairly regular, without strong bending or clumpiness. A clear thick B/PS/X structure in the centre is a distinctive feature of the galaxy \citep{2000A&AS..145..405L} which points to the presence of a strong bar \citep{2006MNRAS.366.1121P}. Also, it is classified as a Sy2/LINER object with significant X-ray emission detected by SWIFT and INTEGRAL \citep[see e.g.][]{2007A&A...462...57S, 2012ApJ...761..184W}. According to \citet{2015A&A...582A..18A}, this galaxy does not show strong disc warping in H{\sc{i}}, nor in optical observations.

\begin{figure*}
\centering
\includegraphics[width=\textwidth]{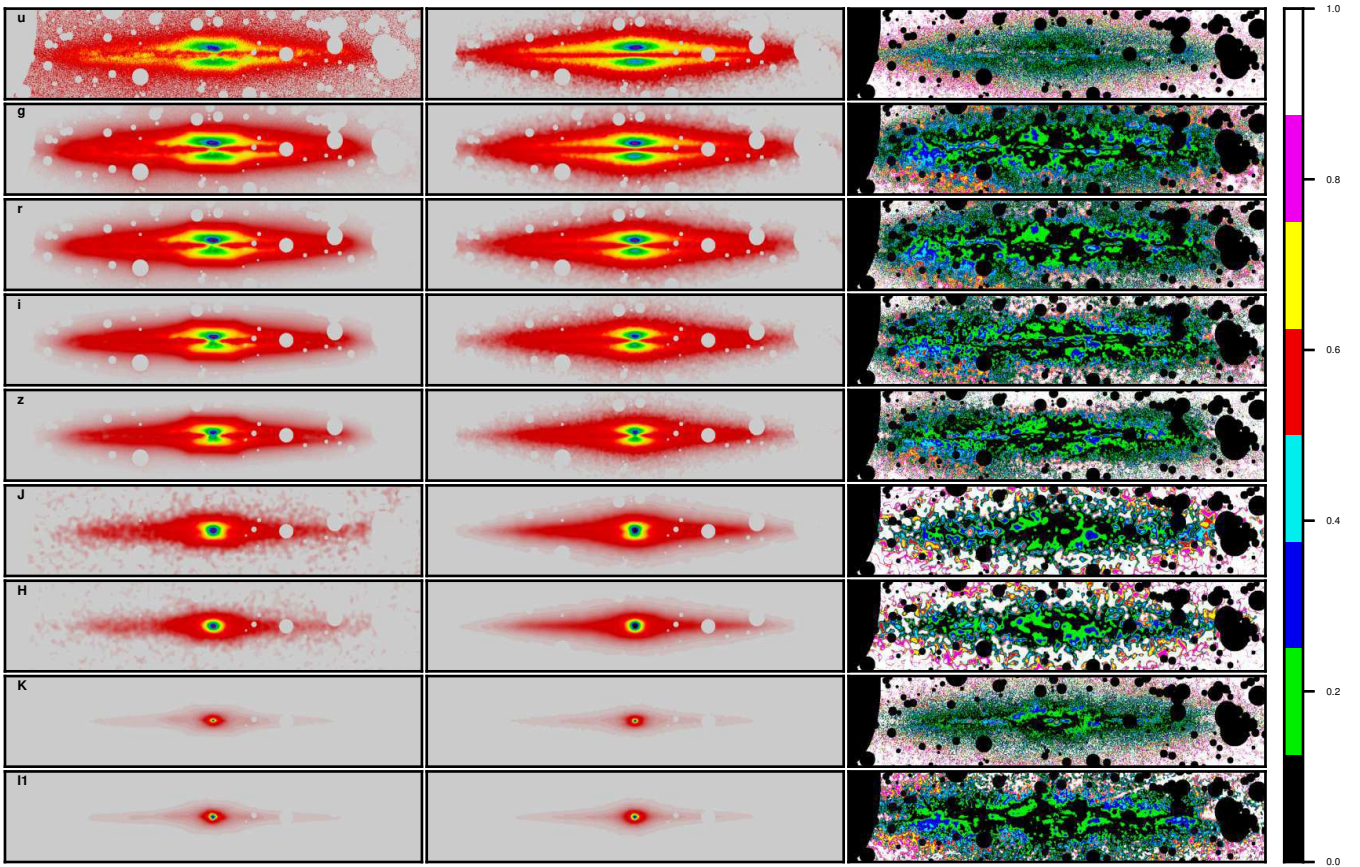}
\caption{Results of the oligochromatic \textsc{fitskirt} radiative transfer fits for NGC\,973. In each panel, the left-hand column represents the observed image, the middle column contains the corresponding fits in the same bands, the right-hand panel shows the residual images, which indicate the relative deviation between the fit and the image (in modulus).}
\label{NGC973_fit_models}
\end{figure*}

\begin{figure*}
\centering
\includegraphics[width=14.5cm, angle=0, clip=]{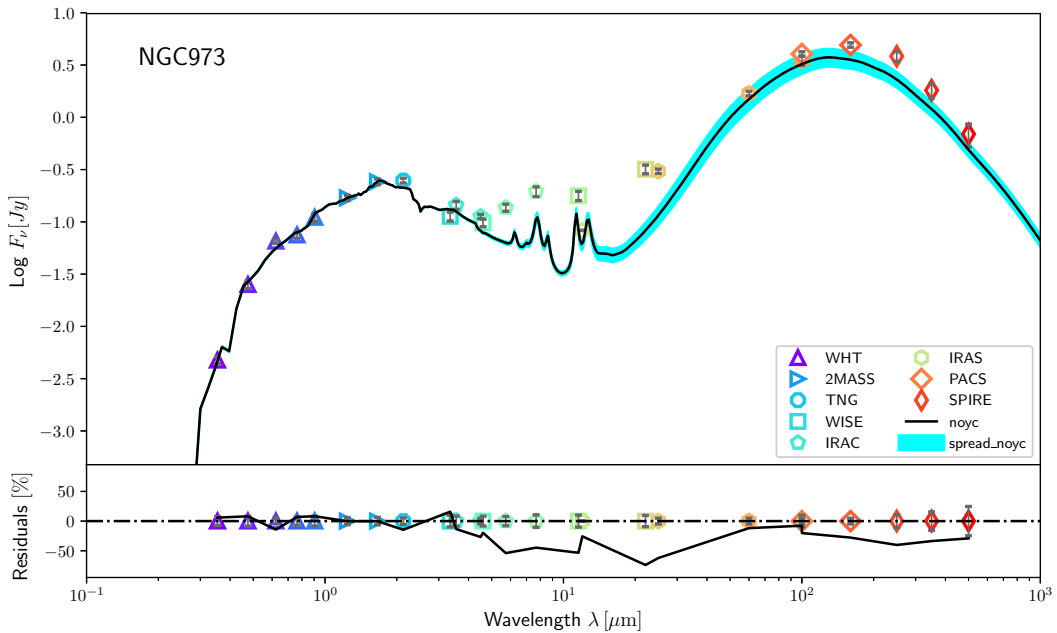}
\caption{The SEDs of NGC\,973 with the THEMIS dust mixture. The coloured markers with error bars correspond to the flux densities listed in Table~\ref{tab:Fluxes}. The bottom panel below the SEDs shows the relative residuals between the observed SED and the model. The cyan spread (denoted as `spread\_noyc') is plotted for the model within one error bar of the best oligochromatic fitting model parameters (the solid black line denoted as `noyc'). }
\label{sed_NGC973}
\end{figure*}

Our \textsc{imfit} model (see Fig.~\ref{imfit_NGC973}) for this galaxy includes three stellar components: thin and thick discs with particularly prominent radial breaks and a bulge (which obviously should be interpreted as a `bulge+bar' component). Surprisingly, the thin disc is remarkably faint compared to the thick disc ($L_\mathrm{t}/L_\mathrm{T}=0.13$). The bulge/bar component fitted with the \ser\ function is flat and has an extended surface brightness profile (the \ser\ index is large, $n_\mathrm{b}=5.48$) with an ordinary effective radius. The galaxy shows the largest fraction of the bulge/bar to the total galaxy luminosity among all sample galaxies ($L_\mathrm{b}/L_\mathrm{tot}=0.56$).

The \textsc{fitskirt} model, which is based on the obtained IRAC~3.6 model for the stellar components plus the dust exponential disc, describes the observations remarkably well, even despite the X-shape structure in the residual images (see Fig.~\ref{NGC973_fit_models}). The ISM appears to be rather optically thick, even in the face-on orientation ($\tau_\mathrm{V}^{f}>1$); by dust mass this galaxy is one of the most massive galaxies in our sample (excluding NGC\,5907). Unfortunately, since this galaxy is rather distant, the \textit{GALEX} observations are very poor for this galaxy, therefore we did not include $NUV$ and $FUV$ fluxes in the global SED. Thus, for this galaxy we did not add a young stellar disc to the model. As follows from Fig.~\ref{sed_NGC973}, our panchromatic simulations somewhat underestimate the observed fluxes in the MIR and FIR domain -- at 160\,$\mu$m, the model predicts 75 per cent of the observed flux. However, this is not as much as for the other galaxies, as we will see below.  
The panchromatic simulations and observations look highly similar in the optical and NIR bands (see Fig.~\ref{map_ngc973}). The WISE $W3$, PACS\,100 and PACS\,160 images reveal a possible spiral arm, or another structure with a high dust concentration, highlighted by a high emission region, which is hard to discern in the SPIRE images. In comparison with our model, all \textit{Herschel} observations show an elongated, radially extended structure, which obviously makes a significant contribution to the total emission in these bands.

\begin{figure*}
\centering
\includegraphics[width=15cm, angle=0, clip=]{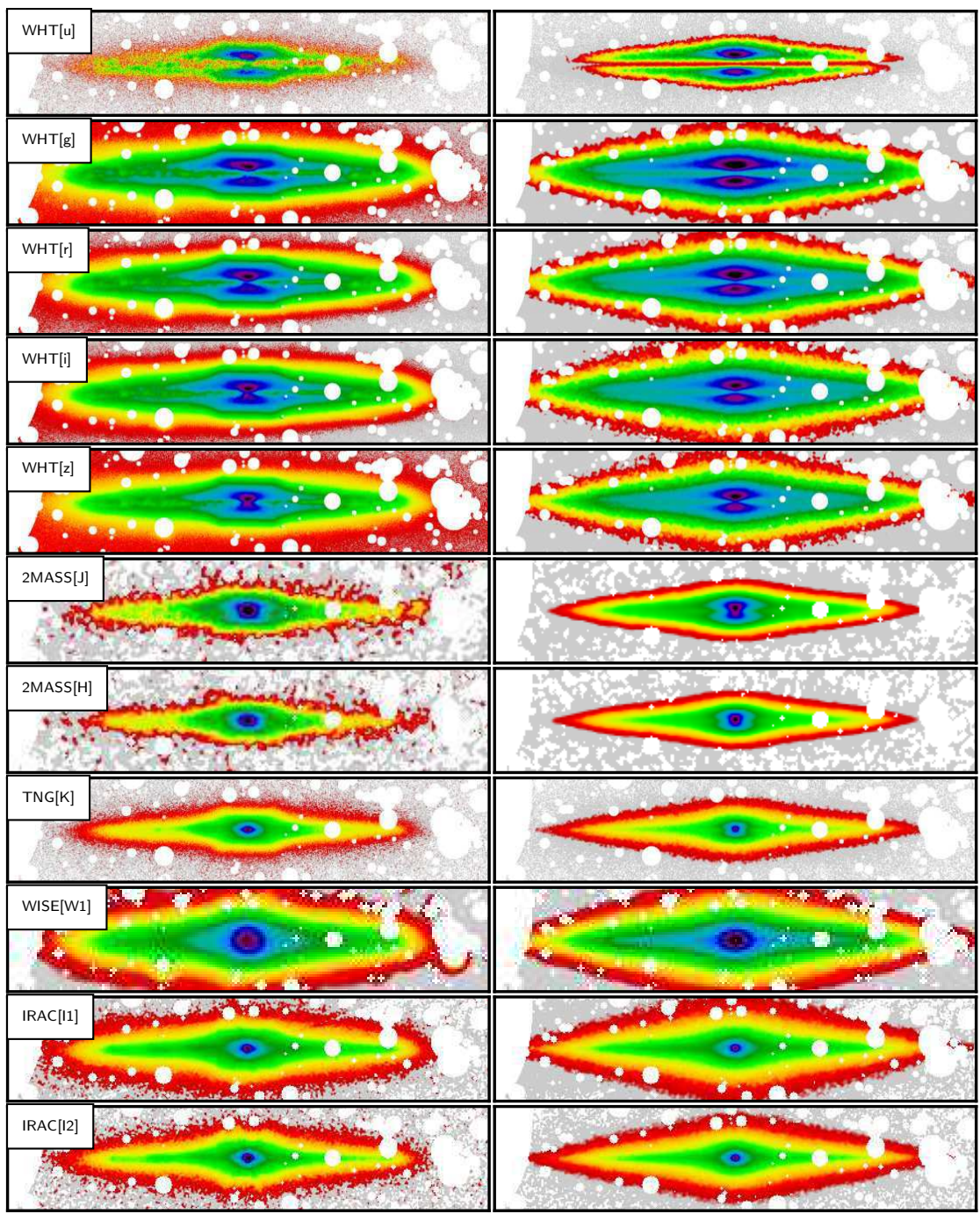}
\caption{Comparison between the observations (left) and panchromatic simulations (right) for NGC\,973. Foreground stars have been masked. Gray-coloured pixels have intensities lower than $2\sigma$ of the background. }
\label{map_ngc973}
\end{figure*}

\addtocounter{figure}{-1}
\begin{figure*}
\centering
\includegraphics[width=15cm, angle=0, clip=]{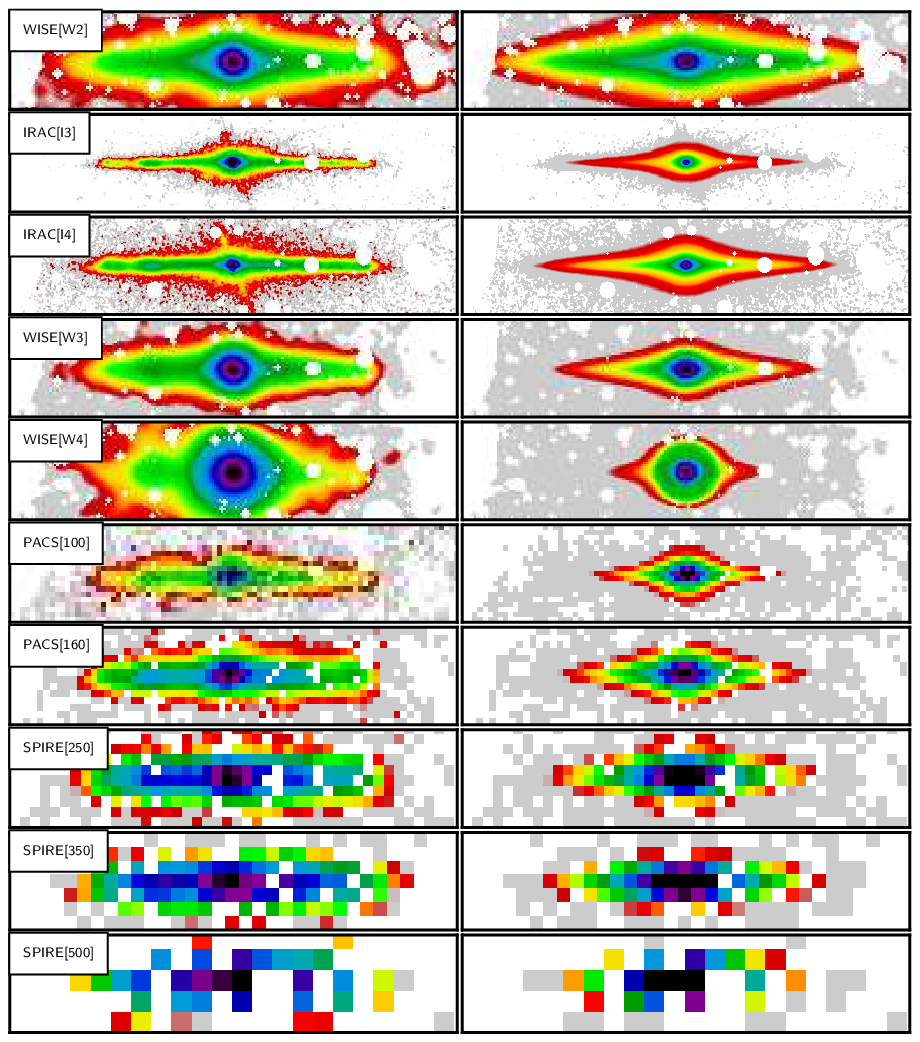}
\caption{(continued)}
\end{figure*}


\subsection{UGC\,4277} 
\label{sec:UGC4277}

UGC\,4277 is a flat Sc/Scd galaxy, one of the largest galaxies (the semi-major axis $sma=36.95$~kpc) in our sample. It is seen almost exactly edge-on; the dust lane is very prominent in the optical. One of the noticeable features is an integral-shaped optical warp. \citet{2015A&A...582A..18A} recently showed that the outer regions of the H{\sc{i}} disc demonstrate a small warp as well, with the position angle deviating at most about 4 degrees from its central value.  UGC\,4277 has a total atomic hydrogen mass of $2.03\times10^{10}\,M_{\odot}$ and, similarly to NGC\,973, the H{\sc{i}} disc has a clumpy and irregular structure. Interestingly, this galaxy is included in the Catalogue of Isolated Galaxies \citep[CIG,][]{1973AISAO...8....3K}, therefore its evolutionary status may differ from those galaxies residing in a group or a cluster.

For the \textsc{imfit} modelling (which was based on the $K$-band image since there is no \textit{Spitzer} data for this galaxy), we used an ordinary (with no breaks since they are likely to be smoothed because of insufficient angular resolution) double-exponential profile for both the thin and thick disc and a \ser\ bulge (see Fig.~\ref{imfit_UGC4277}). Both discs are comparable by luminosity and radial extent. The bulge has a rather large effective radius with a \ser\ index of about 2.6. 

The \textsc{fitskirt} model looks exceptionally good -- only the disc warp is noticeable in the residual images (see Fig.~\ref{UGC4277_fit_models}). The dust attenuation is represented particularly well for the whole set of reference images. It is interesting to note that, photometrically, the bulge shows no signature of an X-shape structure in the centre. However, we can see a prominent B/PS structure in the residual images in the $g$, $r$ and $i$ bands. This is evidence that UGC\,4277 has a bar which is not so extended as in the other \textit{HER}OES galaxies or it is not viewed purely side-on \citep{2006MNRAS.370..753B}.

Unfortunately, the \textit{GALEX FUV} observation of this galaxy is too noisy, therefore we used the \textit{NUV} flux to match the galaxy observation and the model (with an additional young disc) in this band. The model with the young stellar disc (see Fig.~\ref{sed_UGC4277}) describes the observations in the MIR range significantly better, but the discrepancy in the FIR is still present, with an underestimated cold dust emission at wavelengths larger than 200$\mu$m, by a factor of about 2. Consequently, the panchromatic simulations cannot entirely describe the observations. The panchromatic simulations in the optical and MIR are compatible with the observations (see Fig.~\ref{map_ugc4277}). However the comparison between the WISE~$W3$ and the \textit{Herschel} observations with the models indicates that the real observations are again more radially extended, as in the case of NGC\,973. Other structural details (e.g. spirals), which would be present in this galaxy, are washed away because of poor resolution.

\subsection{IC\,2531} 
\label{sec:IC2531}

This galaxy has been described in great detail in \citetalias{2016A&A...592A..71M}, therefore we refer the interested readers to that article.

\subsection{NGC\,4013} 
\label{sec:NGC4013}

NGC\,4013 is the most compact disc galaxy in our sample. It is of the Sb type, with several outstanding features which were studied by different authors. The galaxy has a barely visible optical warp \citep{1991A&A...242..301F}.
\citet{2009ApJ...692..955M} found a giant, arcing stellar structure of low surface brightness around the galaxy. \citet{2011ApJ...738L..17C} produced a decomposition considering three flattened stellar components in this galaxy: thin disc + thick disc + halo.  

The atomic gas content of NGC\,4013 was analysed in great detail by \citet{2015ApJ...808..153Z} with similar techniques as used in \citet{2015A&A...582A..18A}. They found that the gas disc of this galaxy is strongly warped, both along the major axis (in the plane of the sky) and along the line of sight, and shows a lag (decreasing rotation velocity with increasing height above the plane) that is strongest at small radii. Finally, they found no H{\sc{i}} gas at large distances from the plane, despite the presence of extra-planar dust and ionised gas \citep[see e.g,][ and the references therein]{2003ApJS..148..383M}.

Since this galaxy has a bright stellar halo, we used a four-component model `thin disc + thick disc + bulge + halo' (see Fig.~\ref{imfit_NGC4013}). The two discs have a broken double-exponential profile, the bulge and the halo are described by a \ser\ function. The thick disc is the most luminous component of the galaxy, whereas the bulge is compact and round. A foreground star almost coincides with the position of the nucleus of the galaxy, which is also covered by a dark dust lane. \citet{1997ApJS..112..315H} found that this galaxy hosts an AGN in its centre, therefore our bulge profile should contain it as well. The halo is indeed noticeably bright (its fraction to the total luminosity is 15\%) and extended ($r_\mathrm{e,h}\approx7.4$\,kpc), the \ser\ index is close to 1, therefore it can be called the second thick disc. 

Since our \textsc{fitskirt} model does not take into account the X-shape structure, this feature seems a prominent detail in the residuals in Fig.~\ref{NGC4013_fit_models}. Despite this and some unfitted clumped structures on the left side of the galaxy plane, our \textsc{fitskirt} model describes the observations fairly well (the SDSS $u$ band image is obviously too noisy and, thus, the comparison in this band is meaningless).

Our panchromatic simulations (see Fig.~\ref{sed_NGC4013} and \ref{map_ngc4013}) report the same dust energy balance problem -- even with an additional young stellar component, the model emission in the MIR/FIR domain is far below the observed one. 

Since a lot of data are available for this galaxy, it is interesting to compare the observations and the model in detail. In the \textit{FUV} and \textit{NUV} bands, we can see that our galaxy is observed as a patchy disc consisting of a young stellar population with star forming regions, contrary to our model which looks more like an attenuating lane. The $u$ band model is closer to the corresponding observation. All other optical--4.6$\mu$m observations are described exceptionally well by our model. The MIR--FIR observations reveal a probable ring structure (since the inclination angle is almost exactly $90^{\circ}$, we can only assume this by the typical ``shoulders'' for rings on the SB profile) in the plane of the galaxy which strongly adds to the emission in this wavelength range, for the hot as well as for the cold dust.

\subsection{NGC\,4217} 
\label{sec:NGC4217}

NGC\,4217 is another Sb-type galaxy in our sample. Visually, it has a noticeably thick (as compared to the other \textit{HER}OES galaxies) and somewhat warped dust lane, which extends up to the galaxy outskirts.

For the main disc of NGC\,4217, \citet{2015A&A...582A..18A} measured an atomic hydrogen mass of $2.50 \times 10^9\,M_{\odot}$. In addition, they also detected a separate ring of gas outside (and coplanar with) the main gas disc, with an H{\sc{i}} mass of at least $3.81 \times 10^8\,M_{\odot}$. This ring has no optical counterpart and could be the remnant of a recent minor merger. The main gas disc is mostly flat and shows only a small warp in the outer parts where it meets the ring structure. 

Our \textsc{imfit} fitting was performed for the model consisting of two broken thin and thick discs and a bulge (see Fig.~\ref{imfit_NGC4217}). As one can see from Table~\ref{tab:IRAC_dec_res}, the thin disc is less luminous than the thick disc, but they have similar inner disc scale lengths. The bulge is surprisingly compact (the \ser\ index is small) and its fraction to the total galaxy luminosity is minor. The thick disc has the smallest disc scale length-to-scale hight ratio 3.8 -- apparently, this galaxy looks thicker than the others.

The \textsc{fitskirt} model is quite adequate with regard to the observations (see Fig.~\ref{NGC4217_fit_models}). Only some traces of the dust attenuation are left in the residuals in the $r$, $i$ and $z$ bands which may be related to the patchy nature of the dust disc in this galaxy.

Again, the \textsc{skirt} simulations cannot describe the real observations in the whole MIR-FIR domain (see Fig.~\ref{sed_NGC4217}). The typical factor of flux underestimation is 4 to 5. However, the comparison of the panchromatic simulations and real observations for this galaxy is very good (see Fig.~\ref{map_ngc4217}), even for the \textit{FUV} and \textit{NUV} bands. The WISE $W3$ and $W4$ observations reveal some asymmetric structures around the centre, which may be related to a spiral arm. Despite the discrepancy between the observed and model SED, the structure of the emitting cold dust in the \textit{Herschel} observations is particularly well described by our model. It is likely that the characteristic features, which we saw in the FIR in the previous galaxies, are not spatially resolved here. 

\subsection{NGC\,5529} 
\label{sec:NGC5529}

NGC\,5529 is an Sc edge-on galaxy with slightly lower than $90^{\circ}$ inclination ($87.9^{\circ}$). A prominent optical warp and B/PS structure in the centre are the characteristic features of this galaxy.

It is part of a rich galaxy group with at least 16 other members \citep{2007A&A...474..461I}. \citet{2004MNRAS.352..768K} already studied the atomic gas content of this galaxy and discovered that NGC\,5529 is connected to two of its satellites through H{\sc{i}} bridges. They also found a bright H{\sc{i}} ridge in the major axis position-velocity diagram, which could indicate the presence of a strong spiral arm. On the approaching (NW) side the atomic gas disc shows the same warp as the stellar disc, while the receding (SE) side of the gas disc appears more strongly warped. \citet{2015A&A...582A..18A} also found a moderate warp along the line of sight and detected a radial inflow of gas at about 15 km/s in the outer half of the disc. The latter is probably linked to the accretion of gas from the satellites. \citet{2015A&A...582A..18A} measured a total atomic hydrogen mass of $2.69 \times 10^{10}\,M_{\odot}$.

Our \textsc{imfit} model consists of a thin broken disc, a thick disc, and a bulge (see Fig.~\ref{imfit_NGC5529}). The thin and thick discs have an equal contribution to the total luminosity. The bulge/bar structure is rather extended, with the \ser\ index larger than 2 -- a hint that this galaxy may have a classical bulge.

Despite the disc warp (which was masked out during the fitting), the \textsc{fitskirt} decomposition looks quite satisfactory (see Fig.~\ref{NGC5529_fit_models}), with some traces of the dust content in the residuals in the $r$ and $i$ bands. However, the \textsc{skirt} simulations (see Fig.~\ref{sed_NGC5529}) reveal a systematic underestimation of flux densities at all wavelengths larger than 10$\mu$m\ by a factor of 4 to 5, for the models both with and without the young disc. The comparison of the panchromatic simulations with the real galaxy images (see Fig.~\ref{map_ngc5529}) indicates that starting from the WISE $W3$ band with increasing wavelength, the 2D SB distribution of the real observations becomes more complex, with round isophotes (whereas our models have clear discy isophotes at all wavelengths) and apparent existence of spirals (which are distinct emission features on both sides from the centre, from WISE $W3$ to SPIRE 250). Interestingly, our \textit{NUV} model does not look similar to what we see in real -- the absorption lane in the panchromatic simulation versus the emission stripe in the observation.

\subsection{NGC\,5907} 
\label{sec:NGC5907}
NGC\,5907 is the nearest and flattest {\it HER}OES galaxy, also of Sc morphological type. NGC\,5907 is seen within a few degrees of edge-on ($84.9^{\circ}$). Its tiny bulge is almost hidden behind the opaque dust lane. \citet{2008ApJ...689..184M} discovered extremely faint extended arcing loops around the galaxy, which, as they argue, were produced by low-mass satellite accretion in the hierarchical formation scenario. Another interesting feature of this galaxy is an excess of light at the periphery of the minor axis profile, which was interpreted by \citet{1994Natur.370..441S} as an extended faint halo. \citet{2014A&A...567A..97S}, however, showed that this excess can also be explained by diffuse scattered light produced by the extended PSF (we discuss this issue in Sect.~\ref{sec:discussion_stellar}).
 
The atomic gas disc of NGC\,5907 is strongly warped and has a total H{\sc{i}} mass (as measured by \citealt{2015A&A...582A..18A}) of $1.99 \times 10^{10}\,M_{\odot}$. This warp was already noticed by \citet{1976A&A....53..159S}. Later, \citet{1998ApJ...504L..23S} also discovered a small companion galaxy toward the NW of NGC\,5907 in their H{\sc{i}} map. \citet{1976A&A....53..159S} further found that the gas disc of NGC\,5907 is also warped along the line of sight and shows strong asymmetries between the approaching and receding side. An extended reservoir of low-density gas was detected around the main disc.

Our stellar model for this galaxy is `broken thin disc + thick disc + bulge' (see Fig.~\ref{imfit_NGC5907}). The thin disc dominates over the thick disc, which is indeed particularly vertically extended and faint. The small bulge is exponential and flattened, and therefore can be referred to as a pseudo-bulge \citep{2009MNRAS.393.1531G, 2010ApJ...716..942F}.

The {\sc{fitskirt}} model and the real observations are plotted together in Fig.~\ref{NGC5907_fit_models}. In the panel with residuals one can see that for the optical data, our fit does not perfectly restore the observation -- a clear residual lane is well visible. However, the overall model looks good and compatible with the observations.

Nevertheless, if we look at the parameters listed in Table~\ref{tab:FitSKIRT_res}, the {\sc{fitskirt}} model for NGC\,5907 differs from the others. From Table~\ref{tab:FitSKIRT_res} we can see that the dust mass and the face-on optical depth are the largest among all {\it HER}OES galaxies. It follows from our modelling that the ISM of this galaxy is obviously non-transparent. However, if we look at Fig.~\ref{sed_NGC5907}, we can see that our panchromatic simulations systematically underestimate flux densities in the optical domain, whereas at FIR wavelengths our model with an additional young disc is able to predict the SED behaviour exceptionally well (this is the only {\it HER}OES galaxy for which our model follows the observations in the FIR; our panchromatic simulations are compatible with the real observations as well, see Fig.~\ref{map_ngc5907}). However, the serious discrepancy in the optical bands suggests that the {\sc{fitskirt}} model is incorrect due to an overestimation of the dust mass. Perhaps, one of the possible explanations for this erroneous model is that our stellar model is not correct either (e.g. the thick disc is, in fact, not real and produced by scattered light of the PSF).

Interestingly, in the IRAC 4.5~$\mu$m band image there is a prominent detail above the plane of the disc. However, since it is not seen in the other bands, including the WISE $W2$ band close to IRAC 4.5~$\mu$m, we interpret this feature as an artefact.

\subsection{Comparison with other works} 
\label{sec:comparison}
The direct comparison of our results with other works is impossible because in this study we applied a new technique which is based on the two-step decomposition.
This novel strategy has not been applied before, except in \citetalias{2016A&A...592A..71M}. However, for uniformity, we provide some indirect comparison with other authors who carried out their own decomposition analysis of the {\it HER}OES galaxies in the same band(s).

Our {\sc{imfit}} models can be compared to the S$^4$G results \citep{2015ApJS..219....4S} for four {\it HER}OES galaxies: NGC\,4013, NGC\,4217, NGC\,5529, and NGC\,5907. However, some caution should be made. Firstly, our decomposition is performed for 3D stellar discs, whereas the GALFIT code \citep{2010AJ....139.2097P}, which was used in the S$^4$G project, works only with projected intensities. Secondly, for these galaxies the S$^4$G models do not contain a bulge component (for NGC\,4013, NGC\,4217, and NGC\,5529, the central component is a nucleus). In our modelling, all our galaxies have a bulge with an effective radius larger than the FWHM of the PSF (where for IRAC ~3.6$\mu$m the Gaussian PSF FWHM is $2.1\arcsec$). So, even being compact, the bulges in our models should have physical meaning. Thirdly, the GALFIT decomposition from \citet{2015ApJS..219....4S} assumes an isothermal disc geometry with the disc height scale $z_0$, whereas in our decomposition we use a model of a double-exponential disc, although we can compare the height scales of both the exponential and isothermal disc for vertical distances $z \gg h_\mathrm{z}$: $z_0 = 2\,h_\mathrm{z}$. The results of our and the S$^4$G decomposition are presented together in Table~\ref{table4}. From the table we can see that the retrieved disc scale lengths are similar, whereas the disc scale height may differ by a factor of 0.5--2, which can be explained by the reasons listed above.

\begin{table*}[t] 
  \centering
  \caption{Comparison of the results of the photometric decomposition from this work and \citet{2015ApJS..219....4S}. All scales are given in arcsec.} 
  \label{table4}
  \begin{tabular}{lcccccccc}
    \hline \hline\\[-2ex]
Galaxy & \multicolumn{4}{c}{This work} & \multicolumn{4}{c}{S$^4$G} \\
& $h_\mathrm{R,inn}^t$ & $h_\mathrm{z}^t$ & $h_\mathrm{R,inn}^T$ & $h_\mathrm{z}^T$ & $h_\mathrm{R,inn}^t$ & $h_\mathrm{z}^t$ & $h_\mathrm{R,inn}^T$ & $h_\mathrm{z}^T$ \\
\hline\\[-1ex]
NGC\,4013  & 59.3 & 4.0 & 27.1 & 10.9 & ---  & --- & 27.4 & 6.4  \\
NGC\,4217  & 34.8 & 2.3 & 37.2 & 19.6 & 28.9 & 3.6 & 38.2 & 17.7 \\
NGC\,5529  & 39.7 & 2.6 & 31.9 & 10.0 & 27.2 & 2.9 & 28.6 & 7.8  \\
NGC\,5907  & 60.1 & 4.1 & 70.8 & 47.6 & 52.9 & 9.2 & ---- & ---- \\[0.5ex]     
    \hline\\
  \end{tabular}
\end{table*} 

Since radiative transfer models have been obtained before for all {\it HER}OES galaxies, it is interesting to compare the results of this work with the literature. Table~\ref{table5} lists the results taken from \citetalias{1997A&A...325..135X}, \citetalias{1999A&A...344..868X}, and \citetalias{2007A&A...471..765B}. In these works, galaxies have a simple `bulge+disc' model plus a dust disc. The shape of the bulge profile (defined by the \ser\ index) was not fitted. The number of bands, which were used in the analysis, is limited -- from two to five (optical and near-infrared). The fitting procedures are also different in each work. Therefore, this comparison is indirect and serves only for the completeness and reliability of our work. From the comparison, we can see that the inclination angle $i$ is very close for all the works, with the exception of NGC\,5907. For this galaxy, our results slightly differ from \citetalias{1999A&A...344..868X}. The comparison between our results and \citetalias{2007A&A...471..765B} for NGC\,4217 and NGC\,5529 is very good, for all dust parameters. For NGC\,973, our model is very different from that in \citetalias{1997A&A...325..135X}: our dust disc is 50\% less extended and thinner, with a higher optical depth. For UGC\,4277, the comparison for the dust scales is satisfactory. For NGC\,4013, our study, \citetalias{2007A&A...471..765B}, and \citet{2013A&A...550A..74D} provide similar results for the dust scale length and scale height. We should notice that the optical depth for almost all galaxies is rather different in all compared works. The same is true for the bulge-to-total luminosity ratio $B/T$. In addition, it is particularly hard to estimate the uncertainties on these parameters. 

Another reason for the differences with previous results must be also due to differences in the dust models for the optical constants of the
dust grains which were used previously in the literature and in the current work. For example, \citet{2007A&A...471..765B} used the dust scattering properties from the Milky Way dust grain model of \citet{2001ApJ...548..296W}. In this paper, we applied the new THEMIS dust-grain model which is built upon the laboratory-measured properties of interstellar dust analogue materials.  Obviously, even if all dust models are tuned to fit the observed extinction curve of the Milky Way, there will be a different contribution from the different grain components, which constitute these models, to the scattering and absorption, resulting in some changes in the output RT models. This makes the comparison of the RT modelling results even more difficult.

On the whole, taking into account that the works in our comparison used different stellar and dust mixture models, the results for the dust component are compatible.

\begin{table}[h] 
  \centering
  \caption{Comparison of the results for the dust disc taken from different sources. All quantities from the literature are taken for the $V$ band.} 
  \label{table5}
  \begin{tabular}{lccccc}
    \hline \hline\\[-2ex]
 Source & $i$  & $h_\mathrm{R,d}$ & $h_\mathrm{z,d}$ & $\tau_\mathrm{V}^{f}$ & $B/T_\mathrm{V}$ \\
    \hline\\[-1ex]
 NGC\,973 \\
 This work & 89.5 & 8.28 & 0.36 & 1.35 & 0.52 \\
 (1) & 89.6 & 16.34 & 0.59 & 0.52 & 0.63 \\[0.5ex]
 \hline\\[-1ex]
 UGC\,4277\\
 This work & 88.7 & 8.85 & 0.17 & 0.83 & 0.17 \\
 (3) & 88.9 & 12.52 & 0.25 & 0.49 & 0.41 \\[0.5ex]
 \hline\\[-1ex]
 NGC\,4013\\
 This work & 89.8 & 3.09 & 0.15 & 1.29 & 0.13 \\ 
 (4) & 90.0 & 3.00 & 0.19 & 0.97 & 0.54 \\
 (3) & 89.9 & 2.66 & 0.18 & 1.46 & 0.68 \\
 (2) & 89.7 & 3.92 & 0.21 & 0.67 & 0.75 \\[0.5ex]
 \hline\\[-1ex]
 NGC\,4217 \\
 This work & 87.2 & 6.27 & 0.34 & 0.97 & 0.10 \\
 (3) & 88.0 & 6.72 & 0.38 & 1.26 & 0.59 \\[0.5ex]
 \hline\\[-1ex]
 NGC\,5529 \\
 This work & 87.4 & 11.69 & 0.26 & 0.47 & 0.09 \\
 (3) & 86.9 & 11.67 & 0.26 & 0.68 & 0.21 \\
 (2) & 87.4 & 11.87 & 0.54 & 0.65 & 0.36 \\[0.5ex]
 \hline\\[-1ex]
 NGC\,5907\\
 This work & 84.9 & 7.13 & 0.25 & 2.01 & 0.20 \\
 (2) & 87.2 & 7.84 & 0.16 & 0.49 & 0.40 \\[0.5ex]     
    \hline\\
  \end{tabular}
\parbox[t]{90mm}{ {\bf Notes.} (1) \citetalias[Table 2 (de Vacouleurs bulge),][]{1997A&A...325..135X}, (2) \citetalias{1999A&A...344..868X}, (3) \citetalias{2007A&A...471..765B}, (4) \citet{2013A&A...550A..74D}. Inclination angle $i$ is given in degrees. Geometrical scales $h_\mathrm{R,d}$ and $h_\mathrm{z,d}$ are given in kpc (we reconverted original values from these works using the scales from Table~\ref{tab:General_info}). For our study, $B/T_\mathrm{V}$ (except IC\,2531) were calculated by means of interpolation at $\lambda=0.551\mu$m. For (1) and (2), $B/T_\mathrm{V}$ were calculated using the values given in their table~2 and tables~1--6, respectively.}
\end{table}

\section{Discussion} 
\label{sec:discussion}

\subsection{On our stellar models} 
\label{sec:discussion_stellar}

Four galaxies in our sample have quite prominent B/PS structures which are not described in our modelling. As was shown in many studies, B/PS features appear because of the buckling instability of a galactic bar \citep{1981A&A....96..164C,1990A&A...233...82C,1991A&A...252...75P,1999AJ....118..126B,2002MNRAS.330...35A}, and, thus, they point to the presence of another stellar component in the galaxy, a stellar bar. In principle, this might affect the results of our decomposition, especially concerning the thick disc and the bulge. However, as we have stressed before, the bulges in our study are, in fact, a superposition of at least two components at the centre -- a bulge and a bar (if it is present in the galaxy). Also, as the bulge/bar structure is less extended than the stellar discs (we consider mainly late-type spirals with a small contribution of the central component to the total galaxy luminosity), its influence on the retrieved disc parameters should not be significant. 

Recently, a 2D Ferrers profile \citep{ferrers} with a modulated $m=4$ Fourier mode has been applied to describe B/PS structures in 22 edge-on galaxies \citep{2017MNRAS.471.3261S}. A multicomponent decomposition, using which they were able to retrieve disc, bulge, B/PS and, for some galaxies, even ring structural parameters fairly well, also helped to derive geometrical properties of the X-shape structures. For the three galaxies in our sample IC\,2531, NGC\,4013, and NGC\,5529, the mean length of the rays in the X-shape structure were estimated, and they were found not to exceed 36\% of the disc scale length. This proves that, while B/PS structures are not included in our modelling strategy, they should not have a large impact on the decomposition results for the stellar discs. 
Apart from that, a new 3D luminosity-density function based on the Ferrers ellipsoid has been implemented in the last version 1.5 of \textsc{imfit}. For future applications, this component can be used to accurately model 3D bar (B/PS) structures in galaxies. 

Another concern can be the `thin disc+thick disc' model which we used in our fitting. Studies by \citet{2008MNRAS.388.1521D} and \citet{2014A&A...567A..97S, 2015A&A...577A.106S} demonstrated that the extended wings of the PSF can seriously affect our interpretation of different low surface brightness (LSB) features in galaxies, such as LSB haloes and possibly thick discs, which, in fact, may occur as artefacts produced by diffuse scattered light. In particular, \citet{2014A&A...567A..97S} showed for the {\it HER}OES galaxy NGC\,5907 that its faint halo (or the thick disc in our study) can be interpreted as scattered light from a thin disc with a high surface brightness. Unfortunately, there was no possibility to create extended PSFs for the reference frames used in our study. Also, the PSF model should be at least as large as the observed galaxy, therefore PSF convolution in this case would strongly increase the time of the modelling, for both the \textsc{imfit} and \textsc{fitskirt} analysis. We should also note that NGC\,5907 is the most spatially resolved galaxy in our sample, therefore the effects of the scattered light can be seen for it better than for smaller galaxies. Whether the obtained \textsc{imfit} models are physically valid or not, is hard to say without proper extended PSFs for all the reference images. However, recent work by \citet{2018A&A...610A...5C} on a sample of 141 edge-on galaxies convincingly proves that the thick discs `are not an artefact caused by scattered light as has been suggested elsewhere'. They used a robust technique to fit a model consisting of two stellar discs to the SB profiles of these galaxies perpendicular to the galaxy plane. Also, they used a revised PSF for IRAC~3.6 with extended wings up to 2.5\,arcmin. Unfortunately, this work appeared only recently, and it was impossible to use this new kernel earlier. Also, half of our galaxies are larger than their new PSF. Therefore we would not be able to perform a decomposition of all {\it HER}OES galaxies with the same level of accuracy. Anyway, since the thick disc is more extended in the vertical direction and faint in the dust disc plane (the average thick disc-to-thin disc scale height ratio for our sample is larger than 5.5), its influence on the dust emission in the infrared should be minimal.

Two galaxies in our sample (NGC\,973 and NGC\,4013) exhibit some AGN characteristic features. In our analysis, however, we do not include this component because of the following reasons. First, NGC\,973 and NGC\,4013 host low-luminosity AGNs (they contribute no more than several per cent to the total galaxy luminosities in the IRAC 3.6~$\mu$m band, \citealt{2015ApJS..219....4S}). Second, the dusty torus emission has a peak between 10 and 30~$\mu$m (unless a very large, non-physical torus is considered, see \citealt{2006MNRAS.366..767F,2008ApJ...685..160N,2012MNRAS.420.2756S}). After the peak, the emission decreases rapidly with a Rayleigh-Jeans curve. Third, the dust emission from the dust torus, which is heated by photons from the accretion disc, might influence the MIR SED, but in the FIR a very minor contribution is expected from this heating.

In this work, we applied the 3D broken exponential disc geometry to a small sample of edge-on galaxies, probably for the first time. For UGC\,4277, the break radius was not robustly determined. Therefore for this galaxy we used a simple exponential disc. We claim that using the model of the broken exponential disc should give more reliable results for measuring the disc scale length and the broken radius since with this model we are able to describe the observed SB distribution in galaxy discs more accurately. For the other six {\it HER}OES galaxies, the average radius at which the break occurs is $0.64 \pm 0.08\,R_\mathrm{25}$ (or $18.01\pm5.85$~kpc), whereas \citet{2012MNRAS.427.1102M} found it to be slightly larger: $0.77 \pm 0.06\,R_\mathrm{25}$ (or $8\pm1$~kpc). The discrepancy may be caused by the fact that some of the {\it HER}OES galaxies are quite large, of Sb type, whereas most of the galaxies which were considered by \citet{2012MNRAS.427.1102M} are rather small, bulgeless galaxies. The nature of the disc breaks and truncations has been well studied \citep[see e.g.][]{2001MNRAS.324.1074D,2004ASSL..319..713P,2007MNRAS.378..594P,2006A&A...454..759P,2006A&A...455..467F,2012ApJ...759...98C}, and as proposed in \citet{2012MNRAS.427.1102M}, the first phenomenon may be caused by a threshold in the star formation (and, thus, change in the stellar population) at this radius, while the second one can be a real drop in the stellar mass density of the disc, which is associated with the maximum angular momentum of the stars or the protogalactic cloud \citep{1987A&A...173...59V,1988A&A...192..117V}.

\subsection{On our RT models and the dust energy balance problem} 
\label{sec:discussion_oligo}

In our approach, for all the galaxies we applied the same model of a double-exponential dust disc. However, the comparison of our \textsc{skirt} simulations with the observations in the MIR/FIR reveals that for NGC\,4013 and likely NGC\,5907 this model should be replaced by a dust ring. Since one of the aims of this study was to apply exactly the same approach to all {\it HER}OES galaxies using the same dust component, the fitting with a dust ring is beyond the scope of this paper. The same can be said about the spiral arms found in most {\it HER}OES galaxies. 

In our oligochromatic modelling we used a fixed geometry of the stellar components, although early studies \citep{1994A&AS..108..621P, 1996ASSL..209..523C, 1999A&A...344..868X} and recent \citep{2017A&A...605A..18C} show that the disc scale lengths are wavelength dependent. However, according to \citet{2017A&A...605A..18C}, the $R_\mathrm{25}$-normalised scale-lengths $h_\mathrm{R}/R_\mathrm{25}$ in face-on galaxies are almost identical (0.26 for the $g$ band and 0.24 for IRAC 3.6). The dependence of the bulge parameters on wavelength should be stronger \citep{2014MNRAS.441.1340V}. However since only one galaxy in our sample, NGC\,973, has a prominent bulge, we expect that our fixed geometrical model should be satisfactorily applicable at optical wavelengths as well.

A promising result of our oligochromatic fitting is that using additional NIR observations, apart from  optical ones, can better constrain the output dust model. Thus, we found that for the 12 edge-on galaxies from \citet{2014MNRAS.441..869D}, for which the \textsc{fitskirt} model was fitted to the $griz$ images, the uncertainties on the parameters are: 
$\sigma (M_\mathrm{d}) = 20.7 \pm 12.1$\%,
$\sigma (h_\mathrm{R,d}) = 18.7 \pm 9.2$\%,
$\sigma (h_\mathrm{z,d}) = 15.6 \pm 8.6$\%.
Whereas, for the {\it HER}OES galaxies, we found 
$\sigma (M_\mathrm{d}) = 9.9 \pm 3.7$\%,
$\sigma (h_\mathrm{R,d}) = 7.0 \pm 3.2$\%,
$\sigma (h_\mathrm{z,d}) = 11.3 \pm 12.8$\%.
Apparently, one of the reasons why the output oligochromatic models of the {\it HER}OES galaxies are better constrained is that we took into account a larger number of bands within a broader range of wavelengths. In addition, we should notice that the spatial resolution of the {\it HER}OES galaxies is much better than that of the 12 CALIFA \citep{2012A&A...538A...8S} galaxies (by major axis, $364\arcsec \pm 163\arcsec$ versus $116\arcsec \pm 34\arcsec$).
All in all, our \textsc{fitskirt} models adequately describe the observations we used in our fitting.  

In our panchromatic simulations we used the same ages for the same stellar components for all sample galaxies. Obviously, this is not the case since the evolutionary status of each galaxy is different. We tried to vary the ages on a grid of values from 3 Gyr to 8 Gyr for the thin disc and from 5 Gyr to 12 Gyr for the thick disc and the bulge, to find for which ages our model SED better matches the observed SED in the optical and NIR (this is similar to the approach we applied to IC\,2531 in \citetalias{2016A&A...592A..71M}). However, our attempt to find optimal ages was not successful because the recovered ages were always unrealistic (e.g. the age of the thin disc was larger than the age of the thick disc). The precise determination of the ages and metallicities of different stellar populations should be based on spectral observations of the galaxy, and, therefore, this is outside the scope of this paper. However, in our simulations we used different stellar ages from that grid of possible values and did not find a significant influence on the dust emission in the MIR-submm domain. Therefore we set the same ages of the stellar populations for all {\it HER}OES galaxies.

\label{sec:sed_fitting}
\begin{table*}[t] 
  \centering
  \caption{Results of the modified black-body fitting and the fractions of the inner part of the galaxy and the outskirts to the galaxy in the PACS 160~$\mu$m band (the dust masses as derived from the SED fits are denoted as $M_{\mathrm{d}}^{\mathrm{FIR}}$ to make the distinction with the optically determined dust masses).} 
  \label{tab:SED_fitting}
  \begin{tabular}{ccccccccc}
    \hline \hline\\[-2ex]
    Parameter & unit & NGC\,973 & UGC\,4277 & IC\,2531 & NGC\,4013 & NGC\,4217 & NGC\,5529 & NGC\,5907 \\
    \hline\\[-1ex]
$M_\mathrm{d}^\mathrm{FIR}$   & $10^7~M_{\odot}$  & $9.00 \pm 0.60$ & $7.52 \pm 0.69$ & $5.31 \pm 0.44$ & $2.03 \pm 0.14$ & $3.95 \pm 0.28$ & $9.49 \pm 0.72$ & $5.98 \pm 0.40$ \\ [+1ex]
$M_\mathrm{d}^\mathrm{opt} / M_\mathrm{d}^\mathrm{FIR}$ & --- & $0.91 \pm 0.03$ & $0.76 \pm 0.03$ & $0.77 \pm 0.02$ & $0.53 \pm 0.01$ & $0.84 \pm 0.02$ & $0.60 \pm 0.03$ & $1.50 \pm 0.12$ \\[+1ex]
$T_\mathrm{d}$   & K & $20.7 \pm 0.4$ & $18.7 \pm 0.4$ & $19.2 \pm 0.4$ & $22.4 \pm 0.4$ & $22.1 \pm 0.3$ & $20.2 \pm 0.3$ & $20.9 \pm 0.3$ \\[+1ex]
$L_\mathrm{FIR}$   & $10^9~L_{\odot}$ & $17.69 \pm 2.21$ & $11.90 \pm 1.32$ & $9.54 \pm 1.06$ & $8.50 \pm 0.94$ & $15.49 \pm 1.72$ & $22.84 \pm 2.28$ & $17.31 \pm 1.57$ \\[+1ex]
$L_\mathrm{FIR}^\mathrm{SKIRT} / L_\mathrm{FIR}^\mathrm{SED}$ & --- & $0.73 \pm 0.09$ & $0.61 \pm 0.07$ & $0.66 \pm 0.07$ & $0.25 \pm 0.03$ & $0.23 \pm 0.03$ & $0.25 \pm 0.03$ & $0.90 \pm 0.08$ \\[+0.5ex]
    \hline\\[-1ex]

$F_\mathrm{in}^\mathrm{SKIRT} / F_\mathrm{in}$ [160] & --- & $0.69 \pm 0.04$ & $0.63 \pm 0.03$ & $0.60 \pm 0.03$ & $0.26 \pm 0.01$ & $0.21 \pm 0.01$ & $0.28 \pm 0.01$ & $0.83 \pm 0.04$ \\[+1ex]
$F_\mathrm{out}^\mathrm{SKIRT} / F_\mathrm{out}$ [160] & --- & $0.17 \pm 0.01$ & $0.16 \pm 0.01$ & $0.15 \pm 0.01$ & $0.11 \pm 0.01$ & $0.15 \pm 0.01$ & $0.11 \pm 0.01$ & $0.28 \pm 0.01$ \\[+1ex]
$F_\mathrm{out} / F_\mathrm{full}$ [160] & --- & $0.16 \pm 0.01$ & $0.20 \pm 0.01$ & $0.10 \pm 0.01$ & $0.12 \pm 0.01$ & $0.05 \pm 0.01$ & $0.18 \pm 0.01$ & $0.03 \pm 0.01$ \\[+1ex]
    \hline\\
  \end{tabular}
\end{table*} 

In order to quantify the discrepancy of the dust emission between the observations and our panchromatic models, we applied a standard approach, in which the dust SED is well approximated by a simple single modified black-body (MBB) radiation \citep{1983QJRAS..24..267H}. The details, along with the SED fits, are given in the Appendix~\ref{Appendix__SED}. The results of the MBB fitting are presented in Table~\ref{tab:SED_fitting}. As one can see, the mass determined from our dust RT fitting (denoted as $M_\mathrm{d}^\mathrm{opt}$) is typically 1.5--2 times smaller than the dust mass derived from the FIR emission. We confirm the result from \citet{2013A&A...556A..54V} that for NGC\,973  both dust masses are in good agreement. For NGC\,5907, the optically determined dust mass is significantly overestimated compared to its FIR mass.

The average dust temperature $\langle T_{\mathrm{d}}\rangle = (20.6\pm1.4)$~K and dust luminosity $\langle L_\mathrm{IR} \rangle=(14.75 \pm 5.09)\,10^9L_{\odot}$ (obtained by integrating from 70\,$\mu$m to 1000\,$\mu$m) are typical values for the interstellar dust medium in spiral
galaxies \citep[see e.g.][]{2012MNRAS.419.3505D,2012MNRAS.425..763G,2014A&A...565A.128C}.

The ratio $L_\mathrm{FIR}^\mathrm{SKIRT} / L_\mathrm{FIR}$ shows how well the dust FIR luminosity, retrieved from the MBB fitting, and the dust luminosity, computed from our panchromatic simulations (with a young stellar population included), are consistent with each other. We can see that for NGC\,5907, the infrared luminosity from our panchromatic RT modelling and from the MBB fitting are very close. For NGC\,973, UGC\,4277, and IC\,2531, our panchromatic models underestimate the infrared luminosity by a factor 1.3--1.6, whereas for the remaining galaxies we find an underestimation by a factor 3--5. If we carefully look at our panchromatic simulations and compare them with the real observations, we can see that such large discrepancies are found mainly in the galaxies, where a complex dust structure is observed. For example, the underestimation of the infrared luminosity emission in NGC\,973 is 30 per cent, whereas in NGC\,5529 it is 400\%! The latter galaxy has very bright spiral arms which have been detected by \citet{2015A&A...582A..18A} in its H{\sc{i}} maps. NGC\,973, IC\,2531, and NGC\,4217 also show some prominent features in the MIR/FIR maps, which may be related to fainter spiral structures. NGC\,4013 and NGC\,5907 are likely to have some ring structures, which should explain their observed profiles in the MIR/FIR.

Another useful test to study the dust energy balance problem in the {\it HER}OES galaxies is by quantifying how much of the observed dust emission deficit is due to the model not capturing the observations, and how much is related to sub-pixel structures (clumps and dust emissivities). In Table~\ref{tab:SED_fitting}, we also provide the following ratios which characterise these features. For each PACS~160$\mu$m image (at the peak of the dust emission and where the spatial resolution is high enough), we measure the standard deviation $\sigma$ of the image background and determine the outermost isophote of the galaxy at the level $2\sigma$. The same $2\sigma$ level is used to determine the outermost isophote for the \textsc{skirt} simulation, at the same wavelength. For all galaxies in our sample, the latter lies within the former one (the real galaxies are more radially extended than their models, see Figs.~\ref{map_ugc4277}--\ref{map_ngc5907}). Thus, to measure the deficit of light in the galaxy outskirts (not captured by the model), we find the ratio between the \textsc{skirt} flux and the observed flux, both measured within an annulus between the \textsc{skirt} outermost isophote and the real one, denoted in Table~\ref{tab:SED_fitting} as $F_\mathrm{out}^\mathrm{SKIRT} / F_\mathrm{out}$. We also measure the ratio of the \textsc{skirt} and observed fluxes within the \textsc{skirt} isophote (inner isophote), to show how consistent our model and the observation are in the inner part of the galaxy. We can see
that in the inner part, for the three galaxies NGC\,4013, NGC\,4217 and NGC\,5529, the discrepancy between the model and the observation is 4--5 times; for these galaxies we also see the highest discrepancy between the model and observed FIR luminosities. This discrepancy comes from the small- and large-scale structure (bright dust clumps, spiral arms, bars, rings, etc.). For the outer part of the galaxy, we can see that the model severely underestimates dust emission, by a factor 5--10. However, the fraction of the flux between the inner- and outermost isophotes to the total galaxy flux $F_\mathrm{out} / F_\mathrm{full}$ (where $F_\mathrm{full}=F_\mathrm{in}+F_\mathrm{out}$) varies from 3\% (for NGC\,5907) up to 20\% (for UGC\,4277), with an average $12\pm6$\%. Thus, due to its small fraction, the vertically and radially extended emission (with respect to the emission based on the optical model) cannot be the only reason of the observed dust energy balance problem. Nevertheless, the fact that in some galaxies this emission can contribute quite a different fraction to the total dust luminosity is interesting by itself. For NGC\,891, \citet{2007A&A...471L...1K,2016A&A...586A...8B} showed that, apart from the regular dust disc, there is a second spatial dust component which extends up to 1.9~kpc above the galactic mid-plane. Based on GALEX and \textit{Swift} observations, diffuse UV haloes have recently been discovered around several galaxies \citep{2014ApJ...789..131H,2015ApJ...815..133S,2016A&A...587A..86B}. This points to the presence of a substantial amount of diffuse extra-planar dust, which scatters UV radiation emitted from the galactic disc.

\begin{figure}[h]
\centering
\includegraphics[width=9cm , angle=0, clip=]{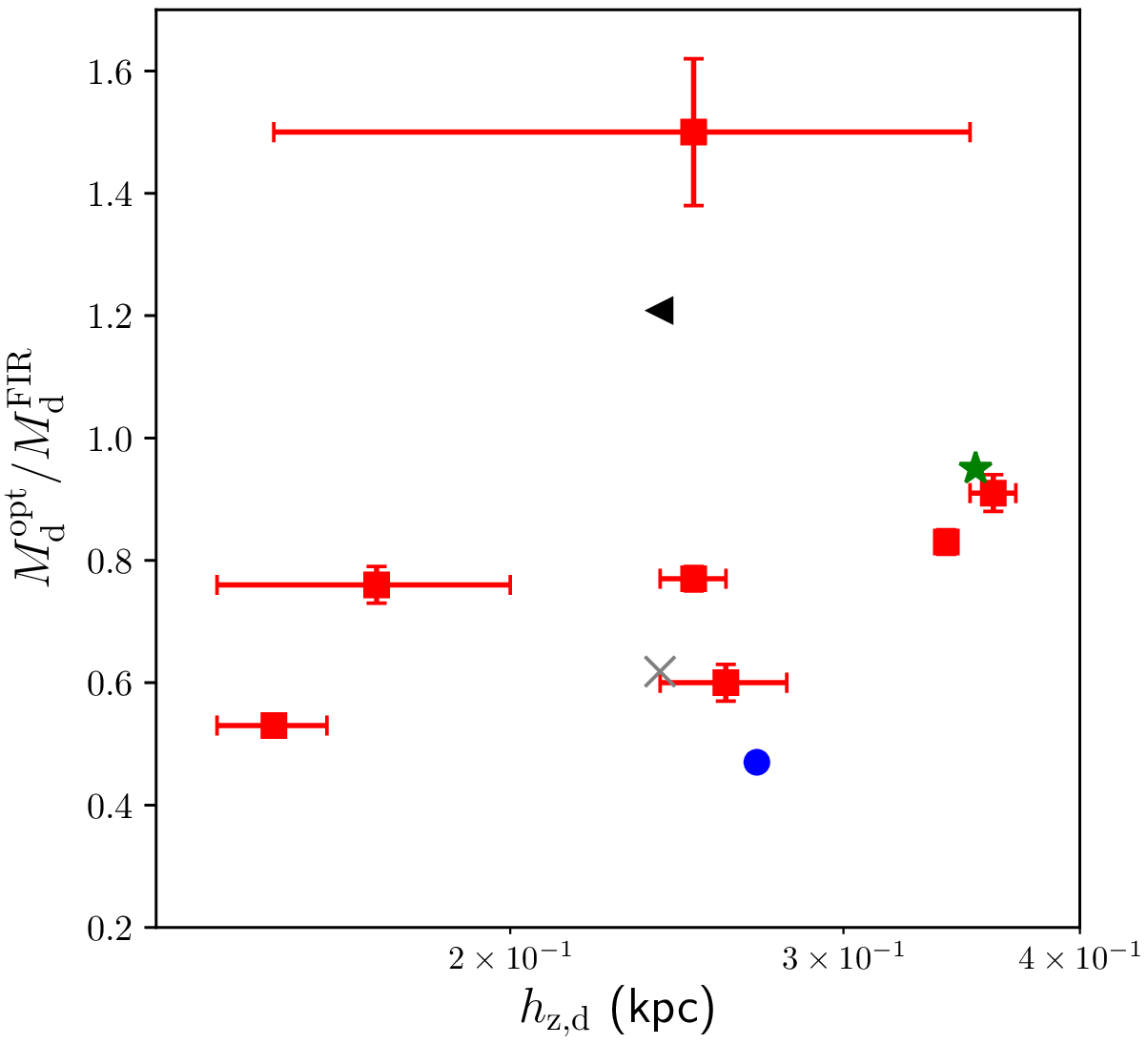}
\caption{Dependence of the dust mass discrepancy on the dust scale height (in logarithmic scale). The red squares are the {\it HER}OES galaxies, the blue circle represents NGC\,891, the green asterisk is NGC\,4565, the black triangle refers to IC\,4225, and the gray cross is NGC\,5166.}
\label{compare_dust_masses}
\end{figure}

An interesting result of our study, which confirms the findings of \citet{2013A&A...556A..54V} (see their fig.~12), is a weak trend between the difference in the values of the dust mass, as computed from our \textsc{fitskirt} radiative transfer modelling and as computed from our black-body fitting, and the dust scale height. To plot this correlation, which is presented in Fig.~\ref{compare_dust_masses} (the regression coefficient $r=0.53$), we added the results for NGC\,891 and NGC\,4565 (as calculated in \citealt{2013A&A...556A..54V} using the results from \citealt{2011A&A...531L..11B} and \citealt{2012MNRAS.427.2797D}, respectively) and the results from \citetalias{2015MNRAS.451.1728D} for IC\,4225 and NGC\,5166. Of course, the number of points is insufficient to make a strong conclusion, but the physical meaning of it may imply that the thinner the dust disc is, the clumpier its structure is. We suppose, that small clumpy regions do not contribute to the global optical extinction at all, and their presence can only be revealed from the thermal emission of the dust they contain, showing up in the FIR \citep[see e.g.][]{2000A&A...362..138P}.

To summarise, in the optical, it is hard to discern any sign of a large or small-scale structure in the ISM, due to line-of-sight projection effects. Smooth, axisymmetric models generally describe the stellar emission and the dust attenuation very well. However, as shown by \citet{2015A&A...576A..31S}, fitting a complex, clumpy 3D structure with simple axisymmetric models can easily underestimate the dust content of a galaxy by a factor of a few.  In the MIR/FIR, such small- and large-scale structure can be observable (if the resolution is high enough). Our observations and modelling support the idea that the inhomogeneous, clumpy ISM, along with various large-scale components, is responsible for the dust energy balance problem.

In the frame of the DustPedia project, special attention will be paid to the study of the dust mass, distribution and heating mechanisms in normal galaxies. A set of face-on galaxies will be considered and a new strategy will be applied to them. The unified model will consist of five components: an evolved stellar disc, a young stellar disc, an ionising stellar disc, a dust disc, and a stellar bulge. Each component will be defined in a special way using available multi-wavelength observations. The main difference of this approach from what we used in this study is that the geometry of each component will be constructed directly from the observed images, by means of the deprojection algorithm available in \textsc{skirt}. Using \textsc{skirt} simulations for the defined model, which has the three remaining fit parameters (the luminosity of the young and ionising stellar components and the total dust mass), the fitting of the simulated model SED to the observed global fluxes will be done. This will allow us to determine the fitted parameters in a self-consistent way. Some initial studies on M\,51 \citep{2014A&A...571A..69D} and the Andromeda galaxy \citep{2017A&A...599A..64V} have been done, and the final strategy has been applied to M81 (Verstocken et al. in prep.) and NGC\,1068 (Viaene et al. in prep). This will help to study the dust energy balance problem in a different way.

\section{Summary}
\label{sec:summary}

In an attempt to address the dust energy balance problem in Milky Way type galaxies, we have carried out a detailed analysis of the distribution of interstellar dust and stellar constituents in the remaining six {\it HER}OES galaxies \citepalias[the prototype galaxy IC\,2531 was described in][]{2016A&A...592A..71M} by computing oligochromatic radiative transfer models and panchromatic simulations. By doing so, we constrained both the stellar and dust distributions and studied the famous dust energy balance problem in the sample galaxies.

In our study, we successfully applied the following technique. First, for each galaxy, a photometric model, consisting of a thin disc, a thick disc, and a bulge, was found using the \textsc{imfit} code applied to the NIR/MIR ($K$ band/IRAC~3.6$\mu$m) images. These stellar models were used in subsequent \textsc{fitskirt} oligochromatic modelling with a fixed stellar geometry and a dust double-exponential disc. In this step, images in different optical and NIR wavebands (from seven to nine bands in total) were fitted simultaneously. The obtained models describe the fitted observations very well and are well-constrained (in comparison to the results from monochromatic or oligochromatic fitting performed in previous works).

The comparison of the results of our modelling with other works on RT fitting of edge-on galaxies shows that, in general, our results are consistent with earlier results.

Using the new THEMIS dust model, we performed \textsc{skirt} simulations with and without an additional young stellar component, which was added to match \textit{FUV}/\textit{NUV} observations. A uniform recipe of modelling was applied to all the galaxies. For five {\it HER}OES galaxies (UGC\,4277, IC\,2531, NGC\,4013, NGC\,4217, and NGC\,5529), the constructed SEDs show the same dust energy balance problem even with an additional young stellar component -- the dust emission in the FIR/submm is underestimated by a factor of 1.5 to 4. For the galaxy NGC\,973, the emission in the FIR/submm is almost completely recovered in our modelling (albeit no young stellar population was added).

For only one galaxy (NGC\,5907) in our sample, we found that the oligochromatic model, consisting of two exponential discs to describe the stellar thin and thick discs, another exponential disc for the dust distribution, and a \ser\ bulge, was not able to accurately reproduce the observations of this galaxy in the fitted wavebands which can be seen in the performed panchromatic simulations. In all other cases, based on the dust attenuation in the opical and NIR, we were able to construct oligochromatic models of the galaxies, which match the observations in these bands very well, and constrain their parameters to an acceptable accuracy. In all of the residual frames for our oligochromatic models more than half of the pixels show deviations of at most 20\%. Parameters of the stellar components, as well as of the dust disc, are determined within $\sim10$\%.  

The comparison of the panchromatic simulations with the real ones reveals a more complex nature of the dust constituent in galaxies -- it is often more extended in the radial and vertical directions than what can be inferred from our optical modelling. Some traces of spirals, rings, dust at high galactic latitudes, and signatures of a clumpy, inhomogeneous medium are clearly present. Thus, we suggest that these small- and large-scale structures may be responsible for the dust energy balance problem. We should notice, however, that the presence of different heating sources, which can be clearly distinguished at different wavelengths in the NIR--FIR domain (as shown in the Appendix~\ref{Appendix__simulations}), suggests that the nature of the dust and stellar constituents and their link in galaxies needs a further investigation.

The average face-on optical depth in the $V$ band for the sample galaxies is $1.07 \pm 0.53$ (for comparison, LSB galaxies with diffuse clumpy dust discs are almost transparent: $\tau_\mathrm{V}^f \lesssim 0.1$, \citealt{2011ApJ...741....6M}). For three galaxies (NGC\,973, NGC\,4013, and NGC\,5907), a significant part of the light is blocked even when the galaxy is seen face-on. 

The results of the decomposition, as well as of the dust RT fitting (including uncertainties
on individual measurements), of all structural parameters are listed in Tables~\ref{tab:IRAC_dec_res} and \ref{tab:FitSKIRT_res}. The \textsc{imfit}, \textsc{fitskirt} and \textsc{skirt} models (with the constructed SEDs) are presented in Appendices \ref{Appendix_IMFITs}, \ref{Appendix_Fit}, \ref{Appendix_SEDs}, and \ref{Appendix__simulations} (available online).

\begin{acknowledgements}
A.V.M. is a beneficiary of a postdoctoral grant from the Belgian Federal Science Policy Office, and also expresses gratitude for the grant of the Russian Foundation for Basic Researches number 14-02-00810 and 14-22-03006-ofi. 

F.A., M.B., I.D.L. and S.V. gratefully acknowledge the support of the Flemish Fund for Scientific Research (FWO-Vlaanderen). 

M.B. acknowledges financial support from the Belgian Science Policy Office (BELSPO) through the PRODEX project “\textit{Herschel}-PACS Guaranteed Time and Open Time Programs: Science Exploitation” (C90370). 

T.M.H. acknowledges the CONICYT/ALMA funding Program in Astronomy/PCI Project No:31140020. He also acknowledges the support from the Chinese Academy of Sciences (CAS) and the National Commission for Scientific and Technological Research of Chile (CONICYT) through a CAS-CONICYT Joint Postdoctoral Fellowship administered by the CAS South America Center for Astronomy (CASSACA) in Santiago, Chile.

J.V. acknowledges support from the European Research Council under the European Union’s Seventh Framework Programme (FP/2007-2013)/ ERC Grant Agreement nr. 291531. 

This work was supported by CHARM and DustPedia. CHARM (Contemporary physical challenges in Heliospheric and AstRophysical Models) is a phase VII Interuniversity Attraction Pole (IAP) program organized by BELSPO, the BELgian federal Science Policy Office. DustPedia is a collaborative focused research project supported by the European Union under the Seventh Framework Programme (2007-2013) call (proposal no. 606847). The participating institutions are: Cardiff University, UK; National Observatory of Athens, Greece; Ghent University, Belgium; Université Paris Sud, France; National Institute for Astrophysics, Italy and CEA (Paris), France.

The Faulkes Telescopes are maintained and operated by Las Cumbres Observatory Global Telescope Network. We also thank Peter Hill and the staff and students of College Le Monteil ASAM (France), The Thomas Aveling School (Rochester, England), Glebe School (Bromley, England) and St David’s Catholic College (Cardiff, Wales). 

This work is based in part on observations made with the {\it Spitzer} Space Telescope, which is operated by the Jet Propulsion Laboratory, California Institute of Technology under a contract with NASA. 

This publication makes use of data products from the Two Micron All Sky Survey, which is a joint project of the University of Massachusetts and the Infrared Processing and Analysis Center/California Institute of Technology, funded by the National Aeronautics and Space Administration and the National Science Foundation. 

This publication makes use of data products from the Wide-field Infrared Survey Explorer, which is a joint project of the University of California, Los Angeles, and the Jet Propulsion Laboratory/California Institute of Technology, funded by the National Aeronautics and Space Administration. 

This study makes use of observations made with the NASA Galaxy
Evolution Explorer. GALEX is operated for NASA by
the California Institute of Technology under NASA contract
NAS5-98034.

The {\it Herschel} spacecraft was designed, built, tested, and
launched under a contract to ESA managed by the \textit{Herschel}/\textit{Planck}
Project team by an industrial consortium under the overall responsibility
of the prime contractor Thales Alenia Space (Cannes), and
including Astrium (Friedrichshafen) responsible for the payload
module and for system testing at spacecraft level, Thales Alenia
Space (Turin) responsible for the service module, and Astrium
(Toulouse) responsible for the telescope, with in excess of a hundred
subcontractors.

SPIRE has been developed by a consortium of institutes led
by Cardiff Univ. (UK) and including: Univ. Lethbridge (Canada);
NAOC (China); CEA, LAM (France); IFSI, Univ. Padua (Italy);
IAC (Spain); Stockholm Observatory (Sweden); Imperial College
London, RAL, UCL-MSSL, UKATC, Univ. Sussex (UK); and
Caltech, JPL, NHSC, Univ. Colorado (USA). This development
has been supported by national funding agencies: CSA (Canada);
NAOC (China); CEA, CNES, CNRS (France); ASI (Italy); MCINN
(Spain); SNSB (Sweden); STFC, UKSA (UK); and NASA (USA).

We acknowledge the use of the ESA {\it Planck} Legacy Archive.

Some of the data presented
in this paper were obtained from the Mikulski Archive for
Space Telescopes (MAST). STScI is operated by the Association
of Universities for Research in Astronomy, Inc., under
NASA contract NAS5-26555. Support for MAST for nonHST
data is provided by the NASA Office of Space Science
via grant NNX13AC07G and by other grants and contracts.

This research makes use of the NASA/IPAC Extragalactic Database (NED) which is operated by the Jet Propulsion Laboratory, California Institute of Technology, under contract with the National Aeronautics and Space Administration, and the LEDA database (http://leda.univ-lyon1.fr).
 \\
\end{acknowledgements}

\bibliographystyle{aa} 
\bibliography{heroes}

\appendix
\section{IMFIT models}
\label{Appendix_IMFITs}

\begin{figure*}
\centering
\includegraphics[width=9cm, angle=0, clip=]{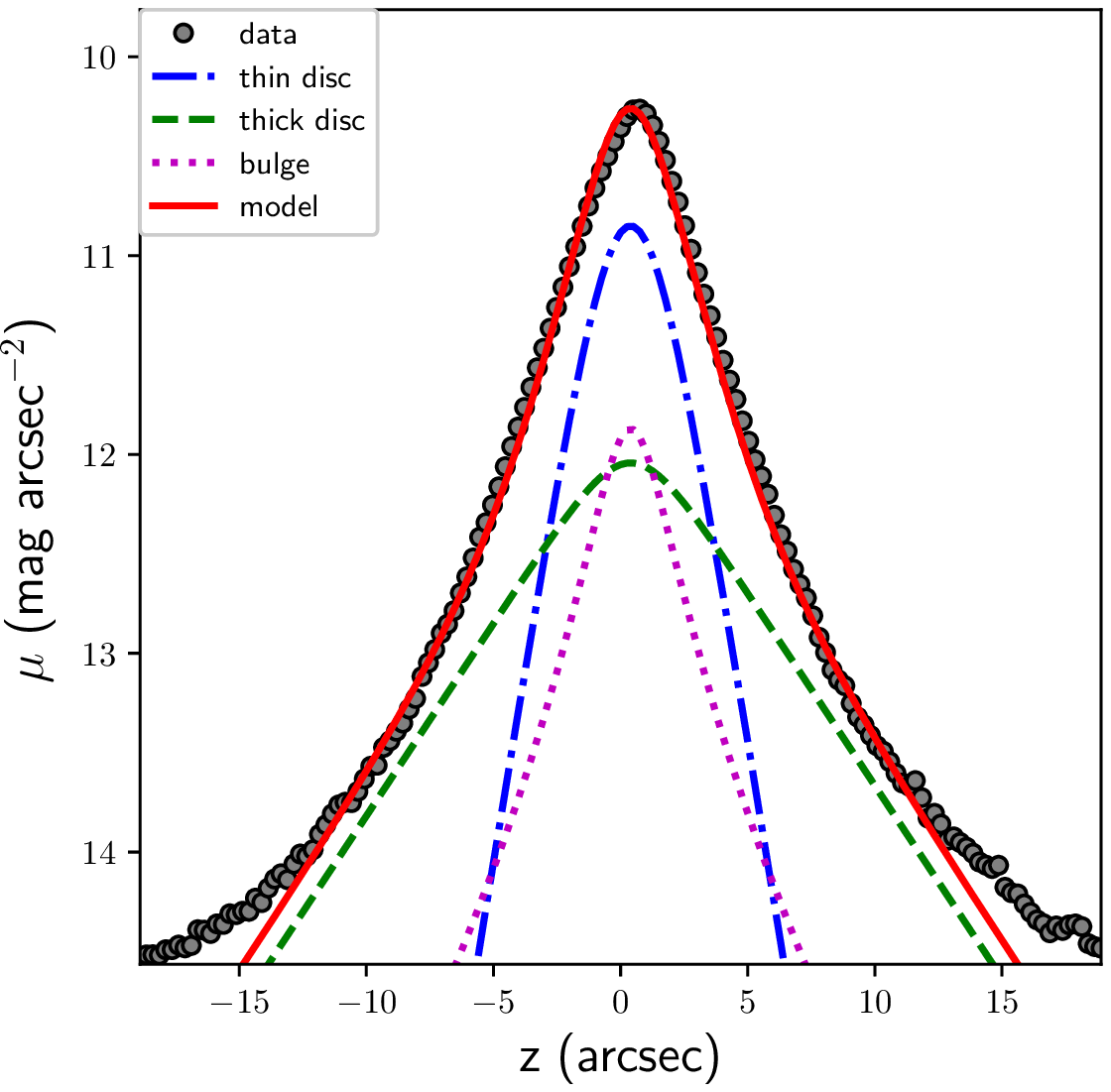}
\includegraphics[width=9cm, angle=0, clip=]{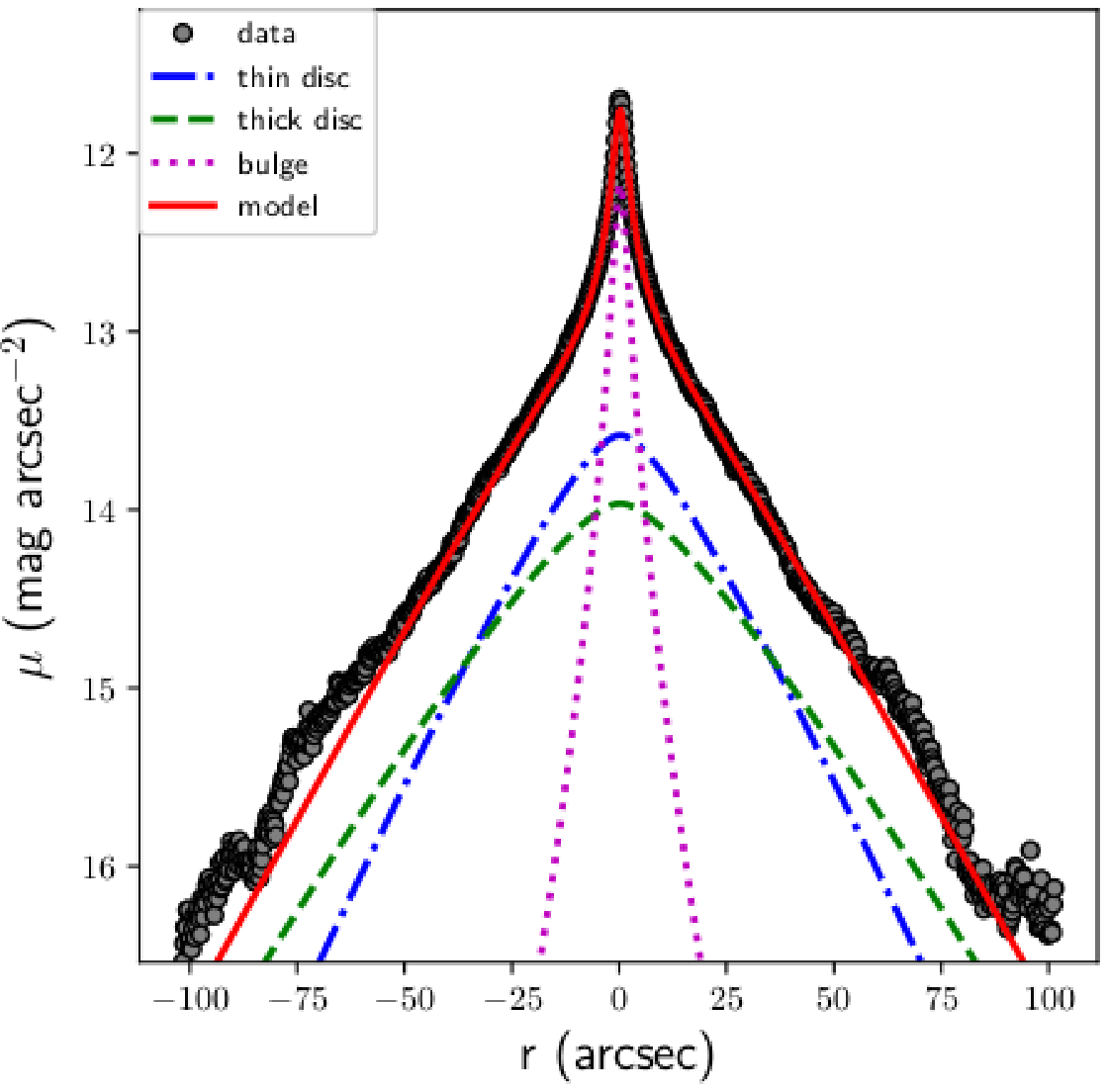}
\caption{Cumulative vertical (top) and horizontal (bottom) profiles of UGC\,4277 plotted for its $K$ band image, with its  overlaid \textsc{imfit} model.}
\label{imfit_UGC4277}
\end{figure*}

\begin{figure*}
\centering
\includegraphics[width=9cm, angle=0, clip=]{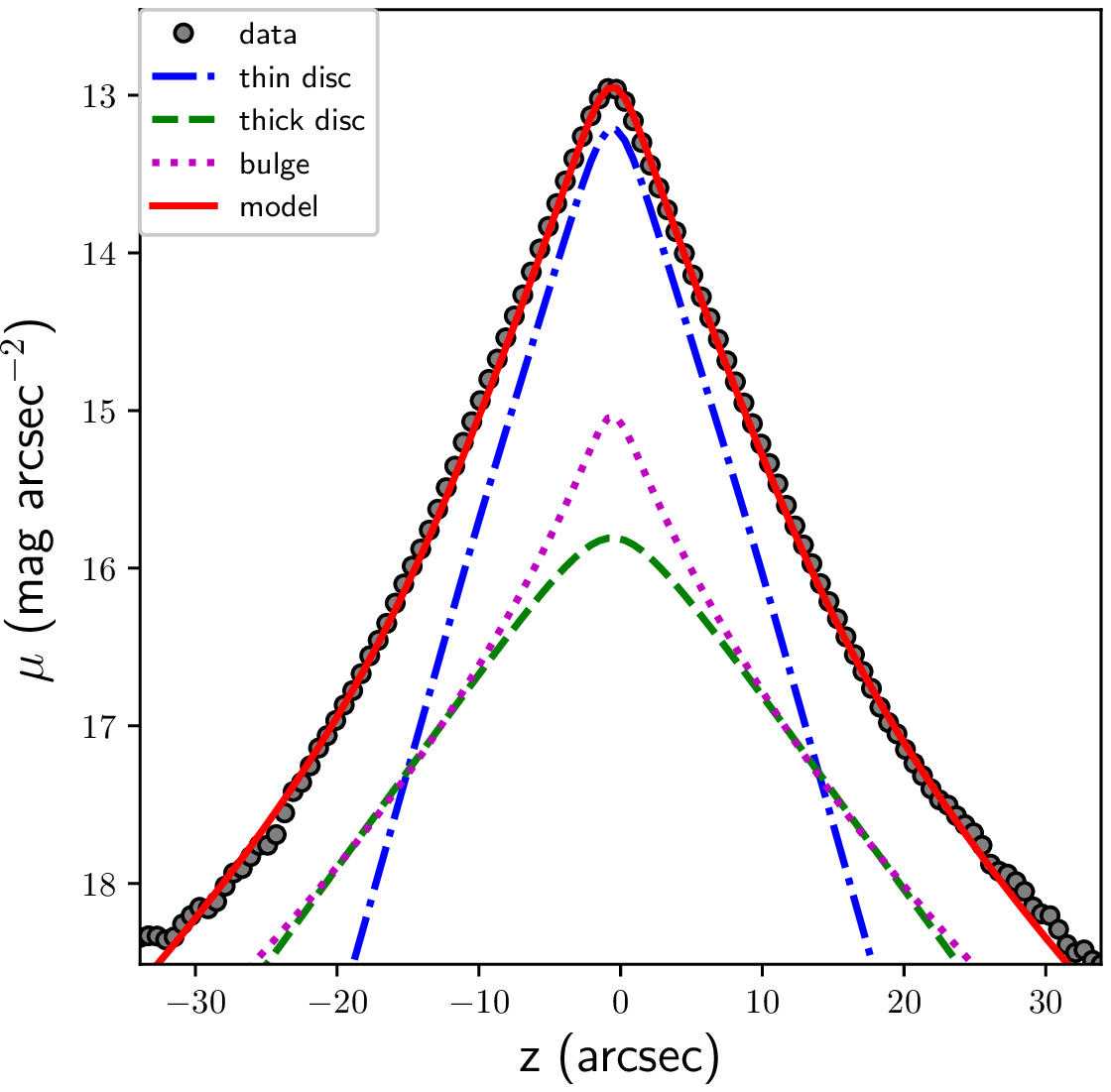}
\includegraphics[width=9cm, angle=0, clip=]{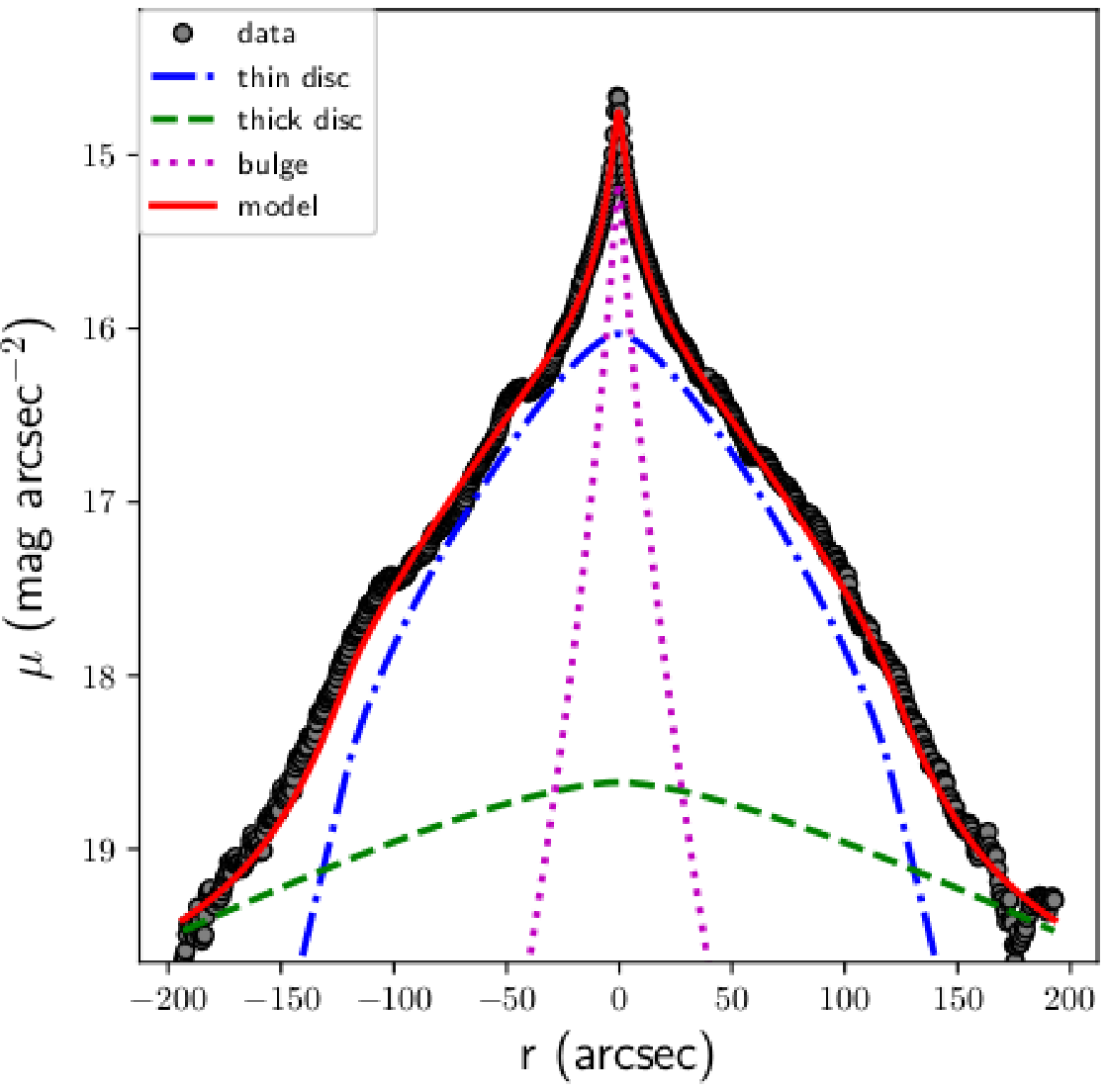}
\caption{Cumulative vertical (lefthand) and horizontal (righthand) profiles of IC\,2531 plotted for its IRAC 3.6\,$\mu$m image, with its overlaid \textsc{imfit} model.}
\label{imfit_IC2531}
\end{figure*}

\begin{figure*}
\centering
\includegraphics[width=9cm, angle=0, clip=]{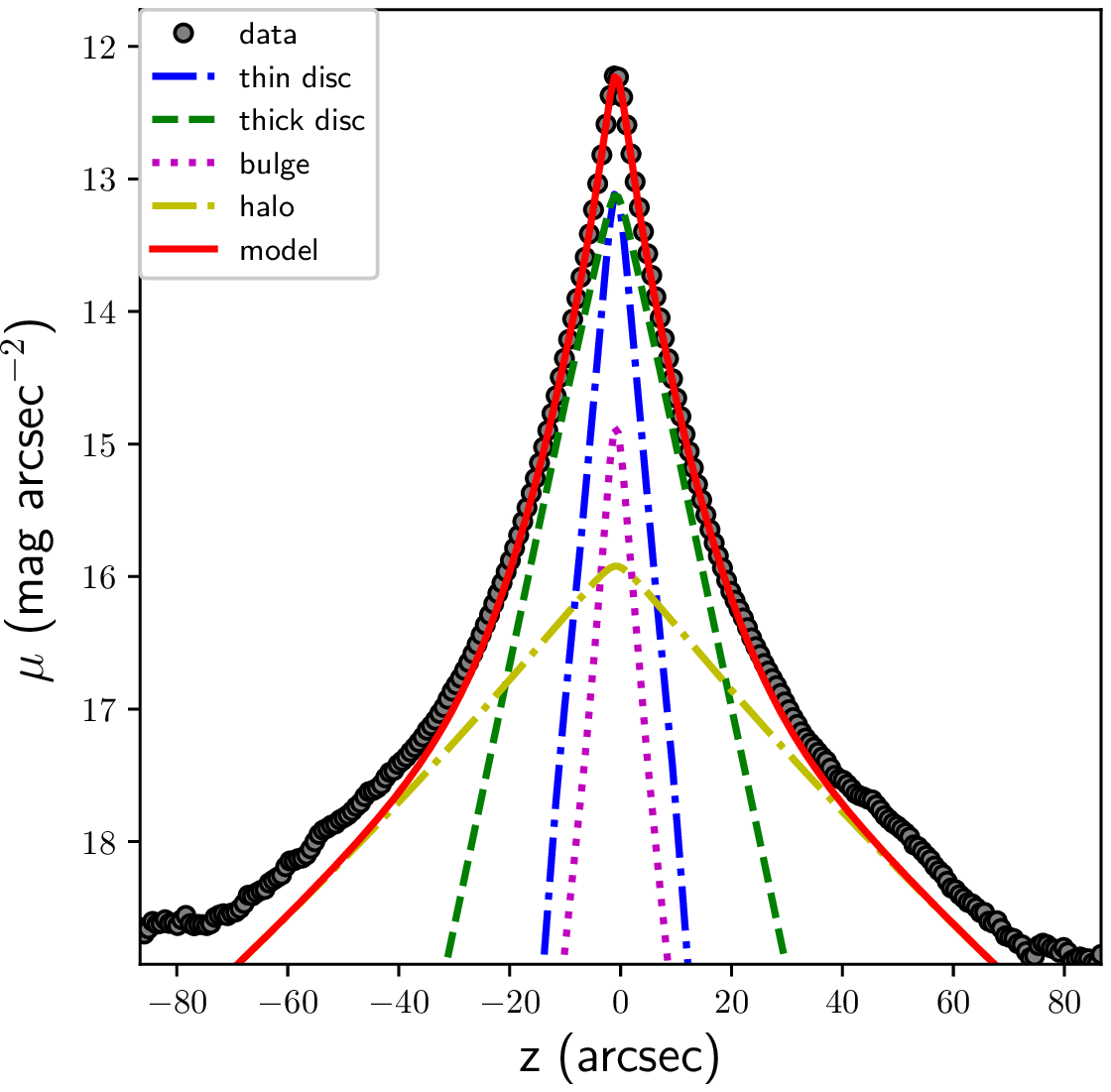}
\includegraphics[width=9cm, angle=0, clip=]{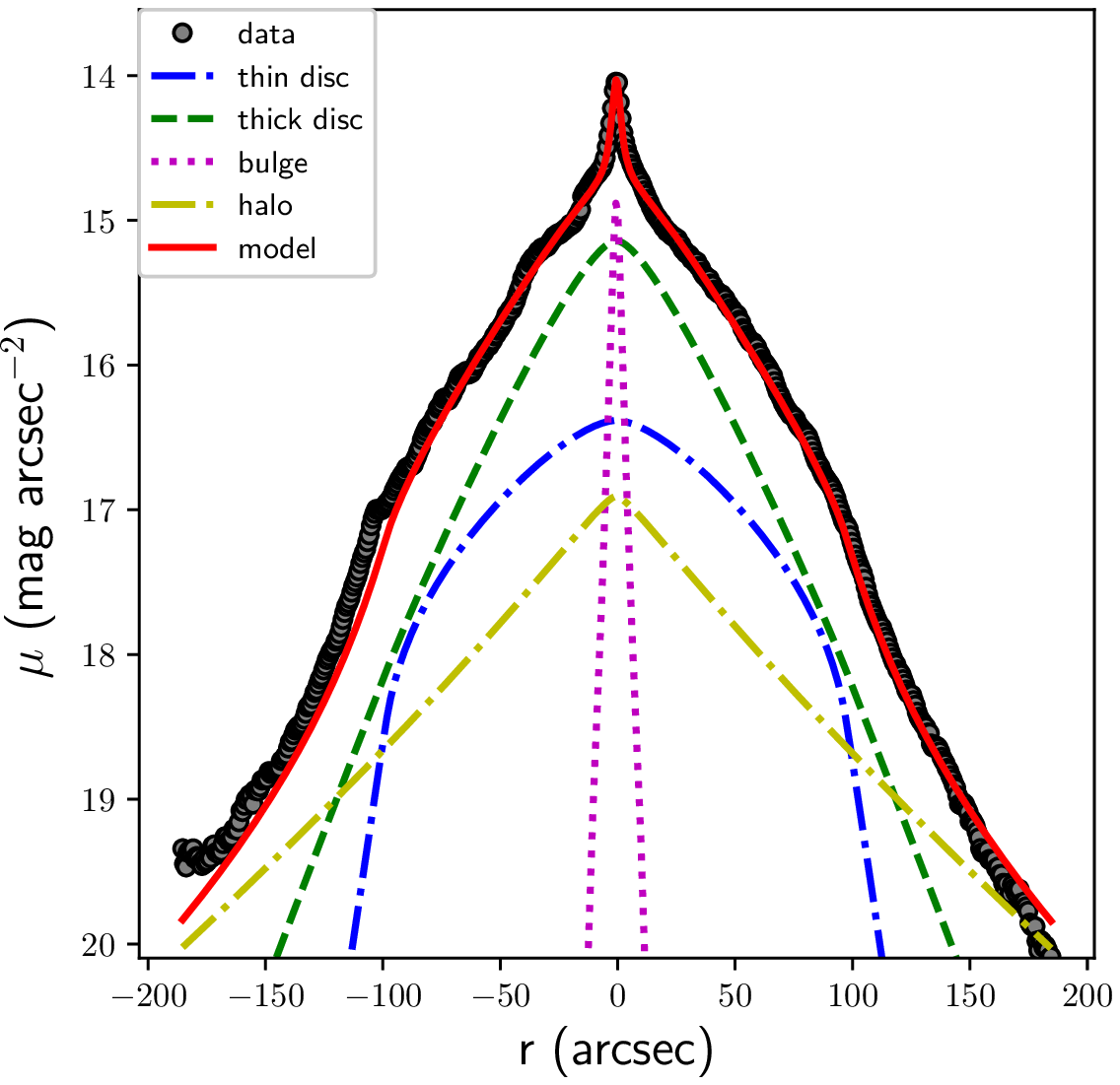}
\caption{Cumulative vertical (lefthand) and horizontal (righthand) profiles of NGC\,4013 plotted for its IRAC 3.6\,$\mu$m image, with its overlaid \textsc{imfit} model.}
\label{imfit_NGC4013}
\end{figure*}

\begin{figure*}
\centering
\includegraphics[width=9cm, angle=0, clip=]{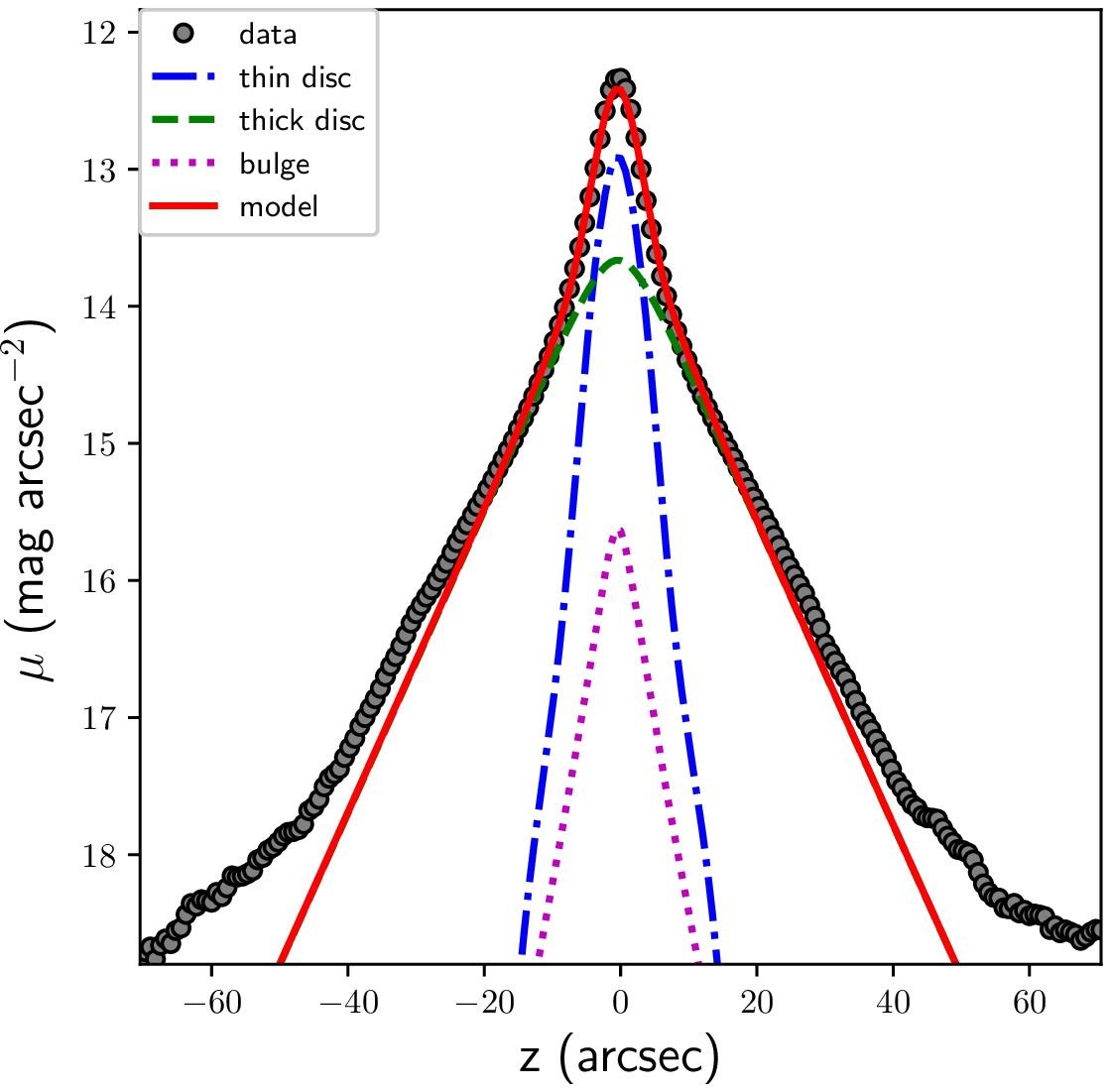}
\includegraphics[width=9cm, angle=0, clip=]{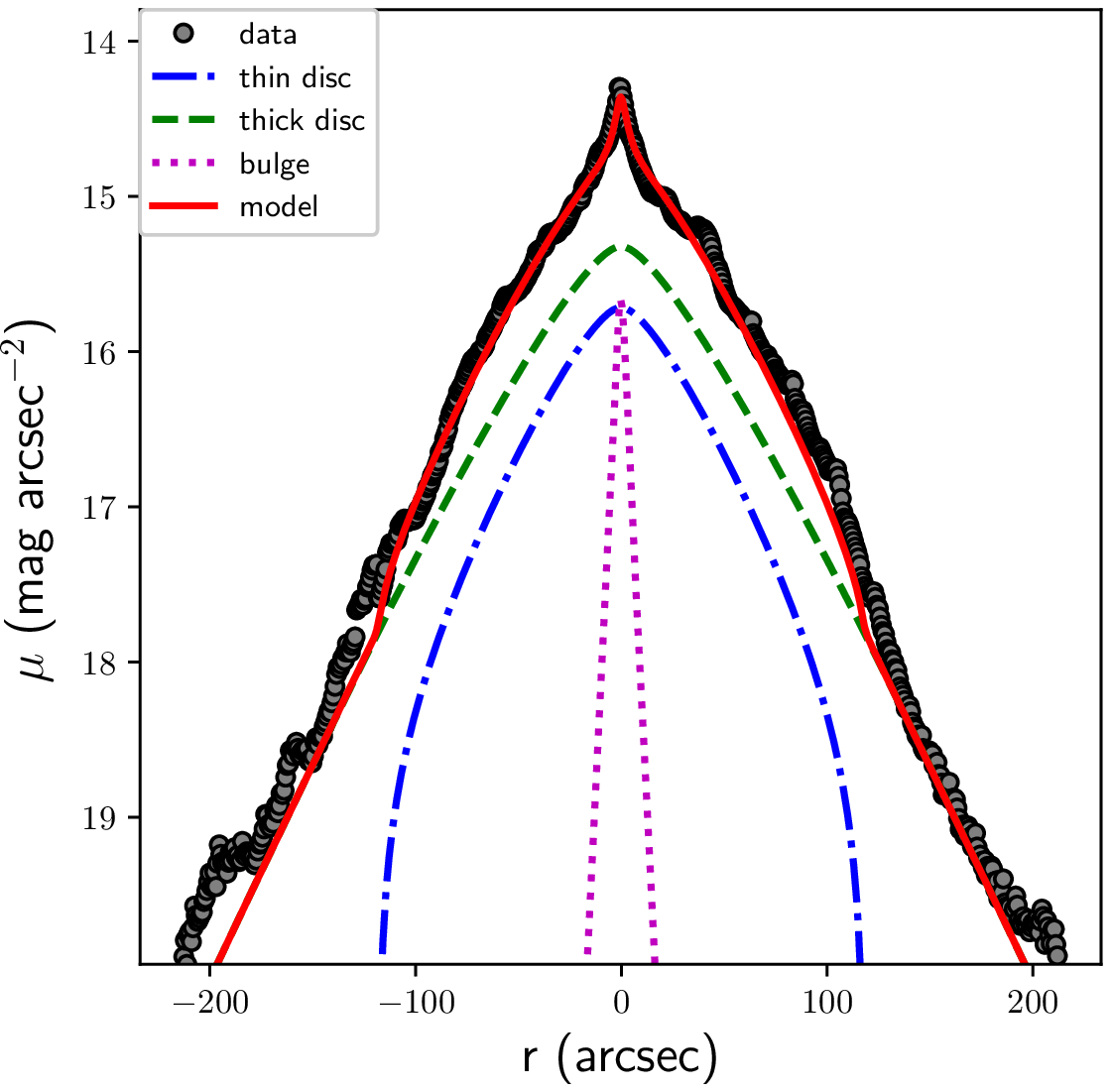}
\caption{Cumulative vertical (lefthand) and horizontal (righthand) profiles of NGC\,4217 plotted for its IRAC 3.6\,$\mu$m image, with its overlaid \textsc{imfit} model.}
\label{imfit_NGC4217}
\end{figure*}

\begin{figure*}
\centering
\includegraphics[width=9cm, angle=0, clip=]{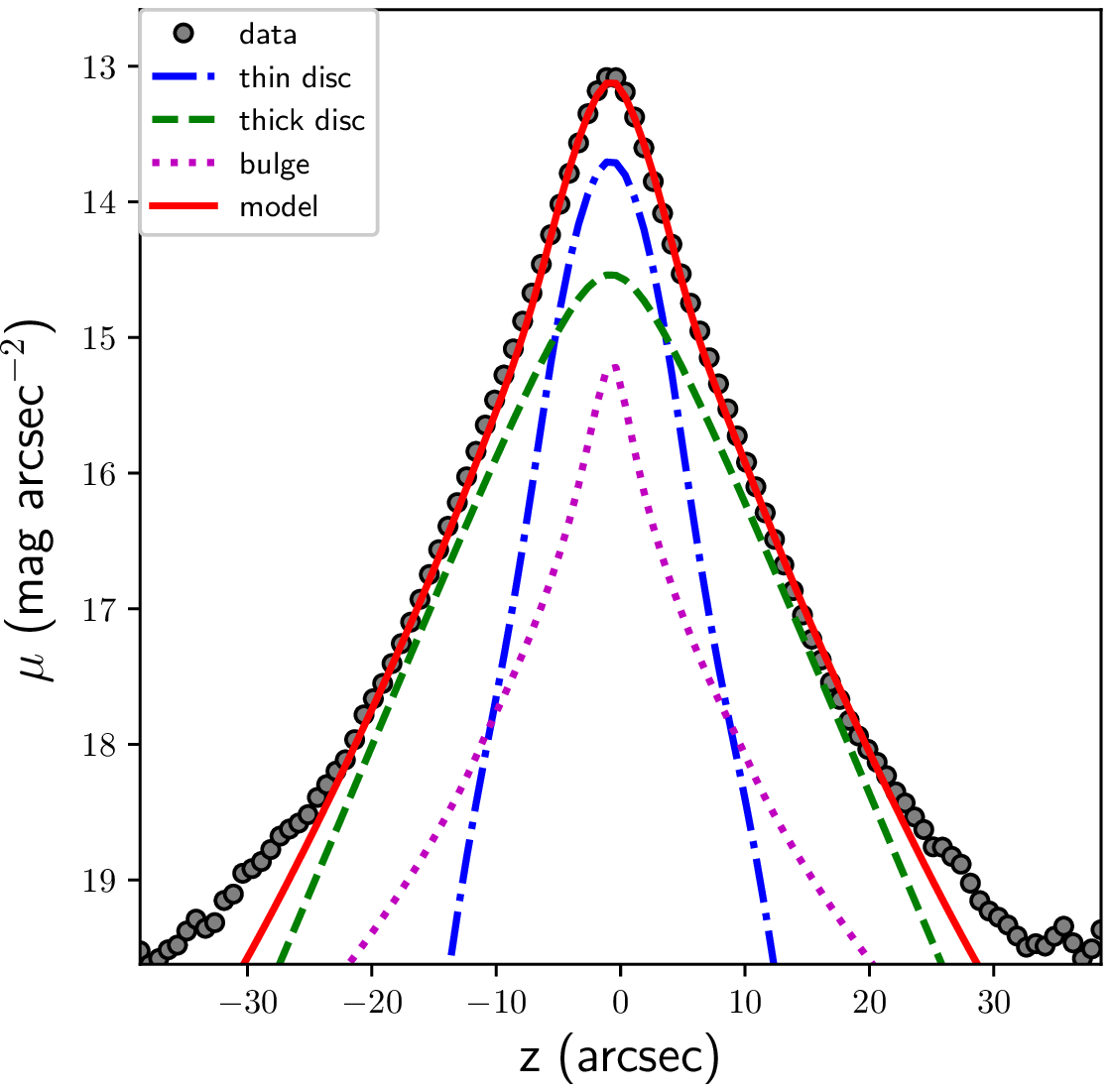}
\includegraphics[width=9cm, angle=0, clip=]{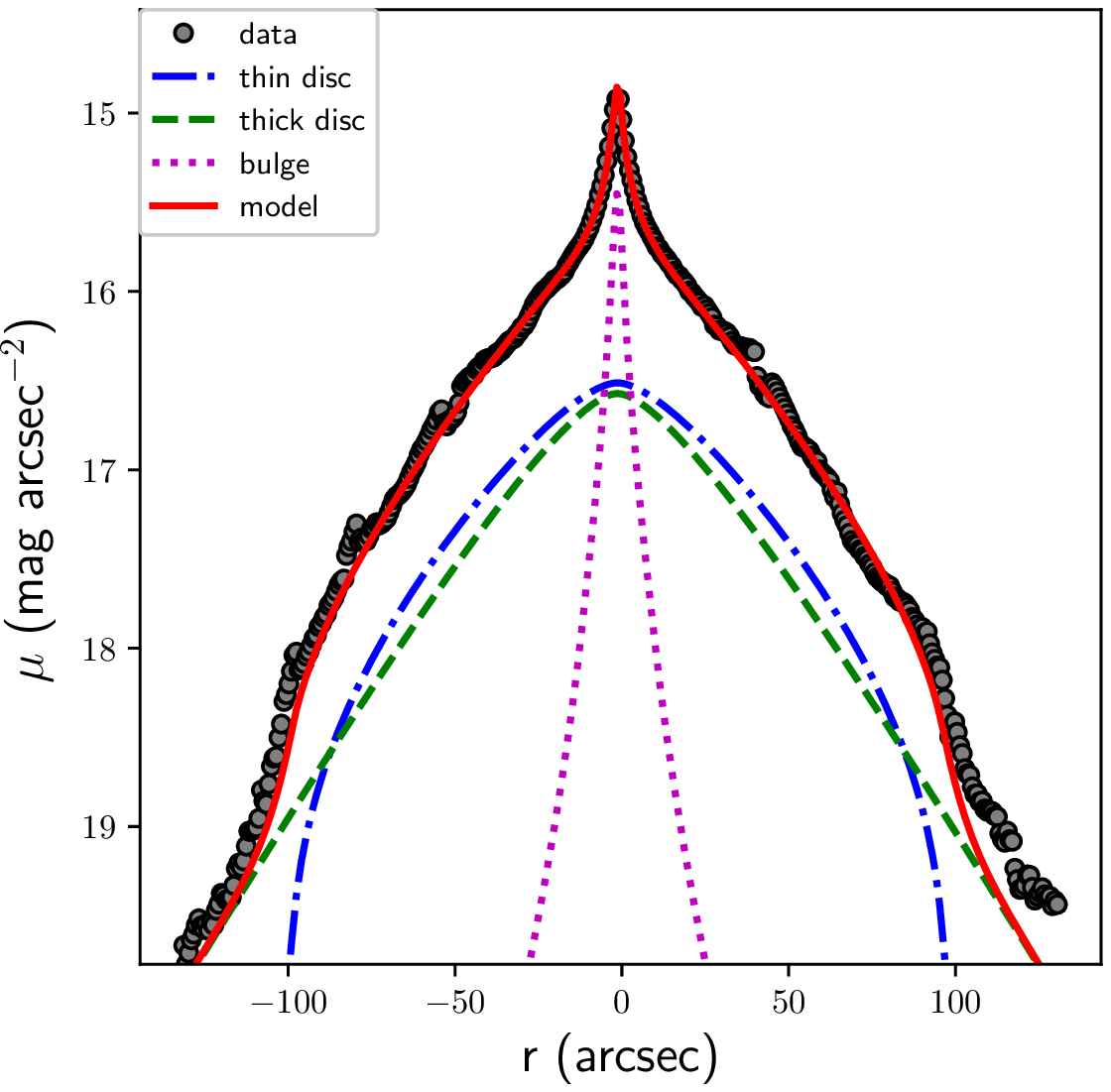}
\caption{Cumulative vertical (lefthand) and horizontal (righthand) profiles of NGC\,5529 plotted for its IRAC 3.6\,$\mu$m image, with its overlaid \textsc{imfit} model.}
\label{imfit_NGC5529}
\end{figure*}

\begin{figure*}
\centering
\includegraphics[width=9cm, angle=0, clip=]{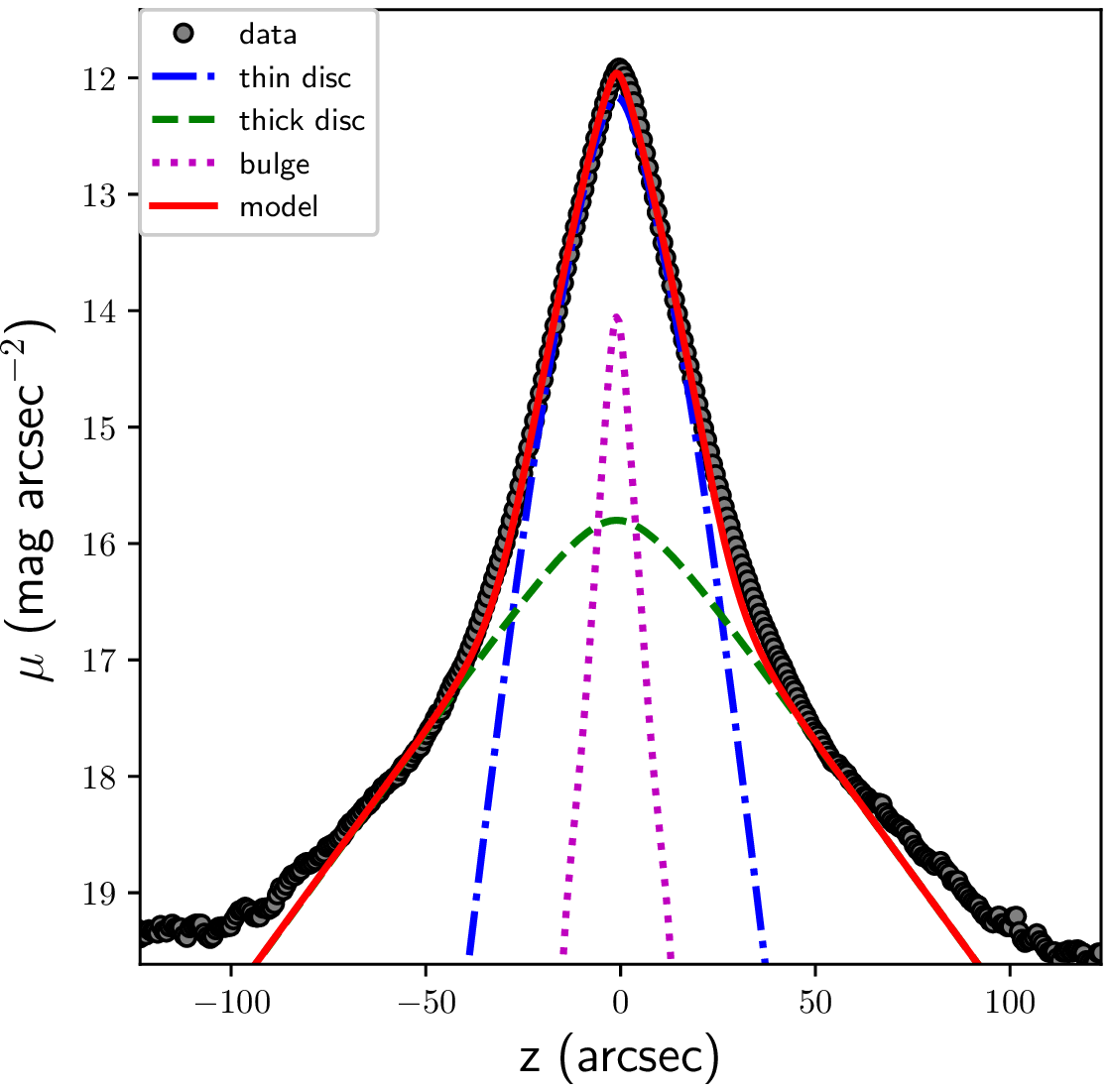}
\includegraphics[width=9cm, angle=0, clip=]{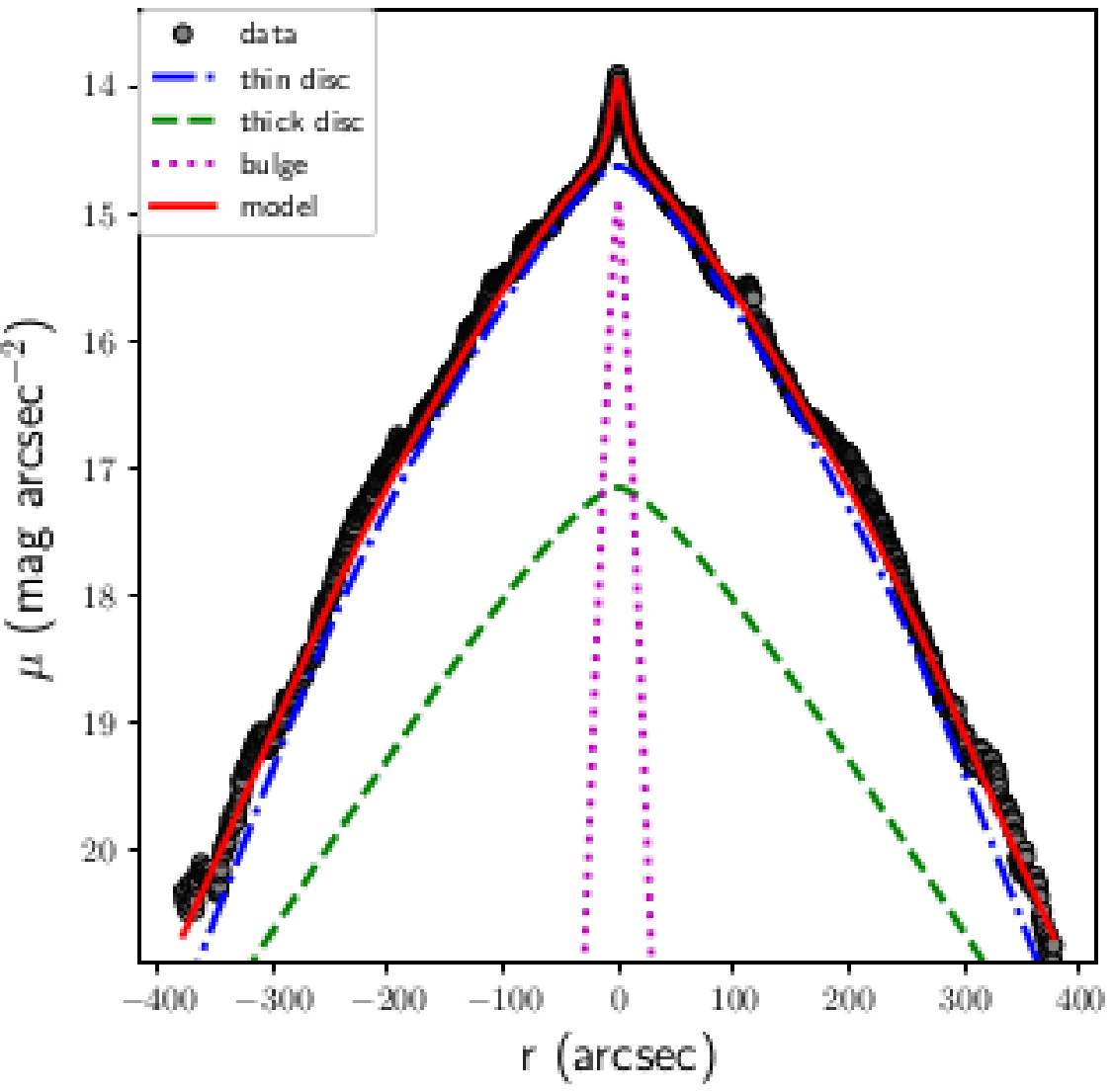}
\caption{Cumulative vertical (lefthand) and horizontal (righthand) profiles of NGC\,5907 plotted for its IRAC 3.6\,$\mu$m image, with its overlaid \textsc{imfit} model.}
\label{imfit_NGC5907}
\end{figure*}


\section{\textsc{fitskirt} models}
\label{Appendix_Fit}

\begin{figure*}
\centering
\includegraphics[width=\textwidth]{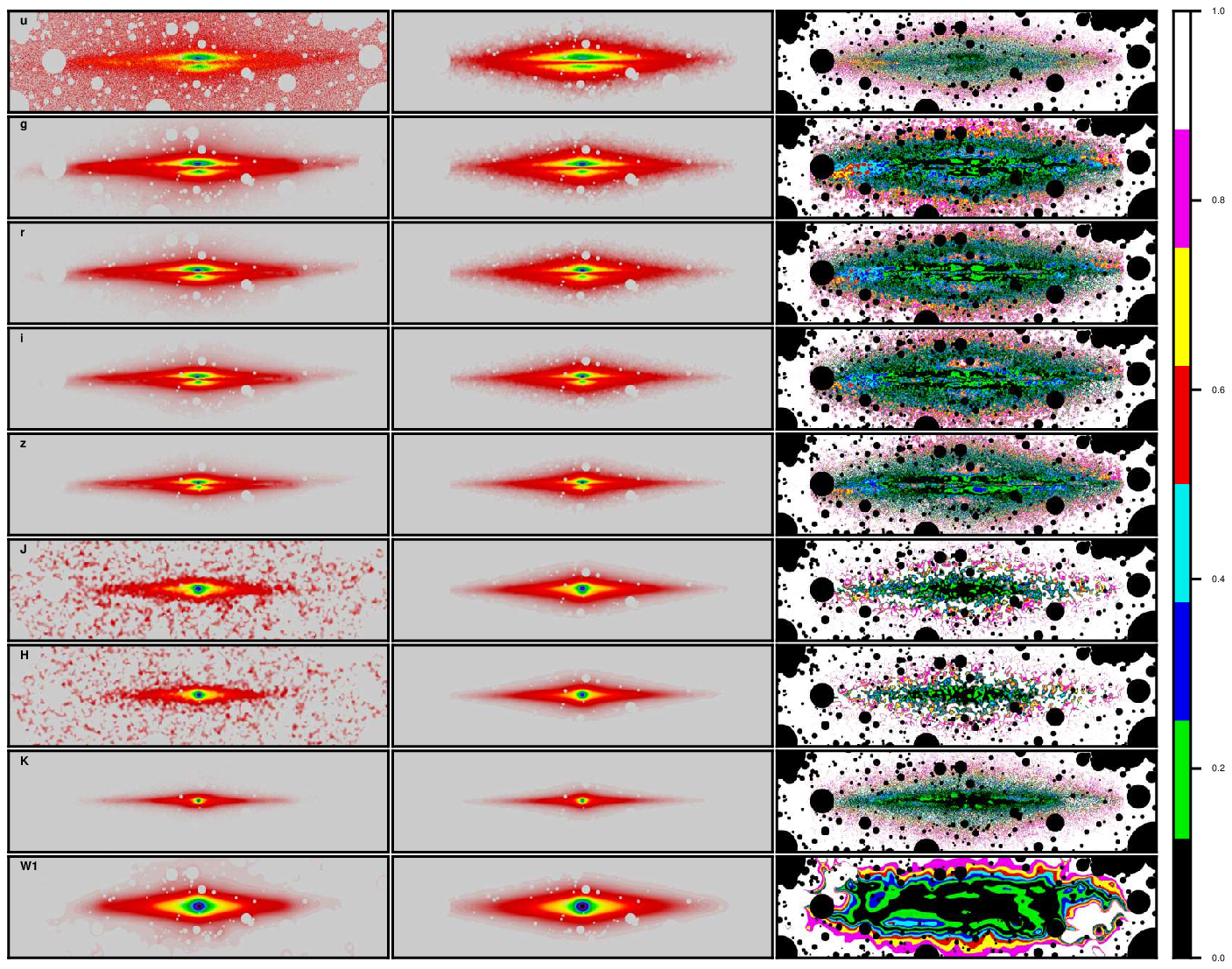}

\caption{Results of the oligochromatic \textsc{fitskirt} radiative transfer fits for UGC\,4277. In each panel, the left-hand column represents the observed image, the middle column contains the corresponding fits in the same bands, the right-hand panel shows the residual images, which indicate the relative deviation between the fit and the image (in modulus).}
\label{UGC4277_fit_models}
\end{figure*}

\begin{figure*}
\centering
\includegraphics[width=\textwidth]{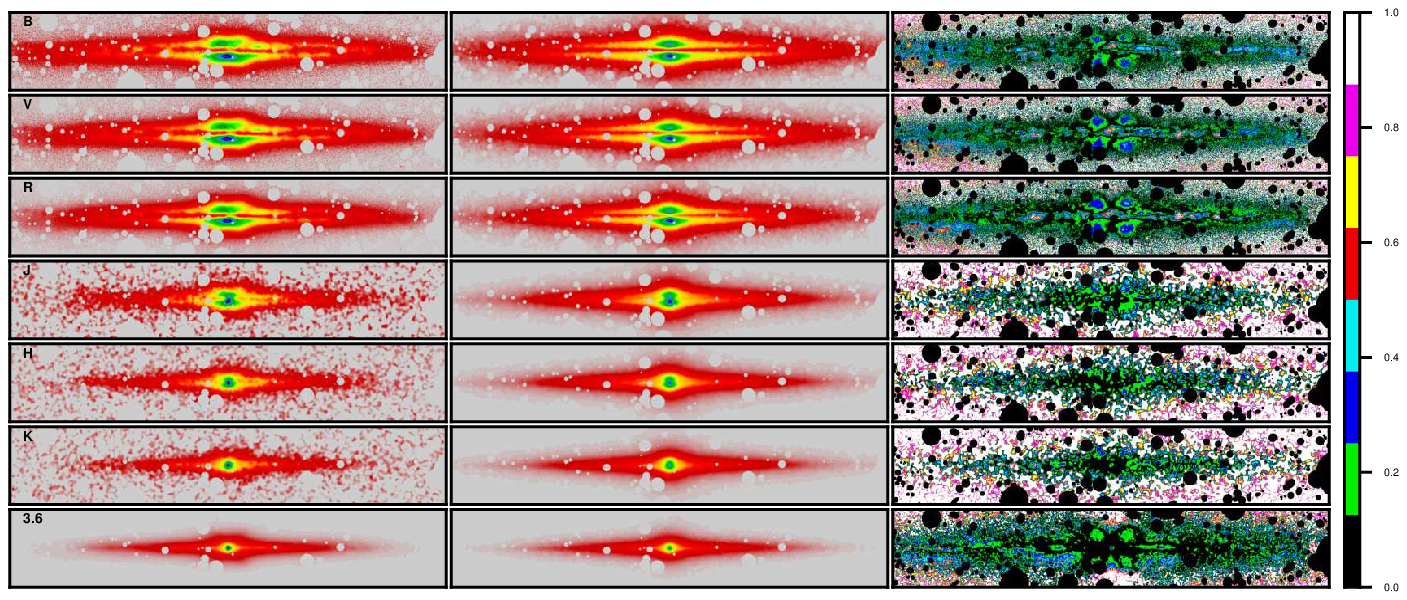}

\caption{Results of the oligochromatic \textsc{fitskirt} radiative transfer fits for IC\,2531. In each panel, the left-hand column represents the observed image, the middle column contains the corresponding fits in the same bands, the right-hand panel shows the residual images, which indicate the relative deviation between the fit and the image (in modulus).}
\label{IC2531_fit_models}
\end{figure*}

\begin{figure*}
\centering
\includegraphics[width=0.85\textwidth]{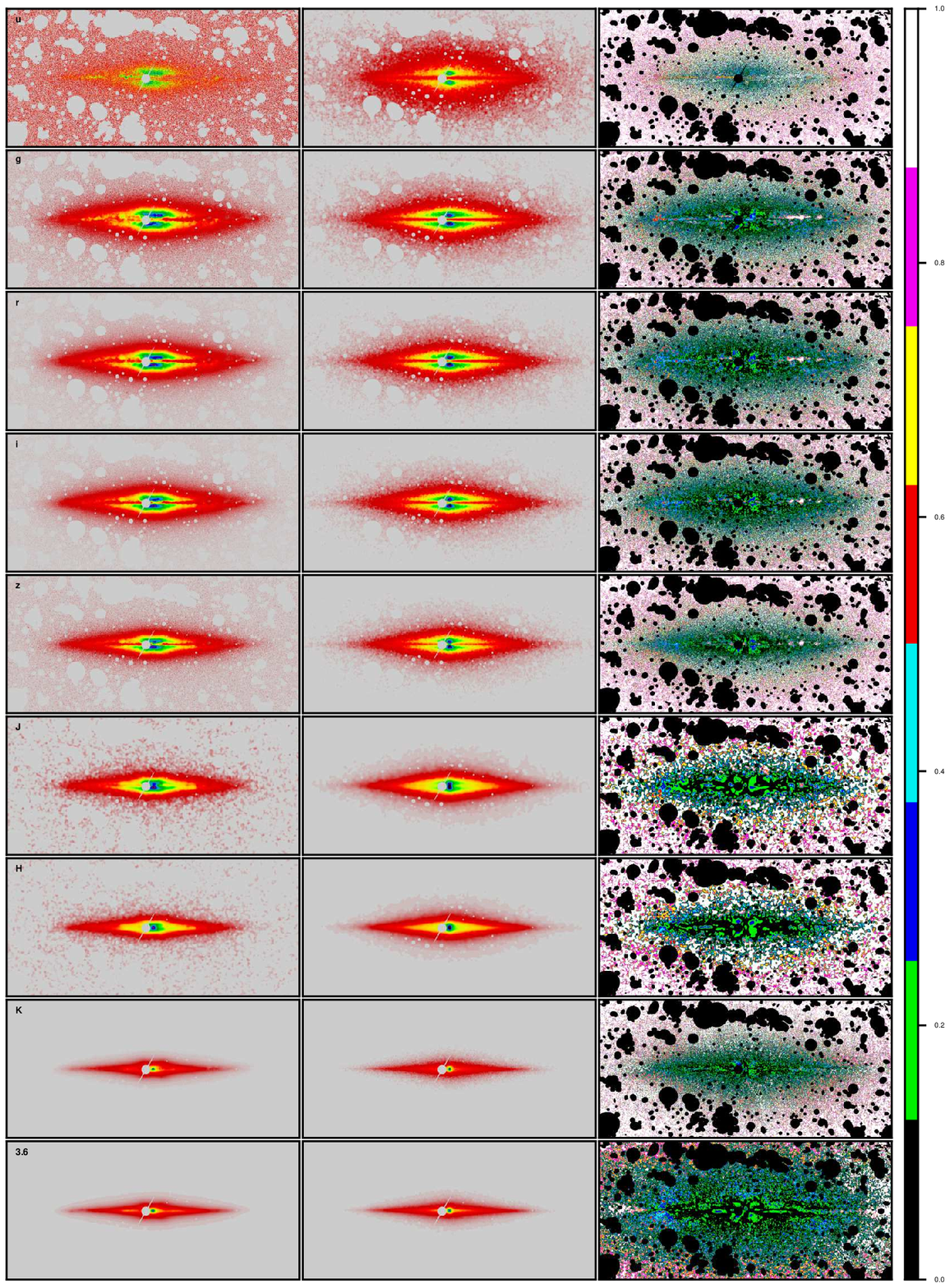}

\caption{Results of the oligochromatic \textsc{fitskirt} radiative transfer fits for NGC\,4013. In each panel, the left-hand column represents the observed image, the middle column contains the corresponding fits in the same bands, the right-hand panel shows the residual images, which indicate the relative deviation between the fit and the image (in modulus).}
\label{NGC4013_fit_models}
\end{figure*}

\begin{figure*}
\centering
\includegraphics[width=\textwidth]{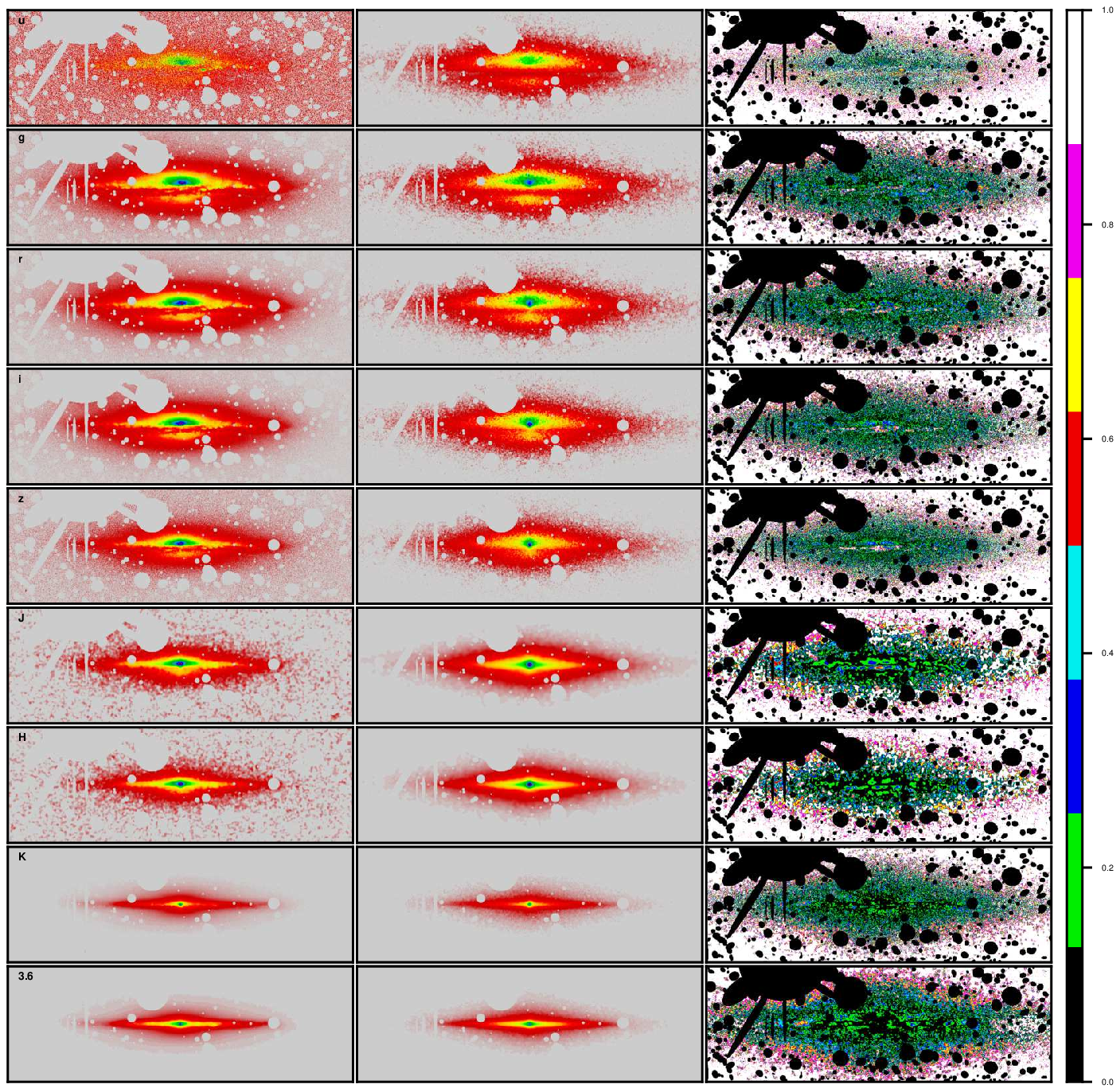}

\caption{Results of the oligochromatic \textsc{fitskirt} radiative transfer fits for NGC\,4217. In each panel, the left-hand column represents the observed image, the middle column contains the corresponding fits in the same bands, the right-hand panel shows the residual images, which indicate the relative deviation between the fit and the image (in modulus).}
\label{NGC4217_fit_models}
\end{figure*}

\begin{figure*}
\centering
\includegraphics[width=\textwidth]{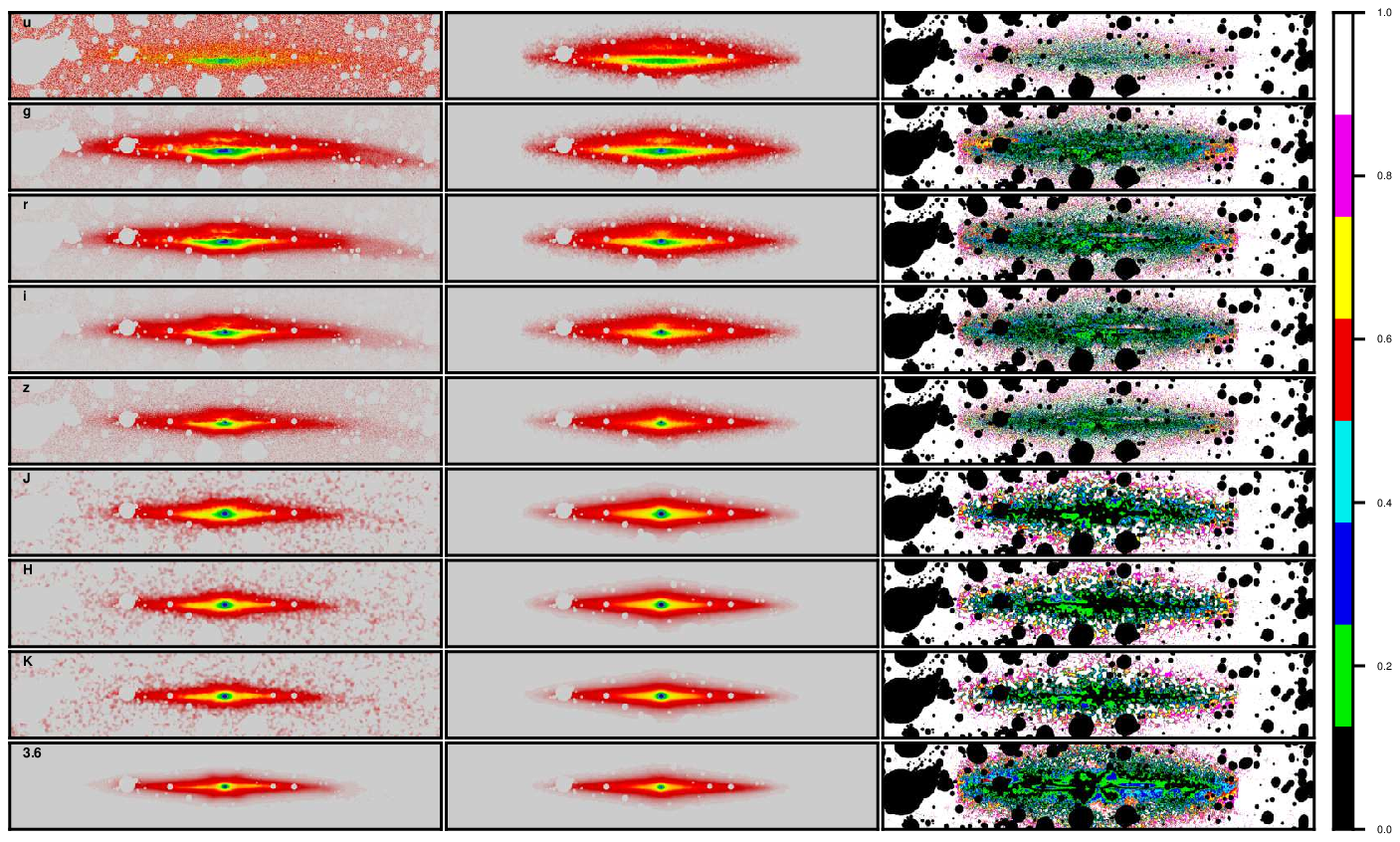}

\caption{Results of the oligochromatic \textsc{fitskirt} radiative transfer fits for NGC\,5529. In each panel, the left-hand column represents the observed image, the middle column contains the corresponding fits in the same bands, the right-hand panel shows the residual images, which indicate the relative deviation between the fit and the image (in modulus).}
\label{NGC5529_fit_models}
\end{figure*}

\begin{figure*}
\centering
\includegraphics[width=\textwidth]{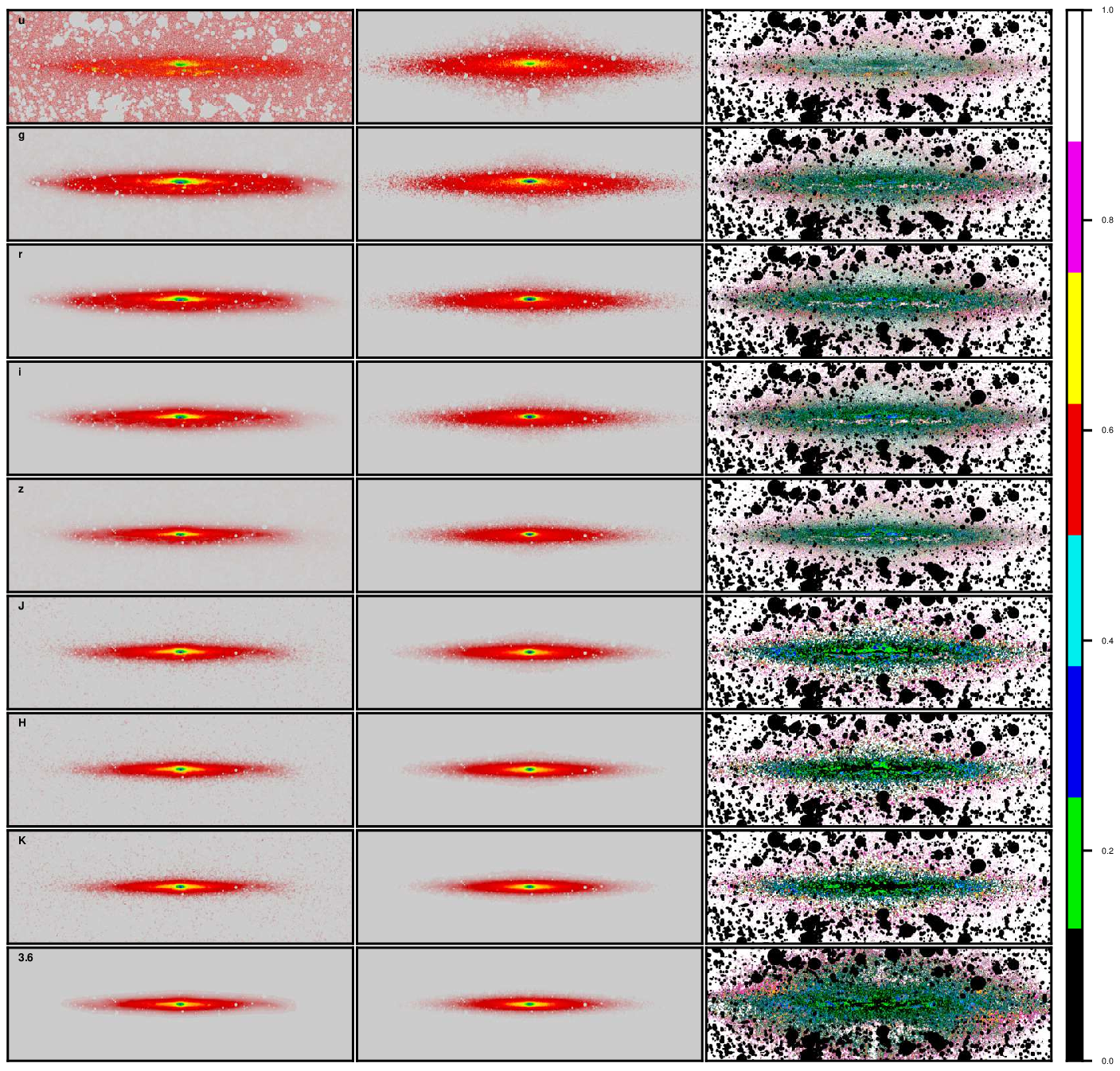}

\caption{Results of the oligochromatic \textsc{fitskirt} radiative transfer fits for NGC\,5907. In each panel, the left-hand column represents the observed image, the middle column contains the corresponding fits in the same bands, the right-hand panel shows the residual images, which indicate the relative deviation between the fit and the image (in modulus).}
\label{NGC5907_fit_models}
\end{figure*}

\section{SEDs}
\label{Appendix_SEDs}

\begin{figure*}
\centering
\includegraphics[width=14.5cm, angle=0, clip=]{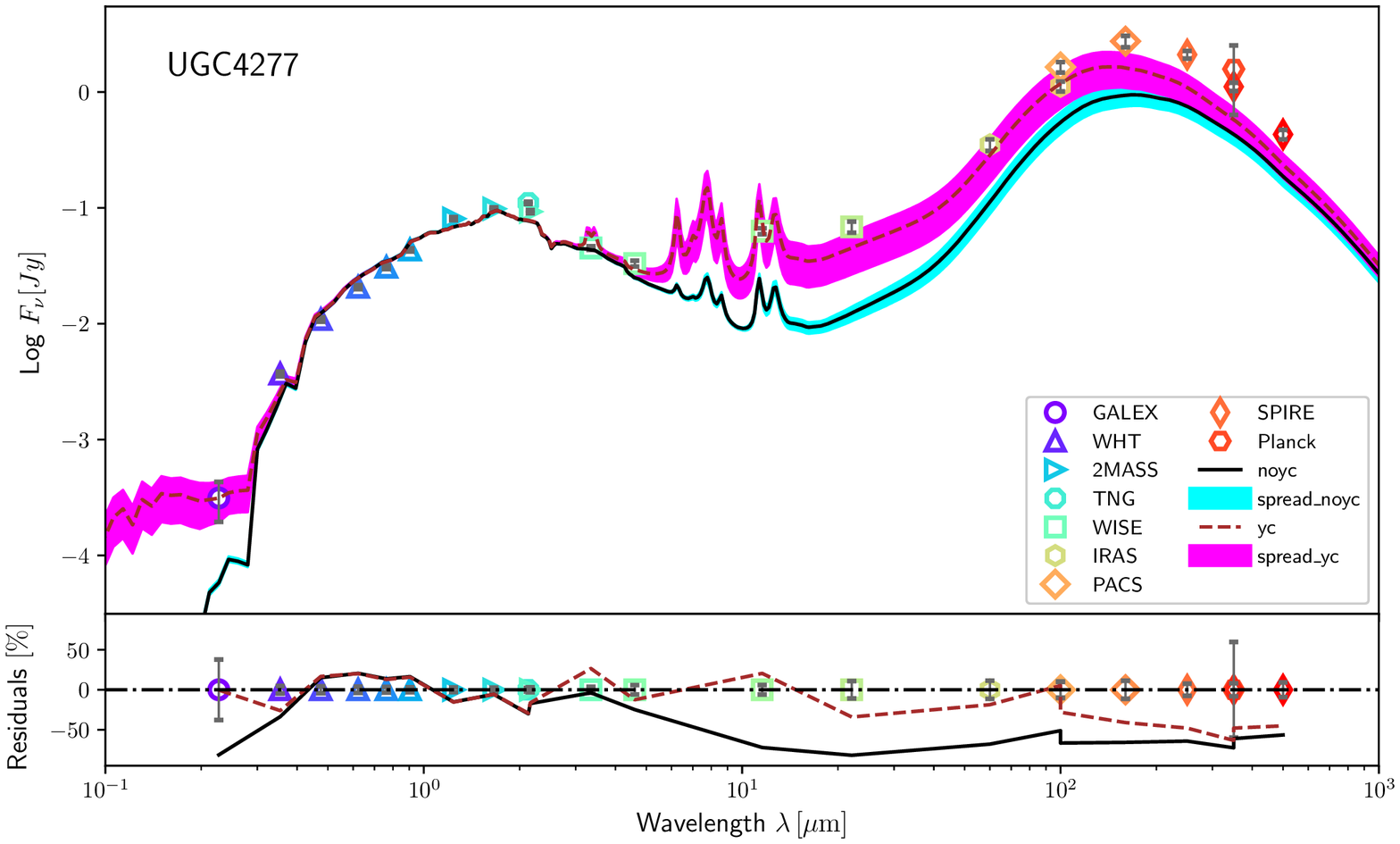}
\caption{The SEDs of UGC\,4277 with the THEMIS dust mixture. The coloured markers with error bars correspond to the flux densities listed in Table~\ref{tab:Fluxes}. The bottom panel below the SEDs shows the relative residuals between the observed SEDs and the models. The black line and the cyan spread correspond to the initial model (without a young stellar component, denoted as `noyc' and `spread\_noyc', respectively), the brown line and the magenta spread refer to the model with an additional young stellar component (denoted as `yc' and `spread\_yc', respectively). The spreads are plotted for the corresponding models within one error bar of the best oligochromatic fitting model parameters.}
\label{sed_UGC4277}
\end{figure*}

\begin{figure*}
\centering
\includegraphics[width=14.5cm, angle=0, clip=]{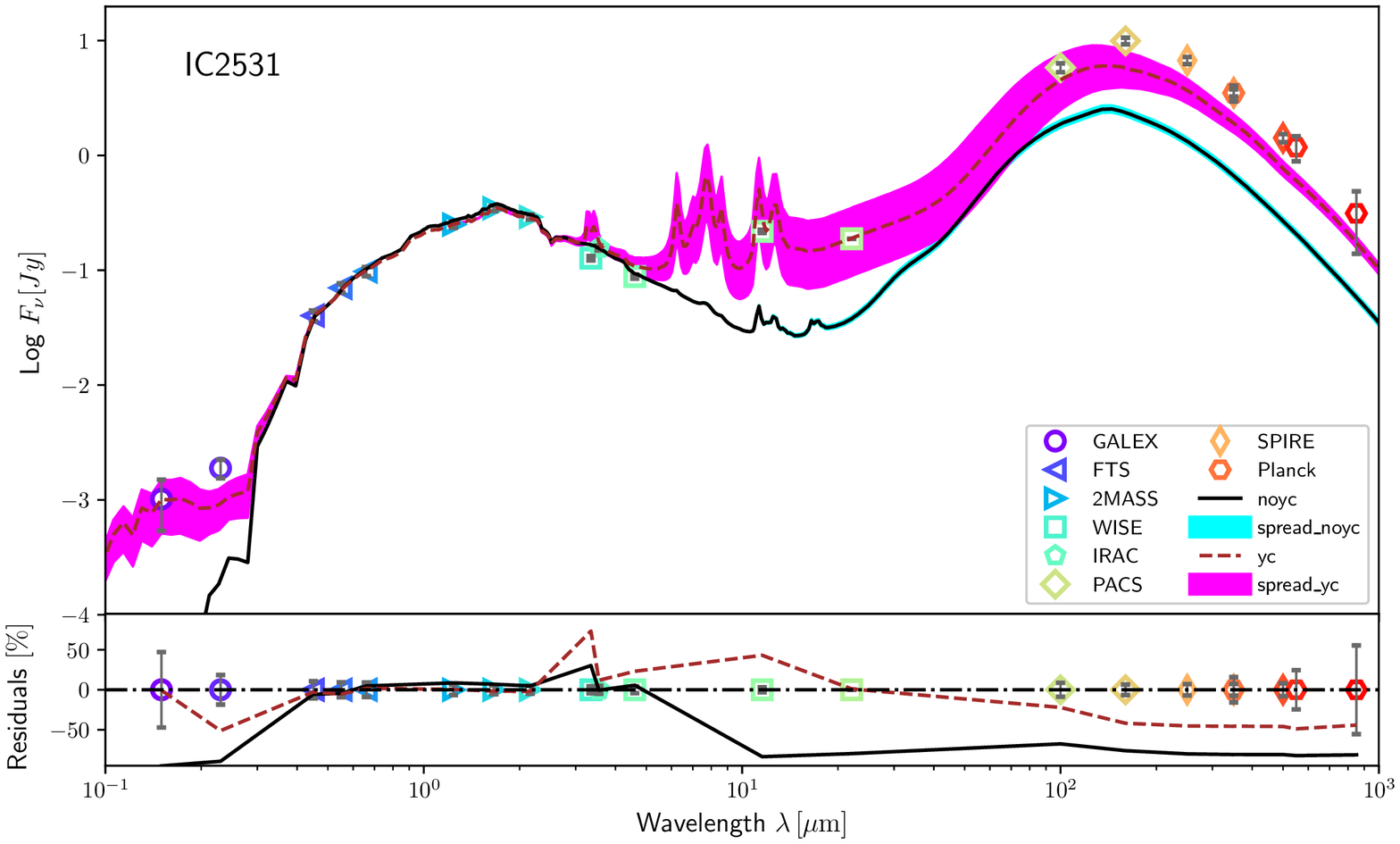}
\caption{The SEDs of IC\,2531 with the THEMIS dust mixture. The coloured markers with error bars correspond to the flux densities listed in Table~\ref{tab:Fluxes}. The bottom panel below the SEDs shows the relative residuals between the observed SEDs and the models. The black line and the cyan spread correspond to the initial model (without a young stellar component, denoted as `noyc' and `spread\_noyc', respectively), the brown line and the magenta spread refer to the model with an additional young stellar component (denoted as `yc' and `spread\_yc', respectively). The spreads are plotted for the corresponding models within one error bar of the best oligochromatic fitting model parameters.}
\label{sed_UGC4277}
\end{figure*}

\begin{figure*}
\centering
\includegraphics[width=14.5cm, angle=0, clip=]{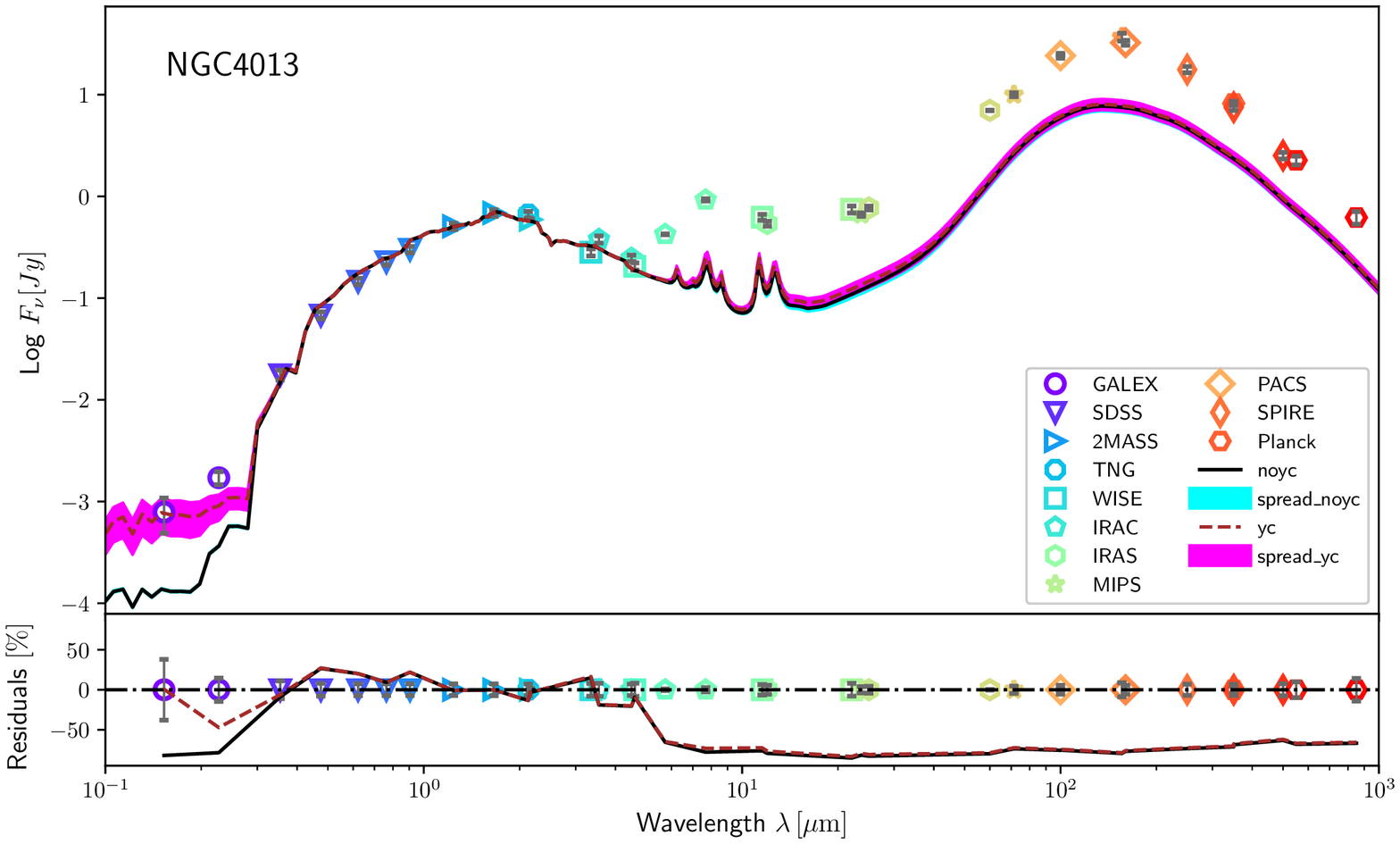}
\caption{The SEDs of NGC\,4013 with the THEMIS dust mixture. The coloured markers with error bars correspond to the flux densities listed in Table~\ref{tab:Fluxes}. The bottom panel below the SEDs shows the relative residuals between the observed SEDs and the models. The black line and the cyan spread correspond to the initial model (without a young stellar component, denoted as `noyc' and `spread\_noyc', respectively), the brown line and the magenta spread refer to the model with an additional young stellar component (denoted as `yc' and `spread\_yc', respectively). The spreads are plotted for the corresponding models within one error bar of the best oligochromatic fitting model parameters.}
\label{sed_NGC4013}
\end{figure*}

\begin{figure*}
\centering
\includegraphics[width=14.5cm, angle=0, clip=]{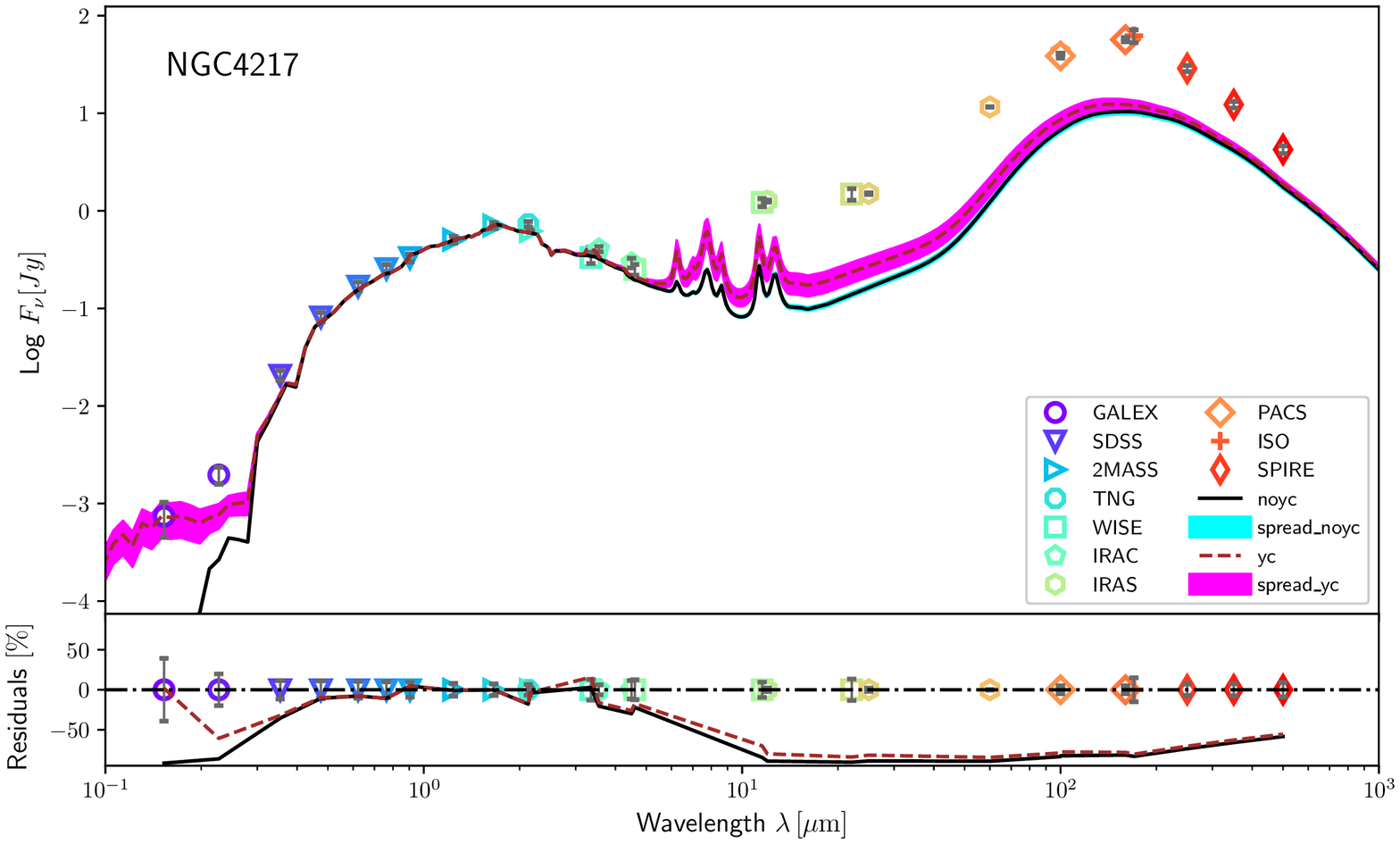}
\caption{The SEDs of NGC\,4217 with the THEMIS dust mixture. The coloured markers with error bars correspond to the flux densities listed in Table~\ref{tab:Fluxes}. The bottom panel below the SEDs shows the relative residuals between the observed SEDs and the models. The black line and the cyan spread correspond to the initial model (without a young stellar component, denoted as `noyc' and `spread\_noyc', respectively), the brown line and the magenta spread refer to the model with an additional young stellar component (denoted as `yc' and `spread\_yc', respectively). The spreads are plotted for the corresponding models within one error bar of the best oligochromatic fitting model parameters.}
\label{sed_NGC4217}
\end{figure*}

\begin{figure*}
\centering
\includegraphics[width=14.5cm, angle=0, clip=]{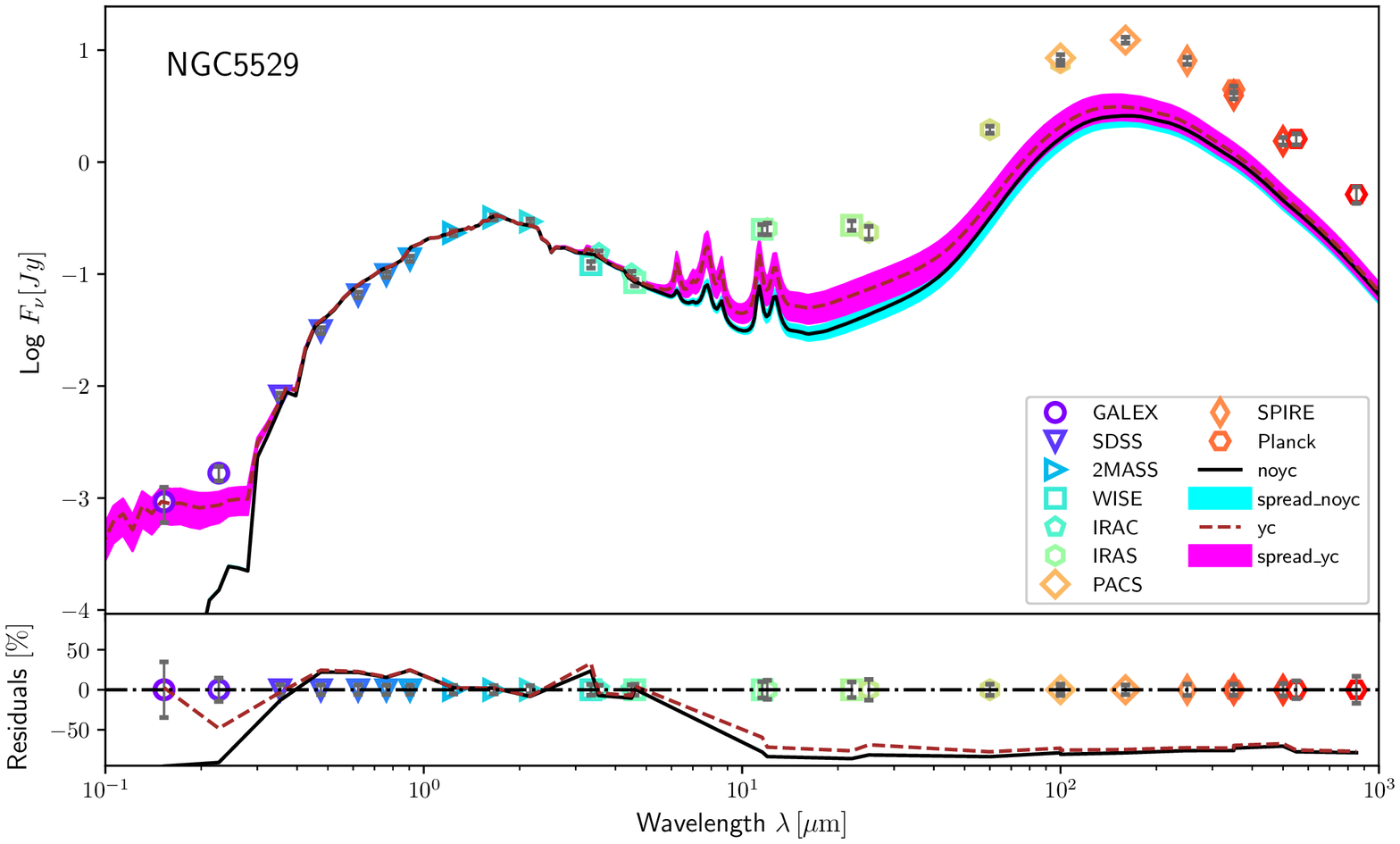}
\caption{The SEDs of NGC\,5529 with the THEMIS dust mixture. The coloured markers with error bars correspond to the flux densities listed in Table~\ref{tab:Fluxes}. The bottom panel below the SEDs shows the relative residuals between the observed SEDs and the models. The black line and the cyan spread correspond to the initial model (without a young stellar component, denoted as `noyc' and `spread\_noyc', respectively), the brown line and the magenta spread refer to the model with an additional young stellar component (denoted as `yc' and `spread\_yc', respectively). The spreads are plotted for the corresponding models within one error bar of the best oligochromatic fitting model parameters.}
\label{sed_NGC5529}
\end{figure*}

\begin{figure*}
\centering
\includegraphics[width=14.5cm, angle=0, clip=]{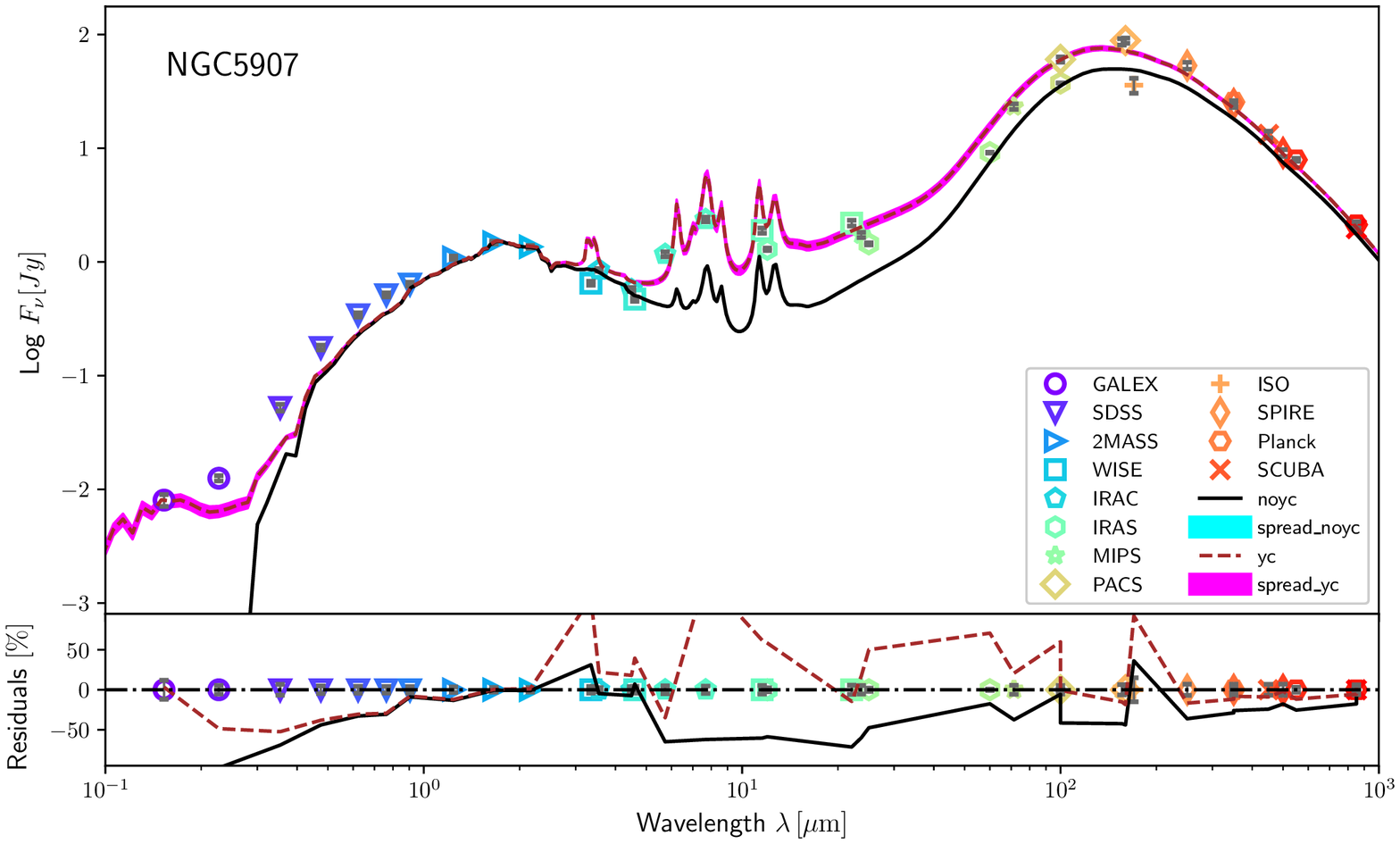}
\caption{The SEDs of NGC\,5907 with the THEMIS dust mixture. The coloured markers with error bars correspond to the flux densities listed in Table~\ref{tab:Fluxes}. The bottom panel below the SEDs shows the relative residuals between the observed SEDs and the models. The black line and the cyan spread correspond to the initial model (without a young stellar component, denoted as `noyc' and `spread\_noyc', respectively), the brown line and the magenta spread refer to the model with an additional young stellar component (denoted as `yc' and `spread\_yc', respectively). The spreads are plotted for the corresponding models within one error bar of the best oligochromatic fitting model parameters.}
\label{sed_NGC5907}
\end{figure*}

\section{Panchromatic simulations}
\label{Appendix__simulations}

\clearpage

\begin{figure*}
\centering
\includegraphics[width=0.9\textwidth, angle=0, clip=]{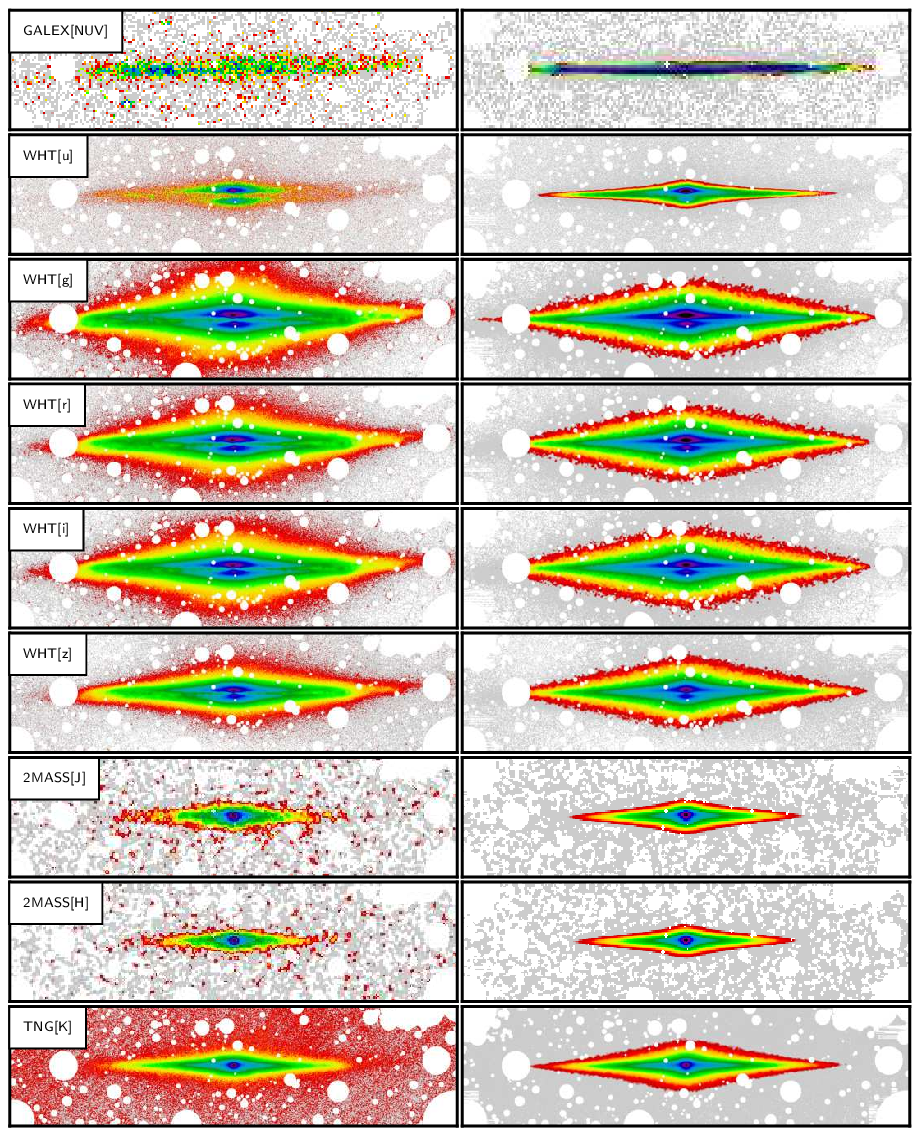}
\caption{Comparison between the observations (left) and panchromatic simulations (right) for UGC\,4277. The model includes the young stellar population disc. Foreground stars have been masked. Gray-coloured pixels have intensities lower than $2\sigma$ of the background.}
\label{map_ugc4277}
\end{figure*}

\addtocounter{figure}{-1}
\begin{figure*}
\centering
\includegraphics[width=0.9\textwidth, angle=0, clip=]{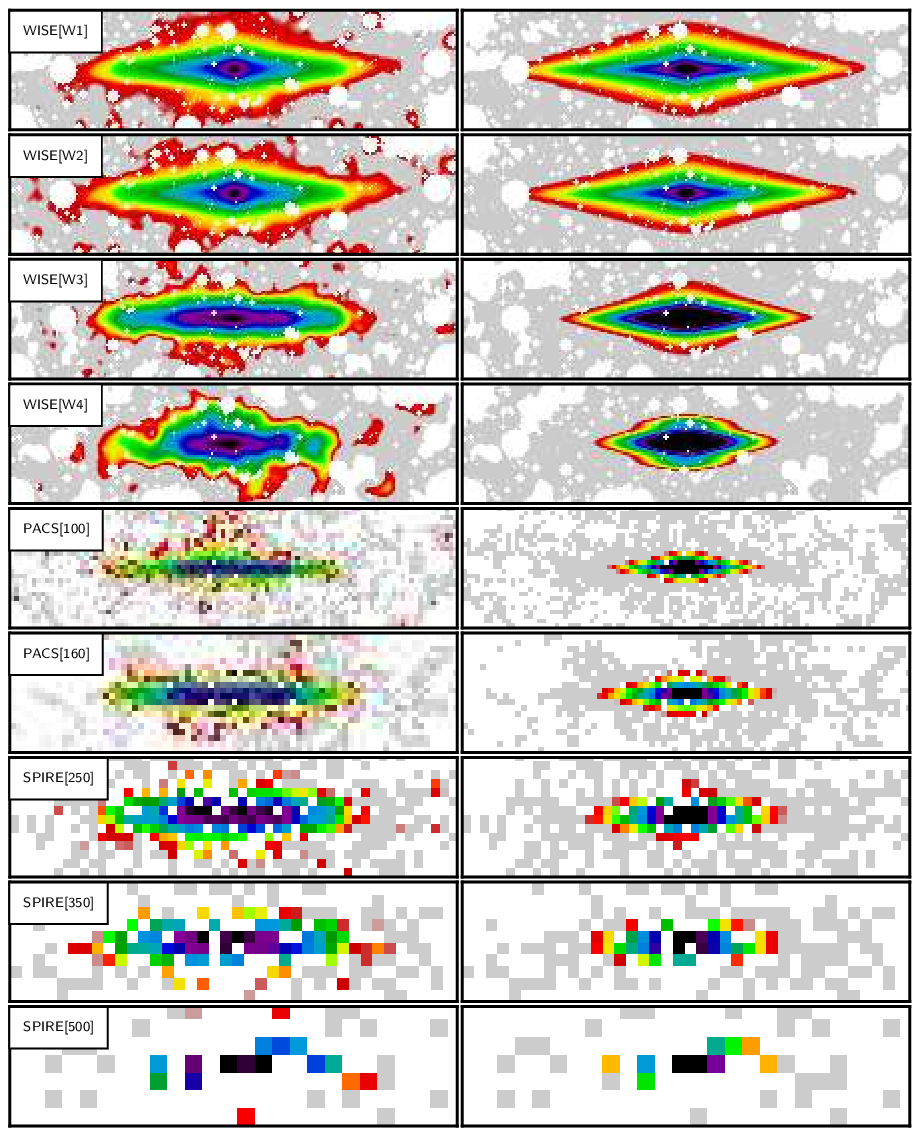}
\caption{(continued)}
\end{figure*}

\begin{figure*}
\centering
\includegraphics[width=0.9\textwidth, angle=0, clip=]{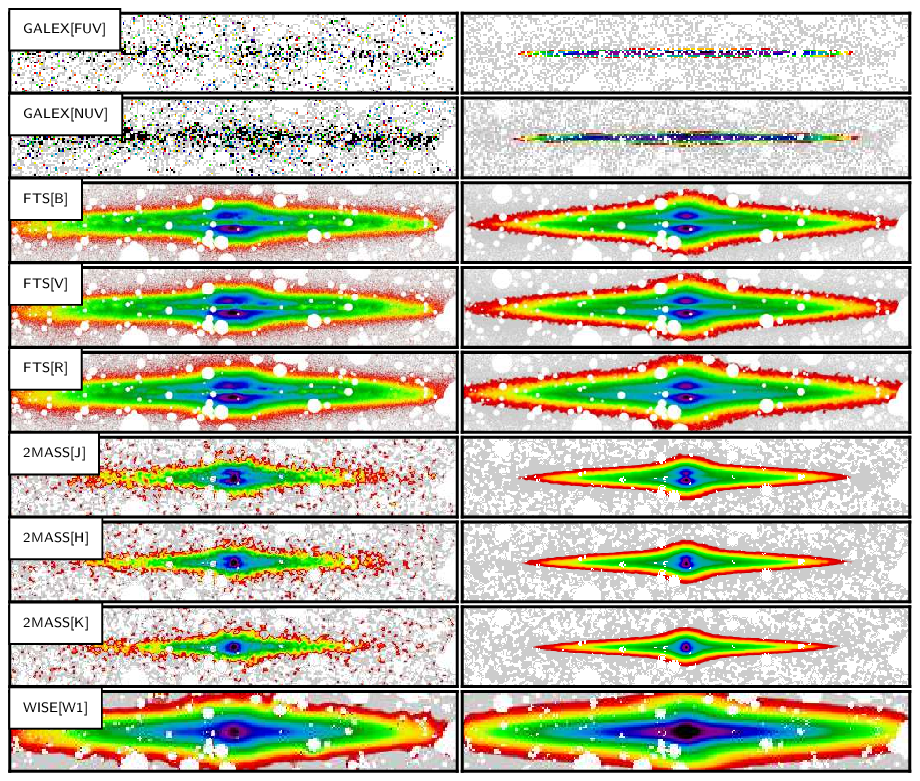}
\caption{Comparison between the observations (left) and panchromatic simulations (right) for IC\,2531. The model includes the young stellar population disc. Foreground stars have been masked. Gray-coloured pixels have intensities lower than $2\sigma$ of the background.}
\label{map_ugc4277}
\end{figure*}

\addtocounter{figure}{-1}
\begin{figure*}
\centering
\includegraphics[width=0.9\textwidth, angle=0, clip=]{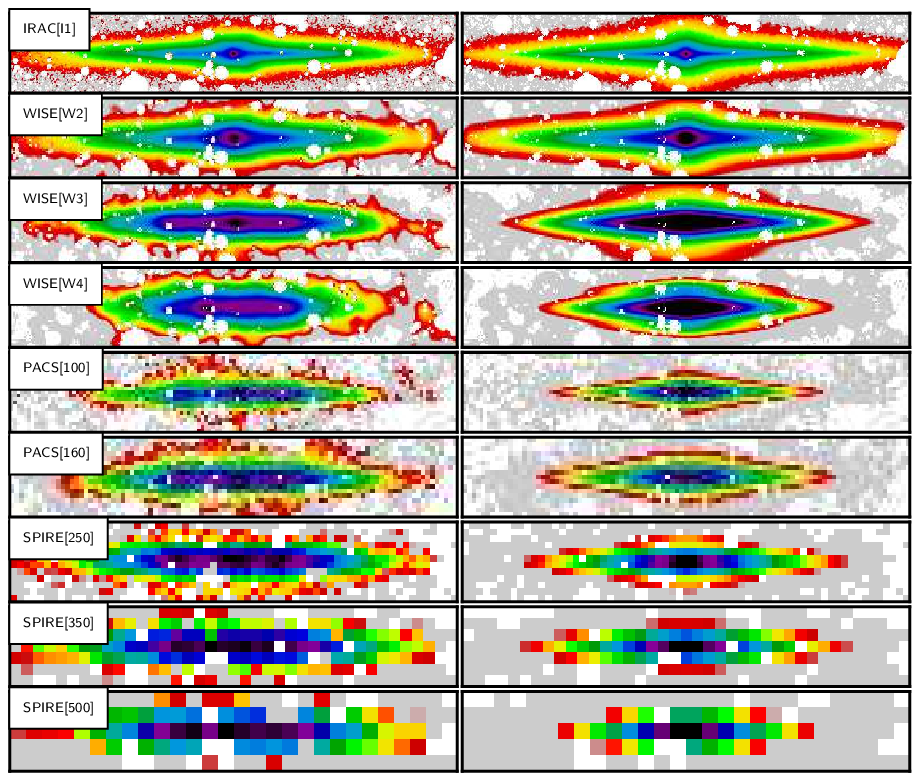}
\caption{(continued)}
\end{figure*}

\begin{figure*}
\centering
\includegraphics[width=0.9\textwidth, angle=0, clip=]{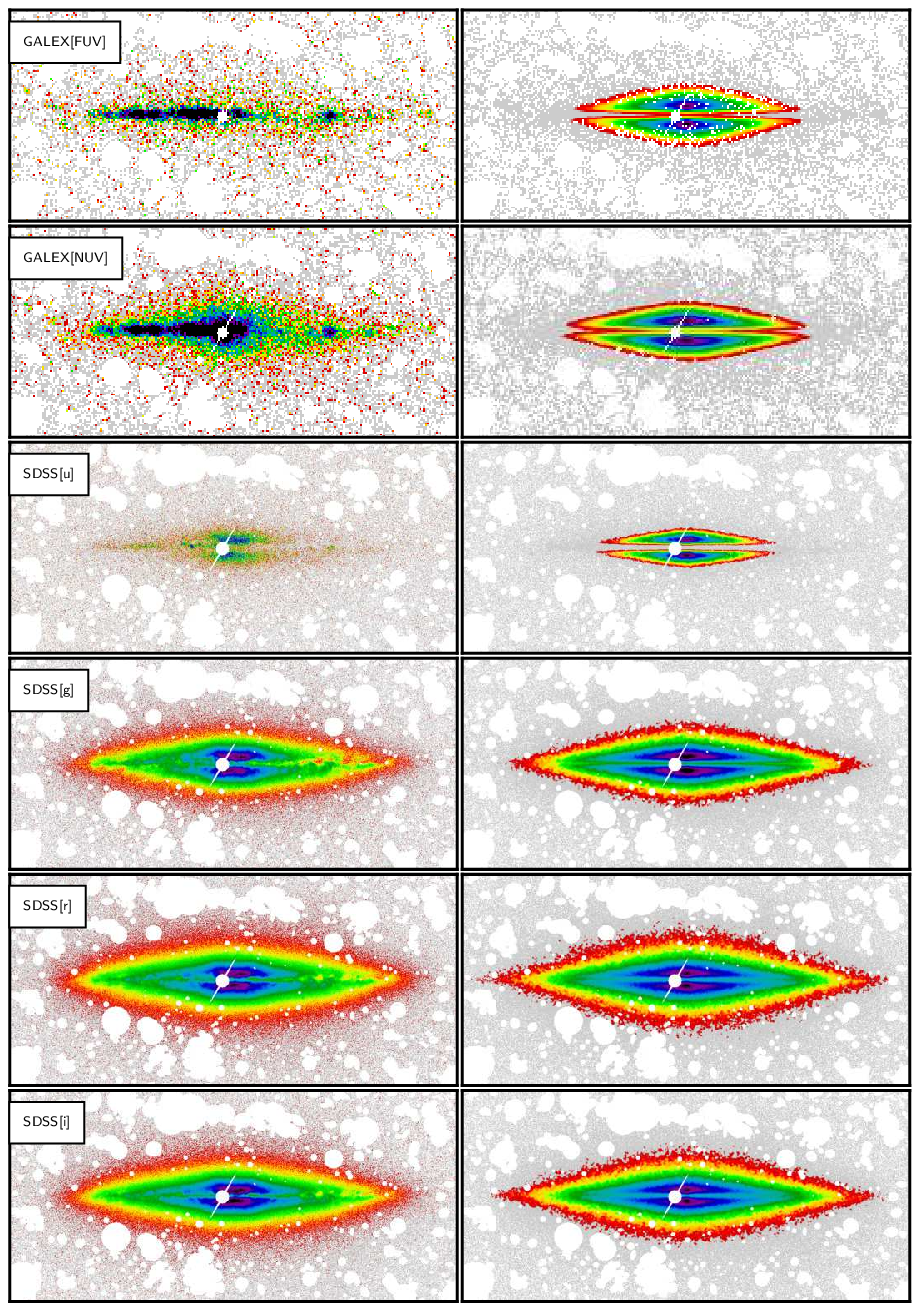}
\caption{Comparison between the observations (left) and panchromatic simulations (right) for NGC\,4013. The model includes the young stellar population disc. Foreground stars have been masked. Gray-coloured pixels have intensities lower than $2\sigma$ of the background.}
\label{map_ngc4013}
\end{figure*}

\addtocounter{figure}{-1}
\begin{figure*}
\centering
\includegraphics[width=0.9\textwidth, angle=0, clip=]{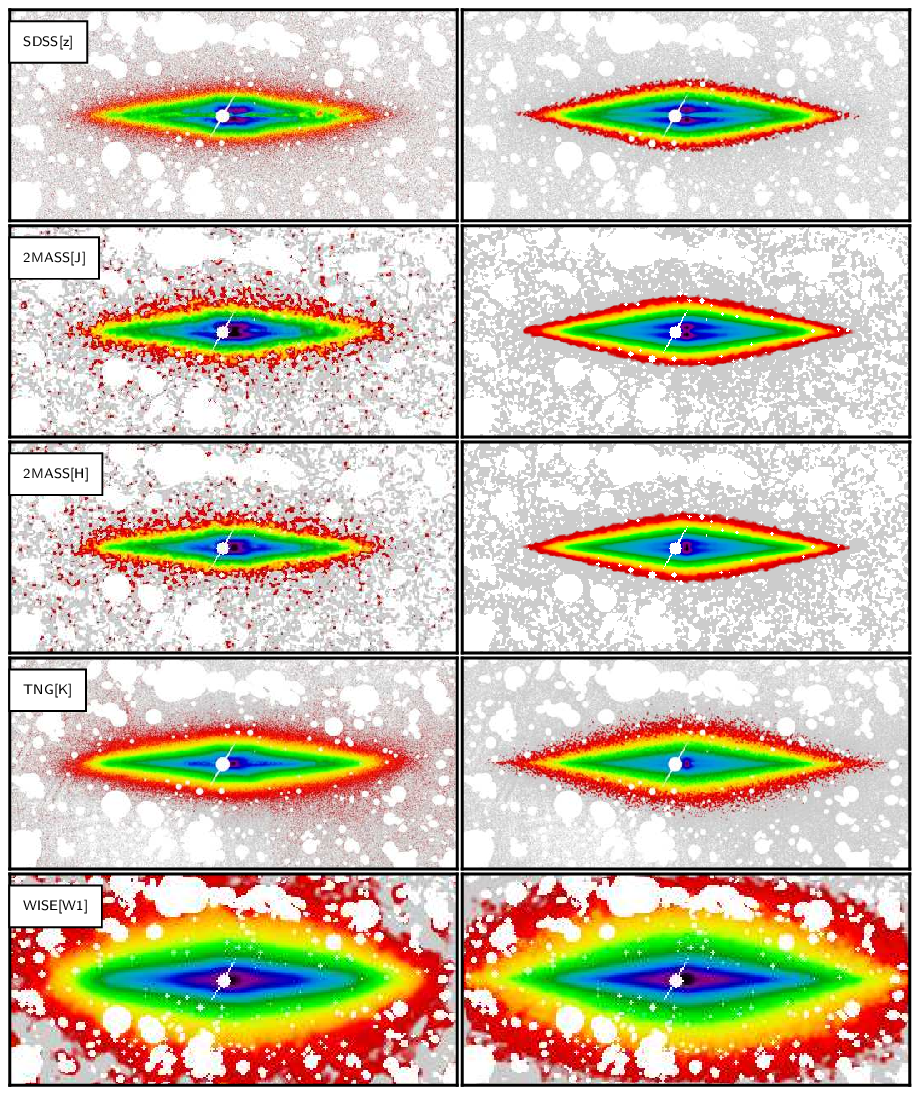}
\caption{(continued)}
\end{figure*}

\addtocounter{figure}{-1}
\begin{figure*}
\centering
\includegraphics[width=0.9\textwidth, angle=0, clip=]{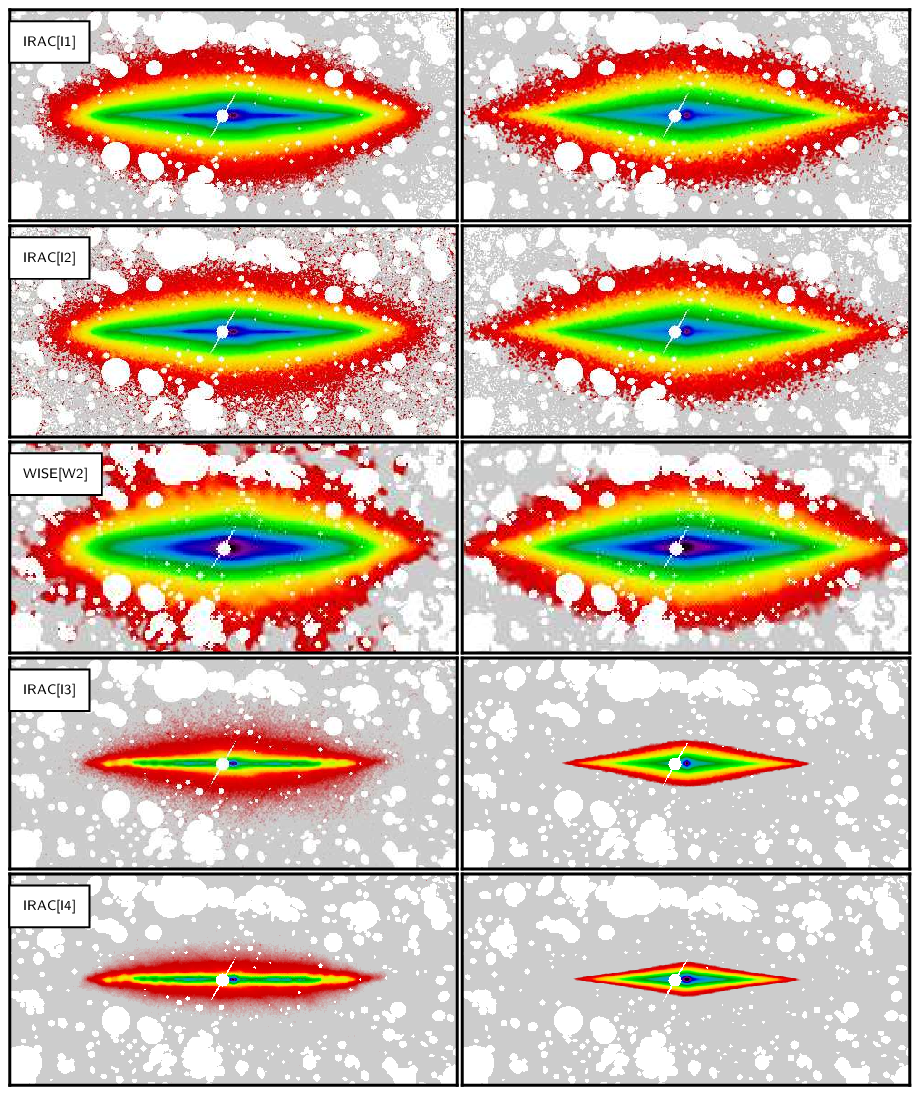}
\caption{(continued)}
\end{figure*}

\addtocounter{figure}{-1}
\begin{figure*}
\centering
\includegraphics[width=0.9\textwidth, angle=0, clip=]{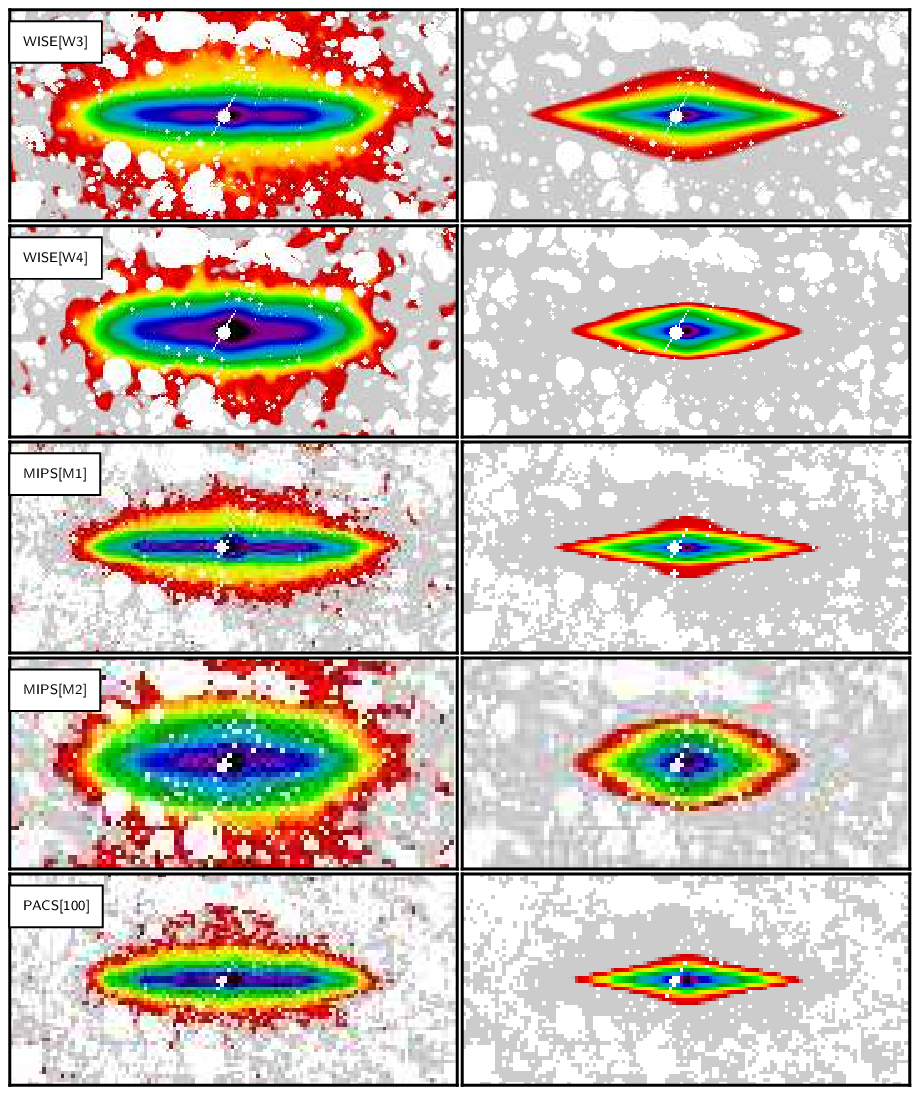}
\caption{(continued)}
\end{figure*}

\addtocounter{figure}{-1}
\begin{figure*}
\centering
\includegraphics[width=0.9\textwidth, angle=0, clip=]{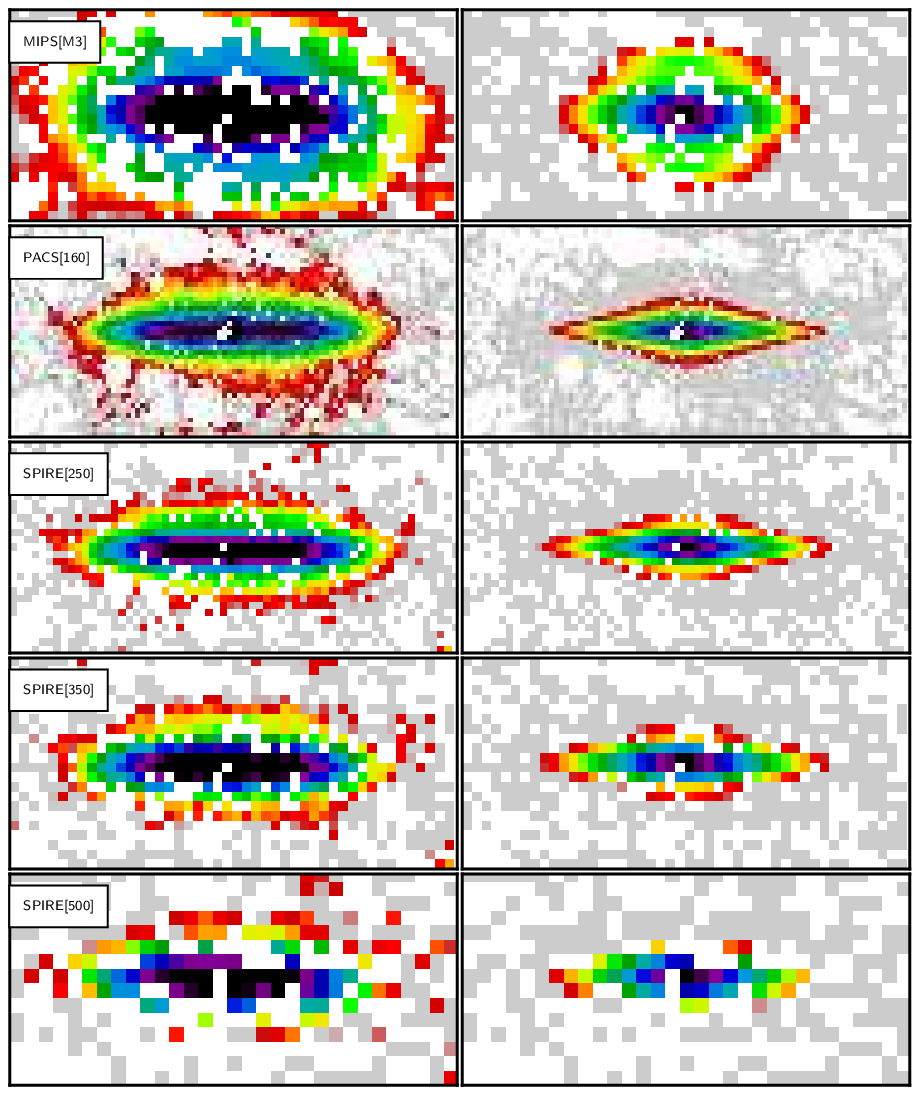}
\caption{(continued)}
\end{figure*}

\begin{figure*}
\centering
\includegraphics[width=0.9\textwidth, angle=0, clip=]{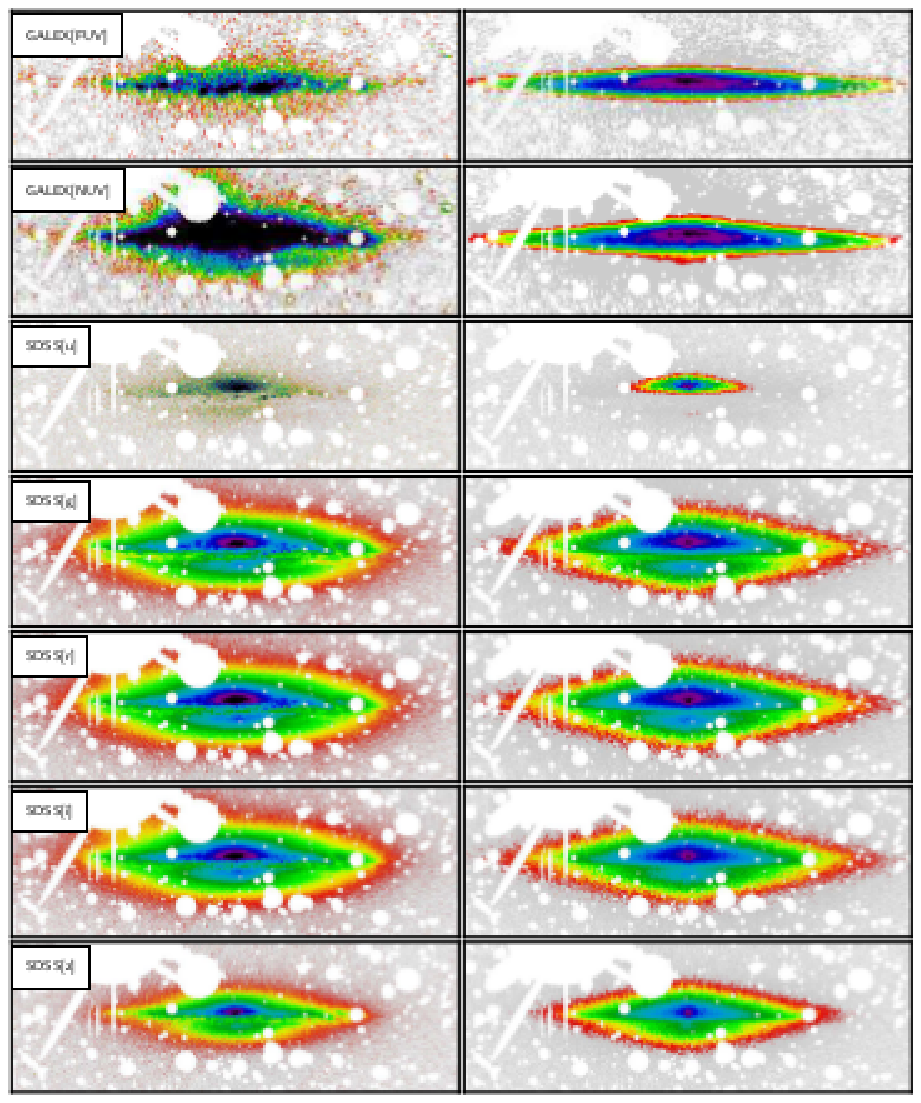}
\caption{Comparison between the observations (left) and panchromatic simulations (right) for NGC\,4217. The model includes the young stellar population disc. Foreground stars have been masked. Gray-coloured pixels have intensities lower than $2\sigma$ of the background.}
\label{map_ngc4217}
\end{figure*}

\addtocounter{figure}{-1}
\begin{figure*}
\centering
\includegraphics[width=0.9\textwidth, angle=0, clip=]{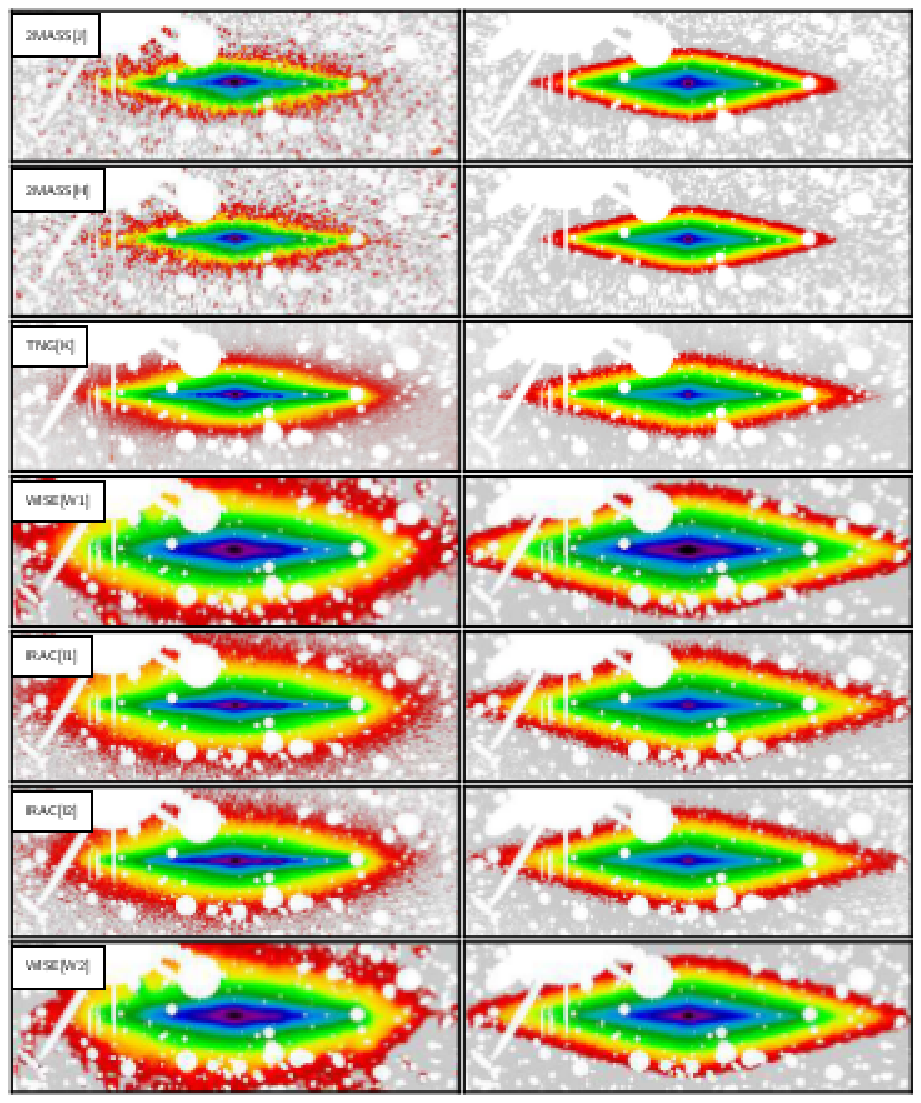}
\caption{(continued)}
\end{figure*}

\addtocounter{figure}{-1}
\begin{figure*}
\centering
\includegraphics[width=0.9\textwidth, angle=0, clip=]{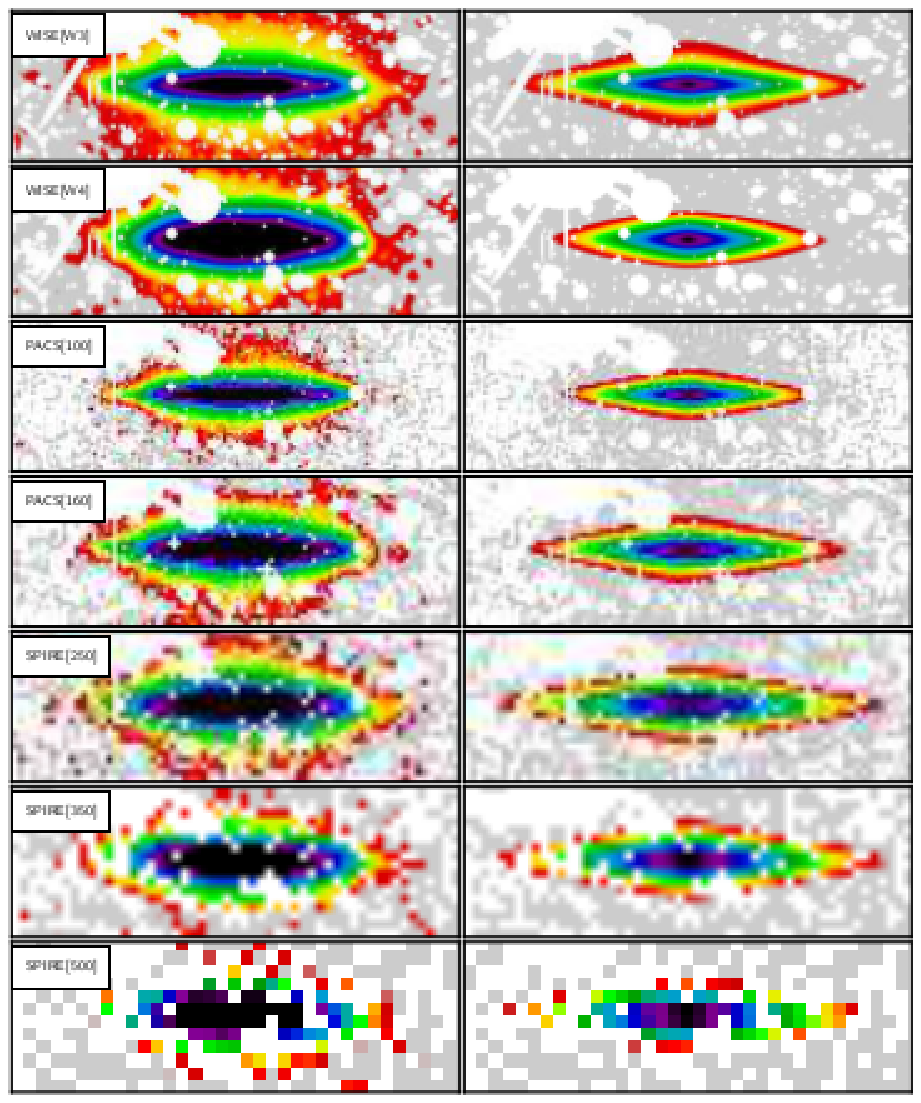}
\caption{(continued)}
\end{figure*}

\begin{figure*}
\centering
\includegraphics[width=0.9\textwidth, angle=0, clip=]{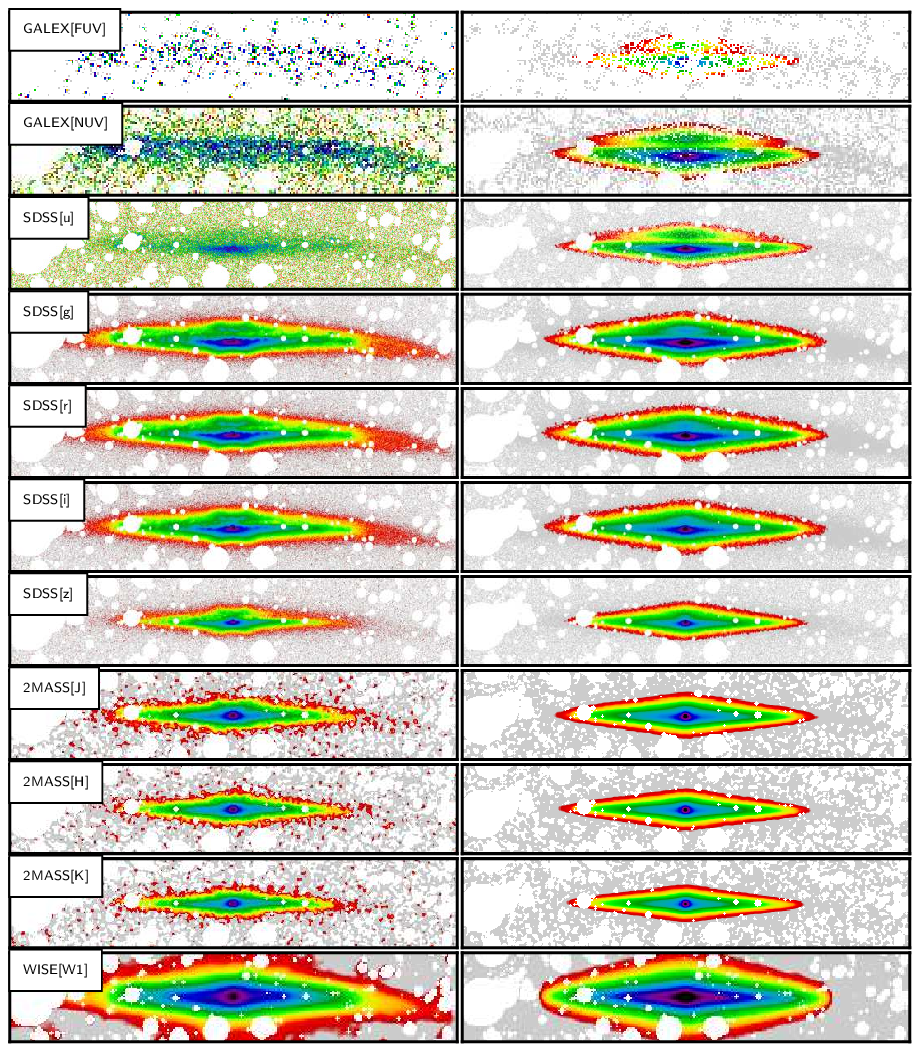}
\caption{Comparison between the observations (left) and panchromatic simulations (right) for NGC\,5529. The model includes the young stellar population disc. Foreground stars have been masked. Gray-coloured pixels have intensities lower than $2\sigma$ of the background.}
\label{map_ngc5529}
\end{figure*}

\addtocounter{figure}{-1}
\begin{figure*}
\centering
\includegraphics[width=0.9\textwidth, angle=0, clip=]{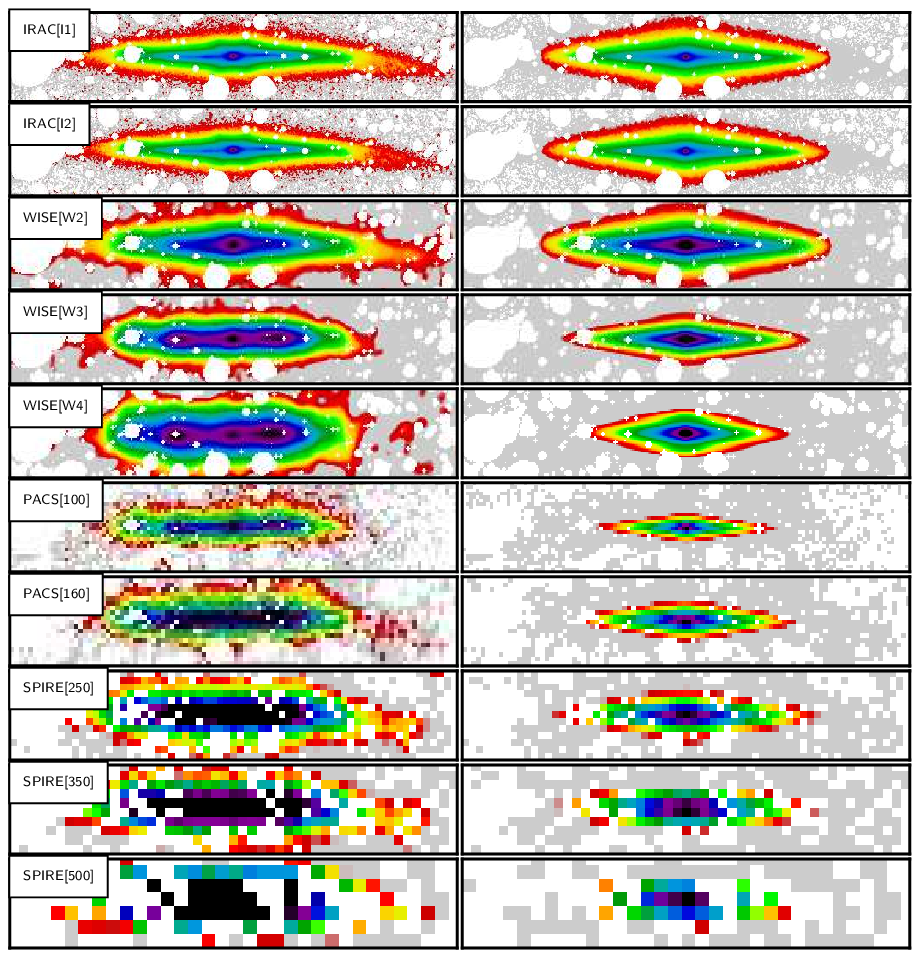}
\caption{(continued)}
\end{figure*}

\begin{figure*}
\centering
\includegraphics[width=0.85\textwidth, angle=0, clip=]{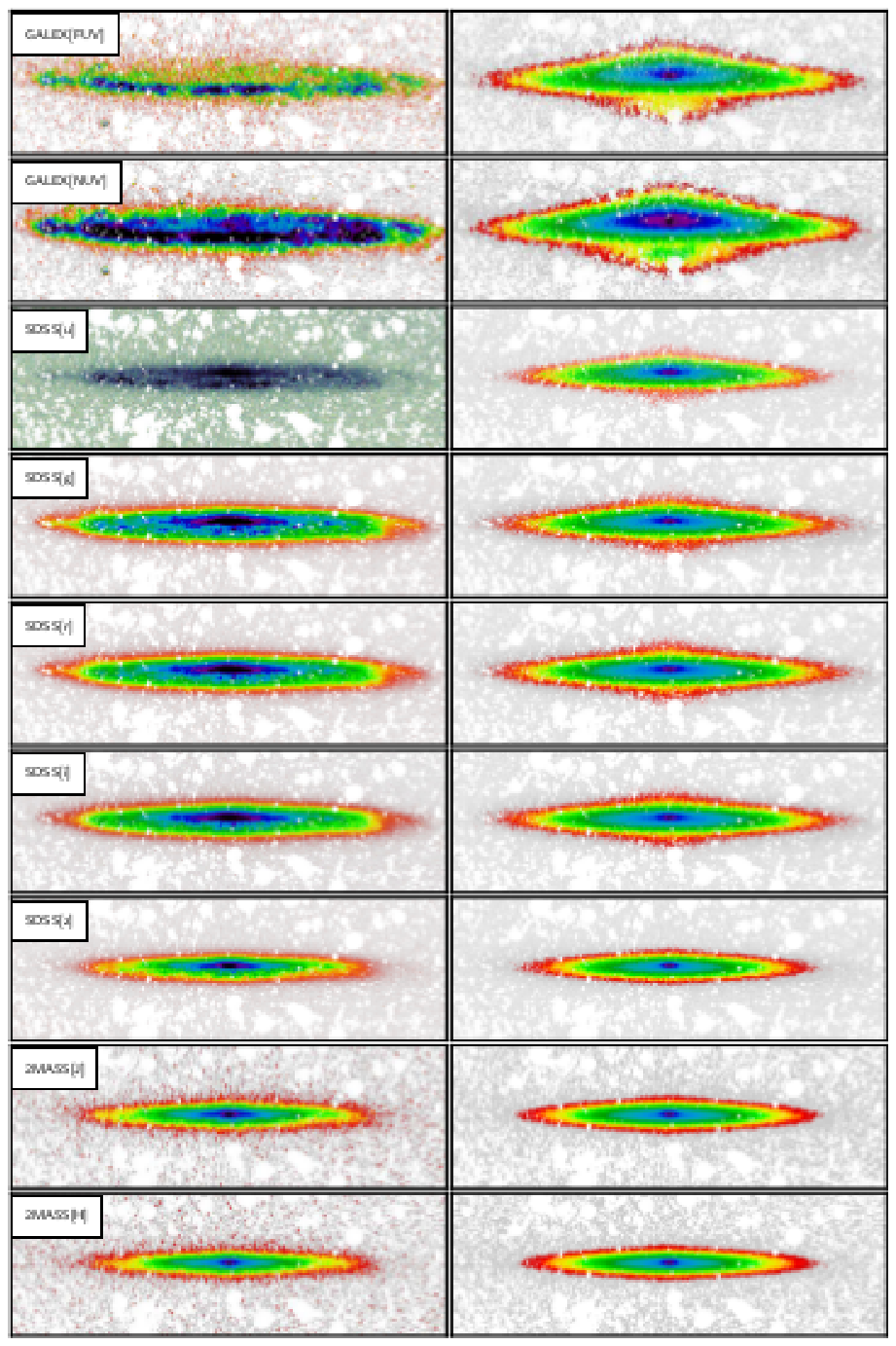}
\caption{Comparison between the observations (left) and panchromatic simulations (right) for NGC\,5907. The model includes the young stellar population disc. Foreground stars have been masked. Gray-coloured pixels have intensities lower than $2\sigma$ of the background.}
\label{map_ngc5907}
\end{figure*}

\addtocounter{figure}{-1}
\begin{figure*}
\centering
\includegraphics[width=0.85\textwidth, angle=0, clip=]{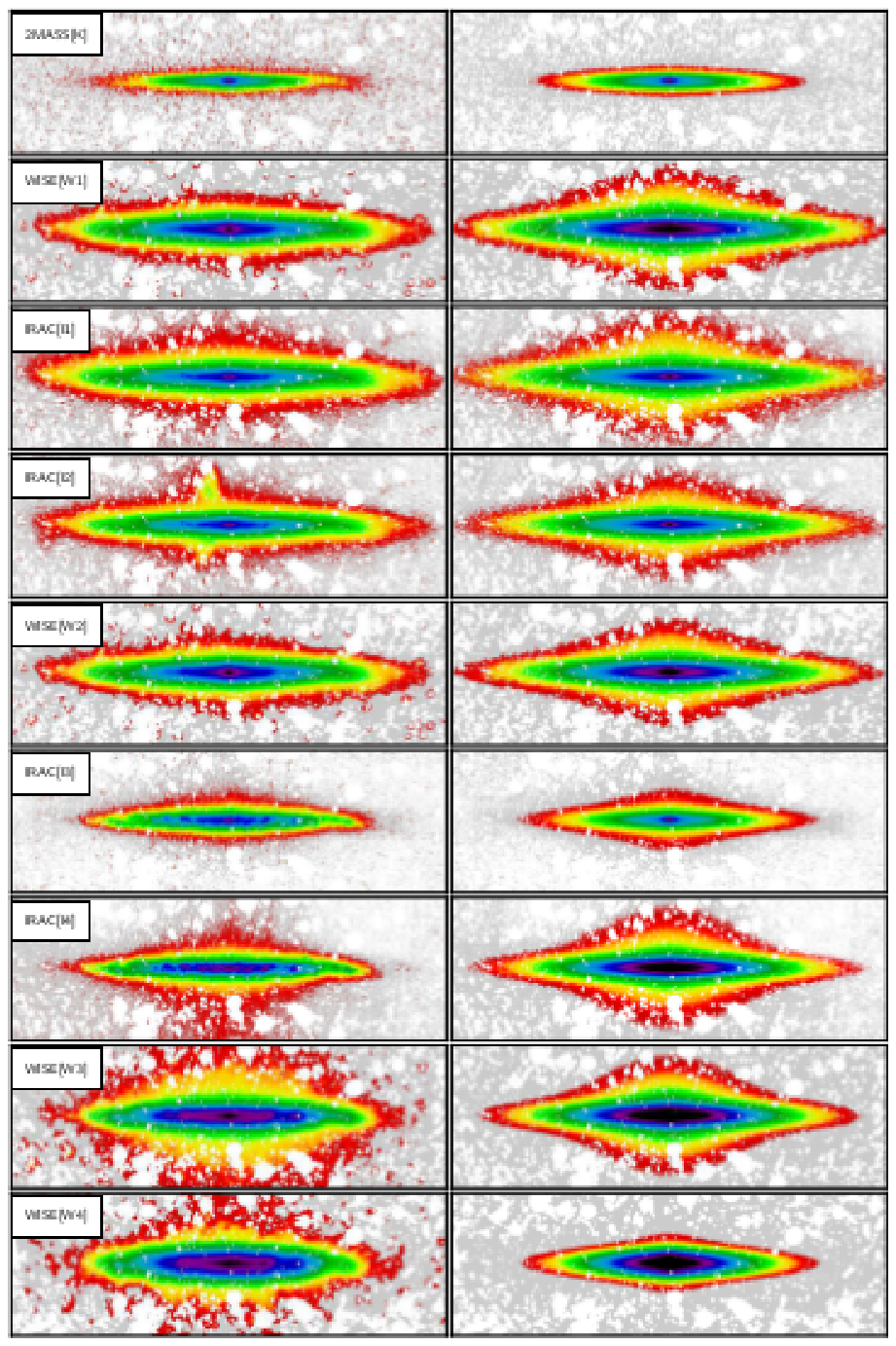}
\caption{(continued)}
\end{figure*}

\addtocounter{figure}{-1}
\begin{figure*}
\centering
\includegraphics[width=0.85\textwidth, angle=0, clip=]{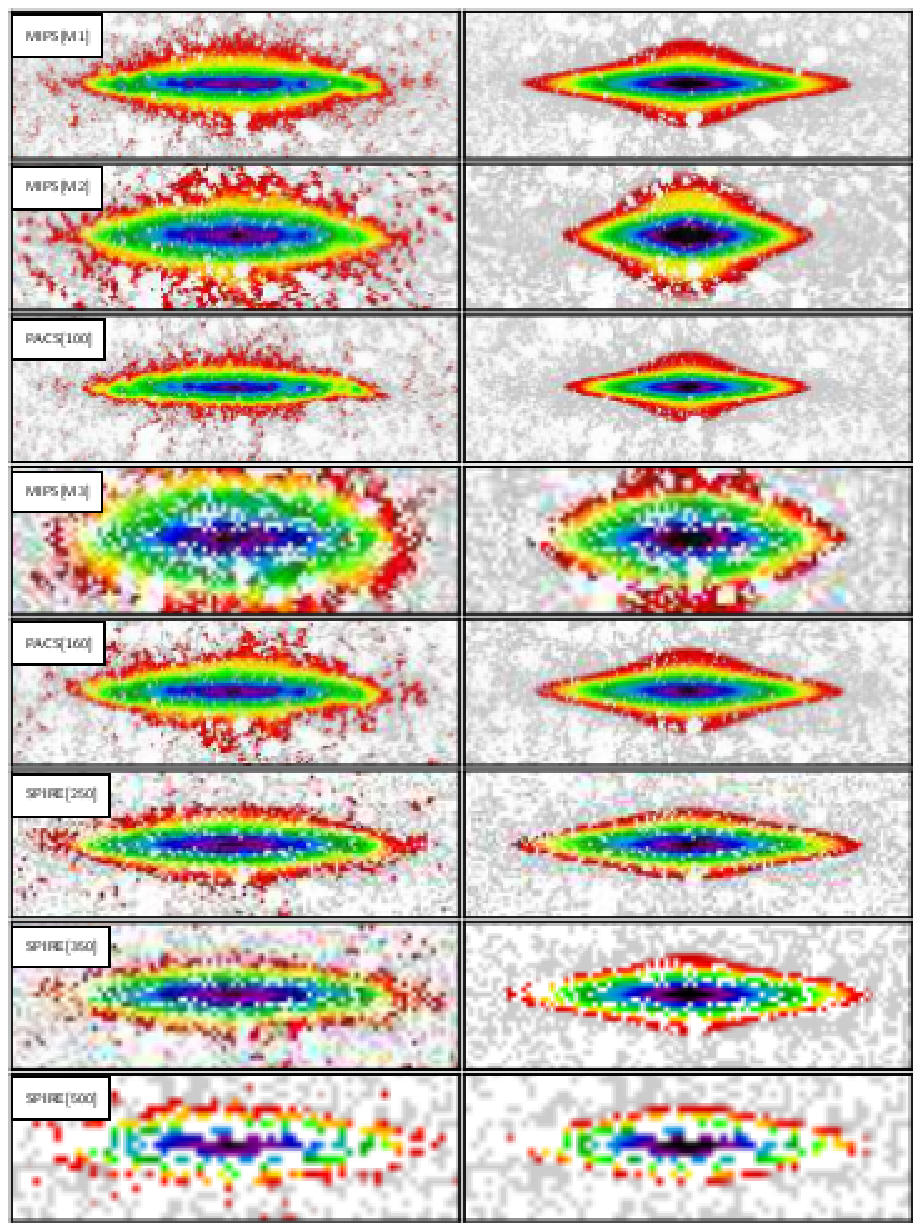}
\caption{(continued)}
\end{figure*}

\section{Global SED fitting}
\label{Appendix__SED}

\begin{figure*} 
  \centering
  \includegraphics[width=0.42\textwidth, height=0.3\textwidth]{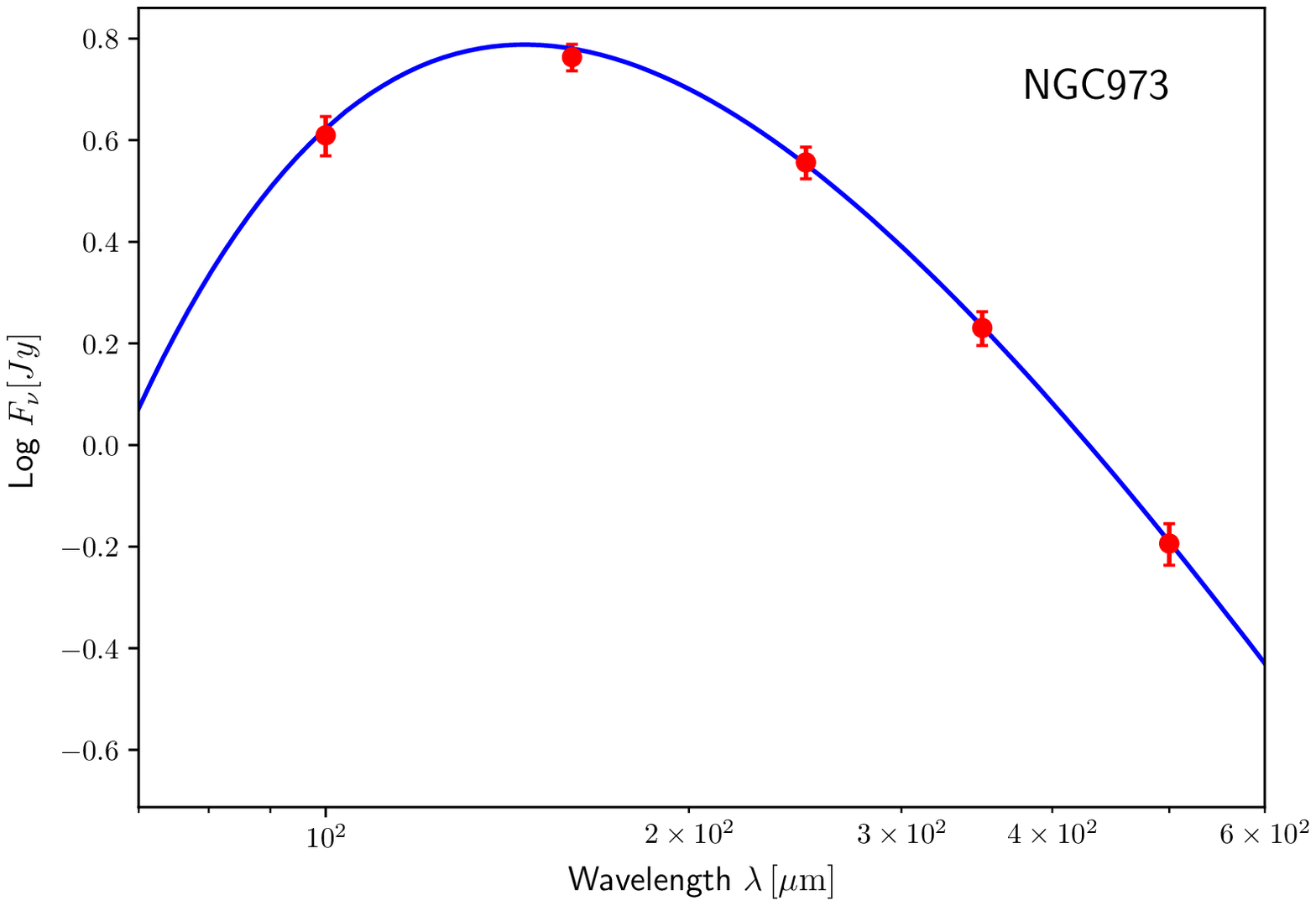}
  \includegraphics[width=0.42\textwidth, height=0.3\textwidth]{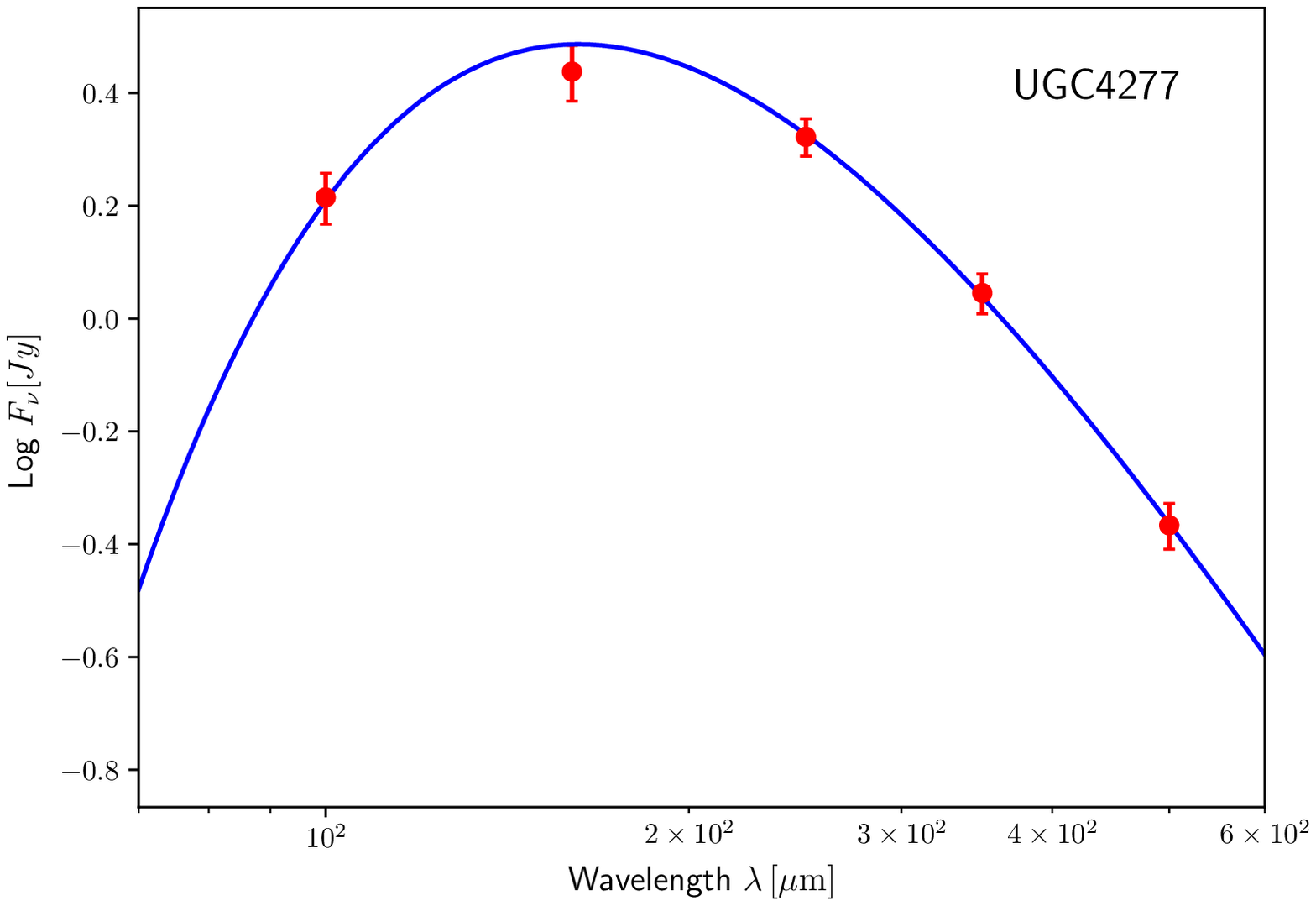}
  \includegraphics[width=0.42\textwidth, height=0.3\textwidth]{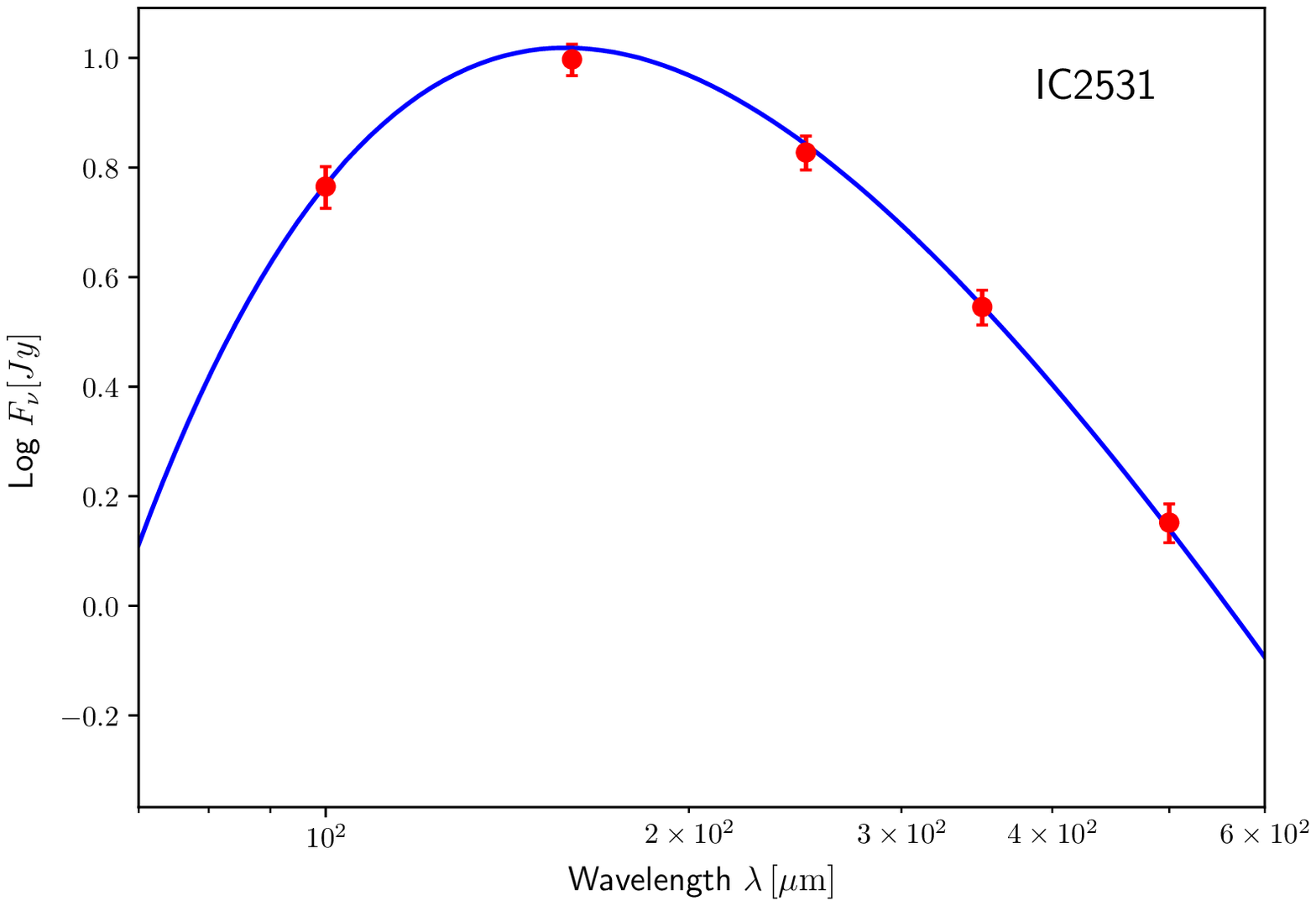} 
  \includegraphics[width=0.42\textwidth, height=0.3\textwidth]{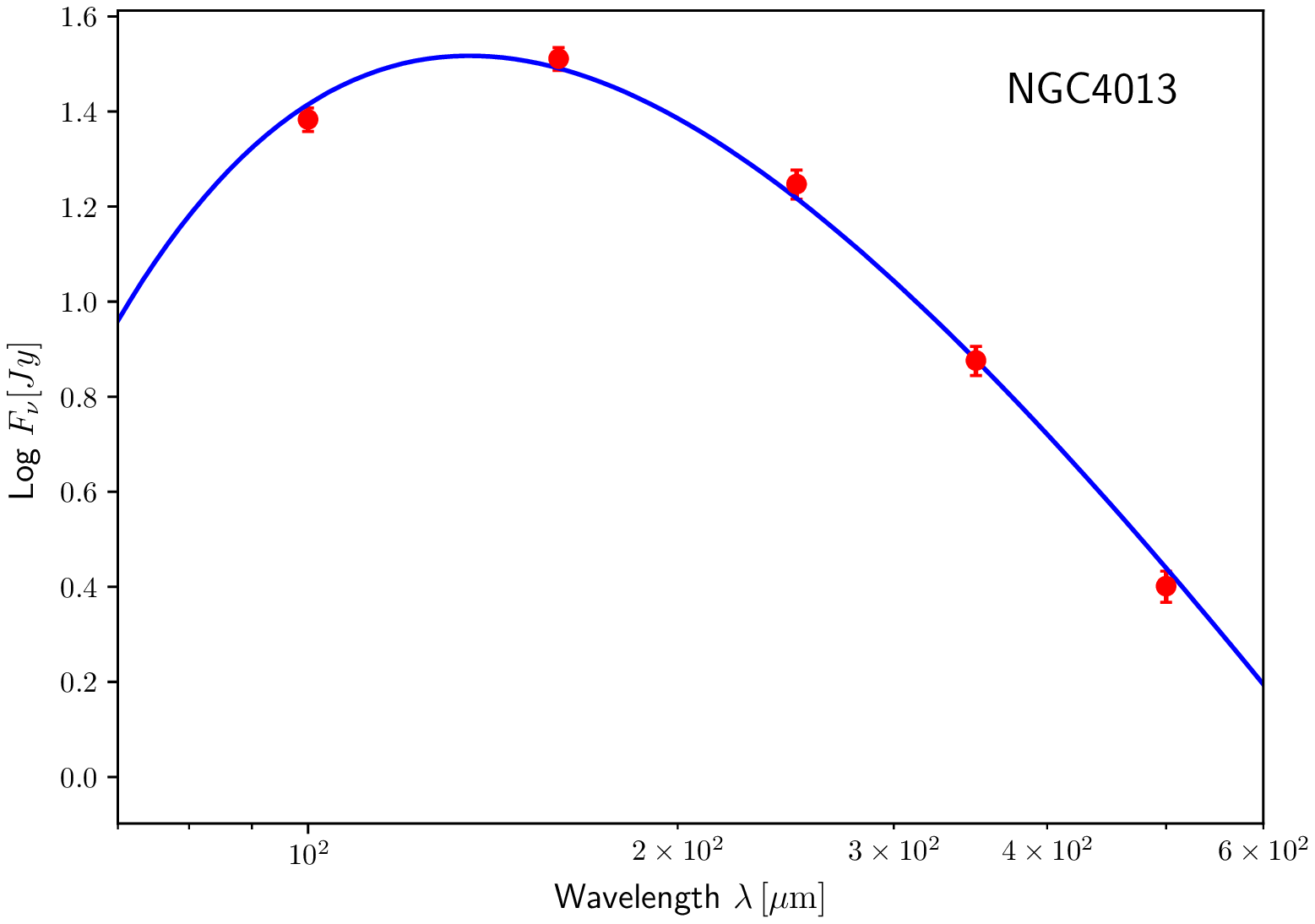}
  \includegraphics[width=0.42\textwidth, height=0.3\textwidth]{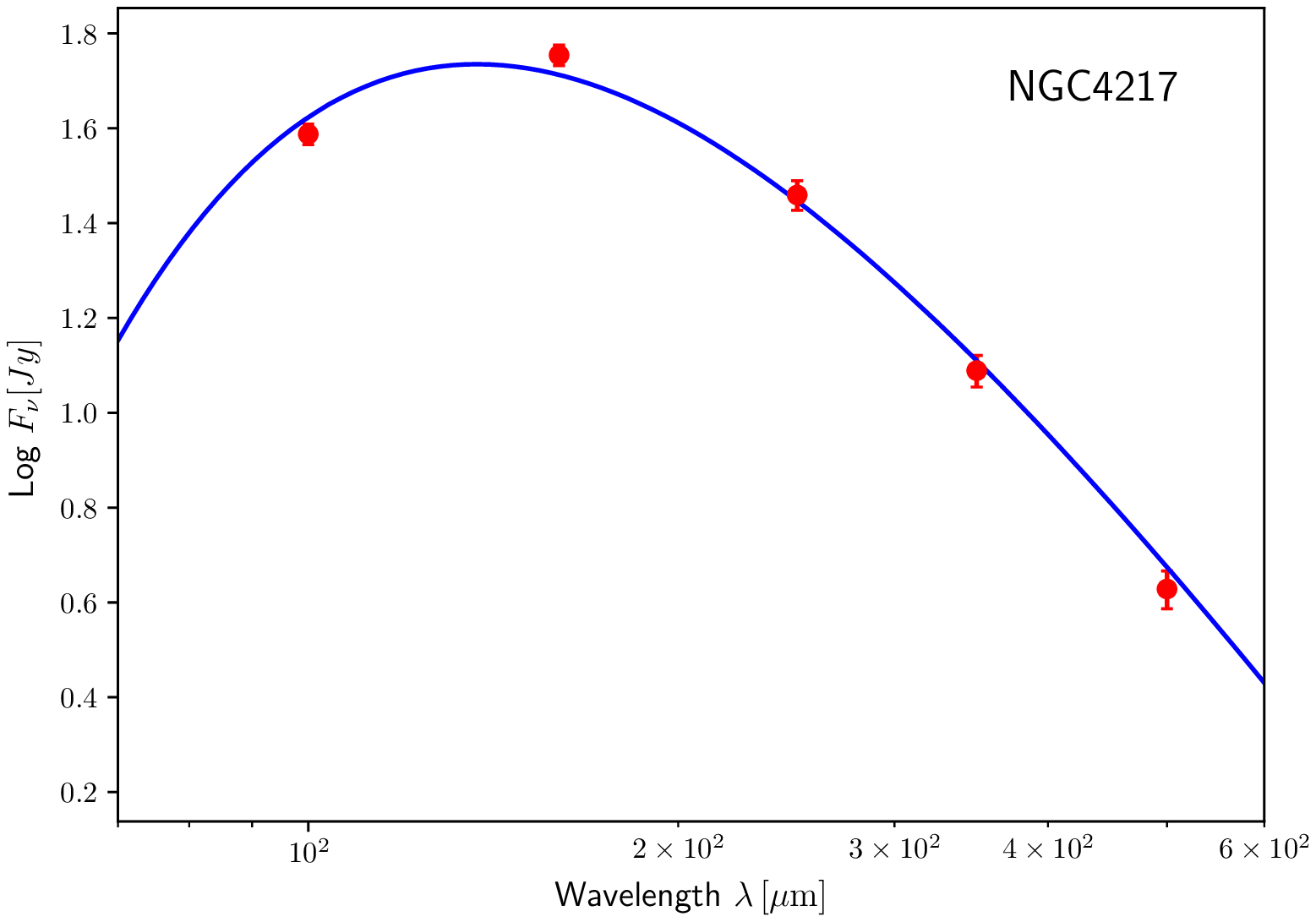}
  \includegraphics[width=0.42\textwidth, height=0.3\textwidth]{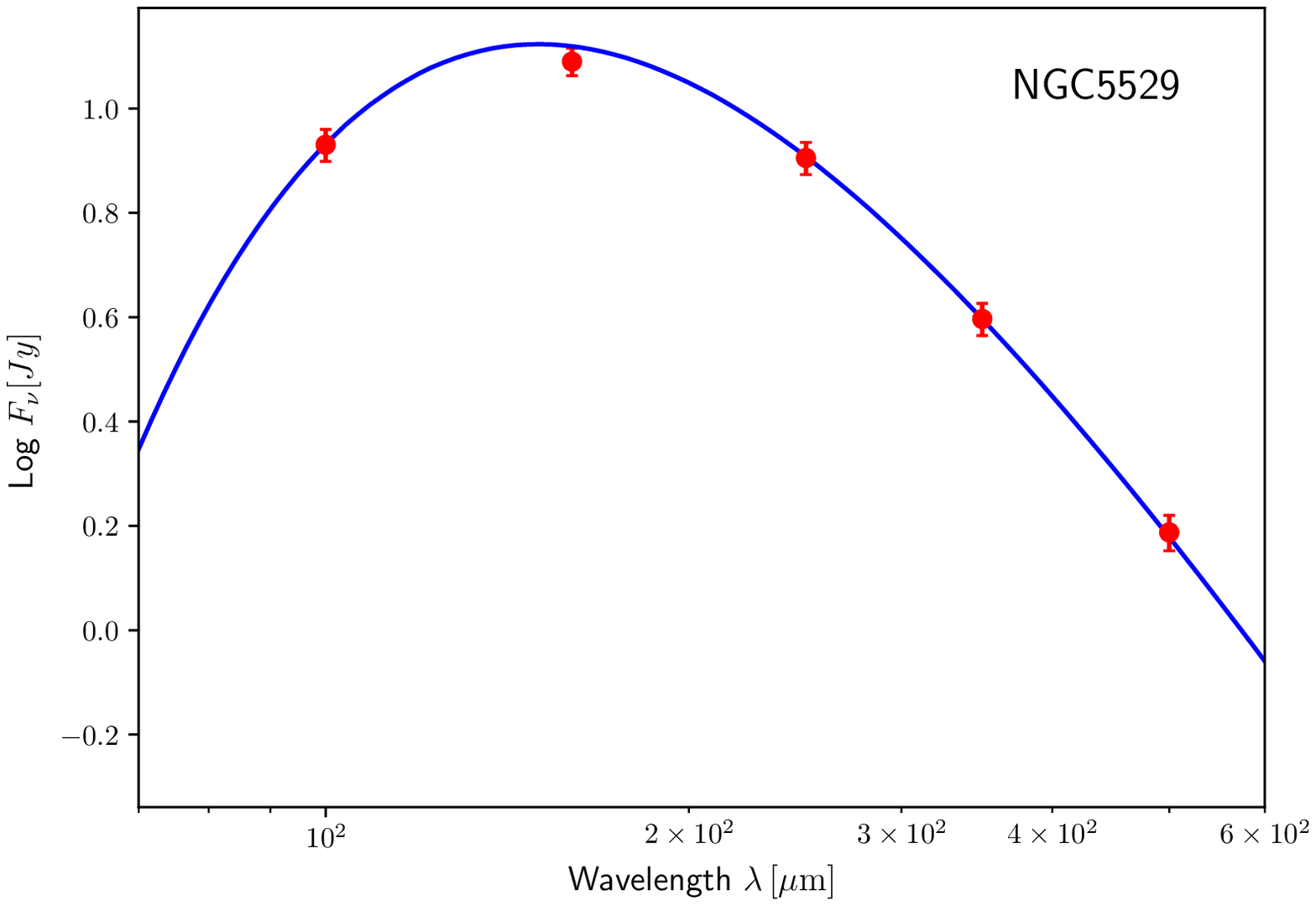}
  \includegraphics[width=0.42\textwidth, height=0.3\textwidth]{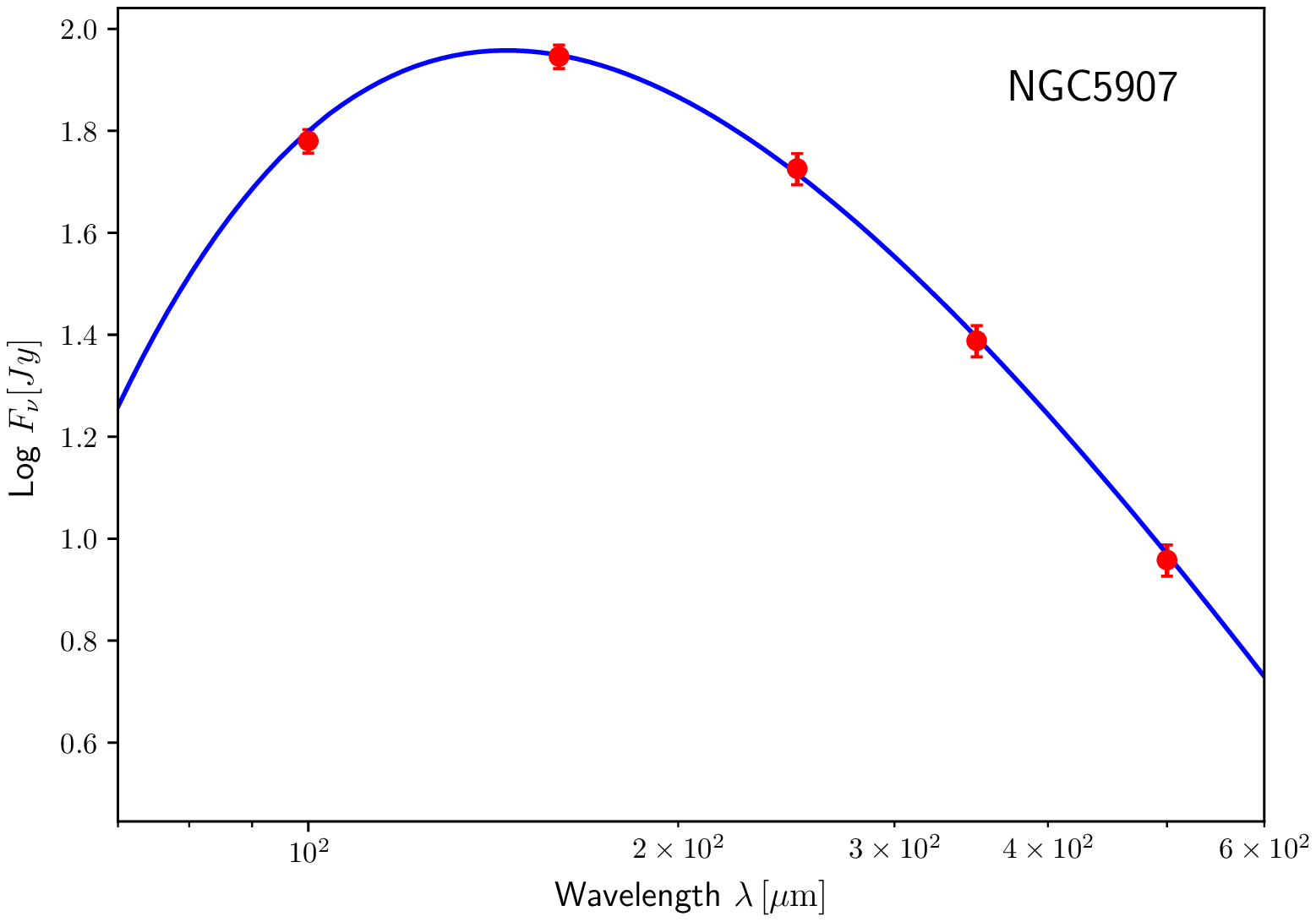}
  \caption{SED fitting to the \textit{Herschel} fluxes for all \textit{HER}OES
    galaxies. A modified, single temperature black-body model (the blue
    line) is adopted. }
  \label{ModBB.fig}
\end{figure*}

We determined the dust masses for the galaxies in our sample by
fitting a simple modified black-body model to the PACS and SPIRE data. The monochromatic luminosity, thus, can be written as\ 
\begin{equation}
  L_\nu(\lambda)
  =
  M_\mathrm{d}\,\kappa(\lambda_0)\,(\lambda_0/\lambda)^{\beta}\times4\pi B_\nu(\lambda,T_\mathrm{d})\,
\end{equation}
where $M_\mathrm{d}$ is the dust mass, $B_\nu$ is the Planck function,
$T_\mathrm{d}$ is the dust temperature, $\kappa(\lambda_0)$ is the dust emissivity at $\lambda_0$, and $\beta$ signifies a power-law dust
emissivity in the FIR/sub-mm wavelength range. In our model, we assume an emissivity of $\kappa(\lambda_0) = 0.64$~m$^2$~kg$^{-1}$ at $\lambda_0=250$~$\mu$m and $\beta=1.79$ adapted to the THEMIS model. The fits were done by using the {\sc idl} procedure \textit{mpcurvefit}.

The results of these modified black-body fits are shown in
Fig.~\ref{ModBB.fig} and listed in
Table~\ref{tab:SED_fitting}. 

\end{document}